\documentclass[fleqn,usenatbib]{mnras}

\usepackage[utf8]{inputenc}
\usepackage{newtxtext,newtxmath,amsmath}
\usepackage[T1]{fontenc}
\usepackage{graphicx}
\usepackage{enumitem}
\usepackage{xcolor}
\usepackage{xspace}
\usepackage{hyperref}
\usepackage{gensymb}
\usepackage{CJK}
\usepackage{tabularx}
\usepackage{orcidlink}
\usepackage{tikz}
\usepackage{float}
\usepackage{soul}

\usepackage{placeins} 

\usepackage{booktabs}
\usepackage{subcaption}

\usetikzlibrary{shapes.geometric, arrows.meta, positioning}
\hypersetup{
    unicode, 
    colorlinks=true,
    linkcolor=linkcolor,
    citecolor=linkcolor,
    filecolor=linkcolor,
    urlcolor=linkcolor,
}
\definecolor{linkcolor}{rgb}{0.0,0.3,0.5}

\newcommand\software[1]{\texttt{#1}}

\newcommand{\teff}{$T_{\text{eff}}$\xspace}
\newcommand{\vsini}{$v$ sin $i$\xspace}
\newcommand{\logg}{log $g$\xspace}
\newcommand{\kms}{\,km\,s$^{-1}$\xspace}

\newcommand{\changed}[1]{{\textcolor{black}{#1}}}
\newcommand{\changedtwo}[1]{{\textcolor{black}{#1}}}

\title[GALAH DR4 Binary Spectroscopy]{Double Trouble: How Unresolved Spectroscopic Binary Stars Skew GALAH DR4 Stellar Parameters}

\author[Lach, Buder, and Traven]{
Yani Lach,$^{1}$\thanks{E-mail: yani.lach@anu.edu.au}\orcidlink{0009-0003-9031-6307}
Sven Buder,$^{1,2}$\thanks{Australian Research Council DECRA Fellow}\orcidlink{0000-0002-4031-8553}
and Gregor Traven$^{3}$\orcidlink{0000-0002-1391-9097}
\\
$^{1}$Research School of Astronomy and Astrophysics, Australian National University, Canberra, ACT 2611, Australia\\
$^{2}$ARC Centre of Excellence for All Sky Astrophysics in 3 Dimensions (ASTRO 3D), Australia\\
$^{3}$Faculty of Mathematics and Physics, University of Ljubljana, Jadranska 19, 1000 Ljubljana, Slovenia
}

\date{Accepted 2026 July 30. Received 2026 July 28; in original form 2026 January 14}

\pubyear{\the\year{}}

\begin{document}
\label{firstpage}
\pagerange{\pageref{firstpage}--\pageref{lastpage}}
\maketitle

\begin{abstract}
Unresolved binary stars introduce systematic biases in stellar parameter estimates when analysed assuming a single star. Using $\sim$917{,}000 spectra from the Galactic Archaeology with HERMES (GALAH) Data Release 4 (DR4), we develop a binary spectral model that fits each observed spectrum as the superposition of two synthetic spectra generated from the same model architecture used by GALAH DR4. These generative spectral models take as input: effective temperature ($T_\mathrm{eff}$), surface gravity ($\log g$), iron abundance ([Fe/H]), projected rotation velocity ($v \sin i$), microturbulence velocity ($v_\mathrm{mic}$), and radial velocity ($v_\mathrm{rad}$). In our binary model, $T_\mathrm{eff}$ and $\log g$ are not fitted directly but are instead interpolated from PARSEC isochrones based on stellar mass and age, under the assumption that both components formed contemporaneously with the same compositions. We fit for the primary and secondary masses, a common age and [Fe/H], as well as individual $v \sin i$, $v_\mathrm{mic}$, and $v_\mathrm{rad}$ for each component. After iterative optimization and conservative quality cuts to ensure robust detections, we identify \changed{$\sim$12{,}911} candidate
binary systems. We quantify typical biases introduced by unresolved binarity: \teff shifts of \changed{$-142$ to $+200$\,K}, \logg shifts of \changed{$-0.070$ to $+0.35$\,dex}, and [Fe/H] shifts of \changed{$-0.055$ to $+0.13$\,dex}. Despite our conservative approach reducing the number of detections, we recover known trends of increasing binary fraction with higher primary mass and lower metallicity, thus reflecting the broad trends of binary populations. Accounting for unresolved binaries is essential for accurate stellar population modelling and for interpreting spectroscopic surveys--particularly in low-metallicity systems or populations with more massive stars.
\end{abstract}


\section{Introduction}
\label{sec:introduction}
Throughout the Milky Way's extensive history, the composition of elements comprising its stars has undergone continuous transformation \citep{freeman_new_2002}. This evolution is primarily attributed to the intricate interplay between the interstellar medium and successive generations of stars. The analysis of the elemental abundances within stars therefore serves as a direct record of the Milky Way's star formation and evolutionary history \citep{freeman_new_2002}. Multiple large-scale stellar spectroscopic surveys are underway to measure this chemical record for a significant number of stars from their spectra such as LAMOST, \textit{Gaia}, GALAH, APOGEE, \changed{and RAVE} \citep[For a comprehensive list see:][]{jofre_accuracy_2019}. To extract the chemical composition from stellar spectra, these spectroscopic surveys \changed{frequently make the implicit assumption} that all spectra are taken of single stars. However, this assumption stands in contrast to the actual process of star formation in the Milky Way, where multiplicity is the norm rather than the exception \citep{duchene_stellar_2013}.

\changed{Up to 50\% of stars within the Milky Way are thought to be in \textit{multiple} systems comprised of two or more stars \citep{batten_binary_1973,shu_star_1987,lada_stellar_2006,raghavan_survey_2010,duchene_stellar_2013,el-badry_gaias_2024}. Large stellar surveys inevitably observe many of these as unresolved pairs, modelling composite light as though it originates from a single source. The resulting stellar properties --- effective temperatures (\teff), surface gravities (\logg), and chemical abundances ([X/H]) --- will be systematically biased, with direct consequences for survey science: a 300\,K error in \teff{} propagates to a $\sim$0.3\,dex error in derived abundances \citep{jofre_accuracy_2019}, sufficient to confuse members of chemically distinct stellar populations and invalidate chemical tagging, given that open clusters exhibit intrinsic [Fe/H] scatters as small as 0.04\,dex \citep{ness_homogeneity_2022}. Biased surface gravities displace stars from their true evolutionary sequences, and systematically low metallicities distort the chemical maps used in galactic archaeology. Critically, this contamination is not uniform across a survey: the close binary fraction increases with primary mass and anti-correlates strongly with [Fe/H] \citep{badenes_stellar_2018, moe_close_2019, price-whelan_close_2020}, so binary-induced bias is most severe precisely in the metal-poor regime where surveys aim to trace the earliest enrichment events of the Galaxy. \citet{price-whelan_close_2020} demonstrated this directly using multi-epoch APOGEE radial velocities, finding that the single-lined binary (SB1) fraction rises significantly toward lower [Fe/H]. This underscores that any survey treating all stars as single will produce metallicity-dependent systematic errors. Beyond their impact on survey data quality, binary systems are also of fundamental astrophysical interest; their population properties --- including the mass ratio distribution ($q = m_2/m_1$), orbital separation, and the excess of near-equal-mass ``twin'' binaries ($q \gtrsim 0.95$; \citealt{el-badry_wide_2019}) --- encode the conditions of star formation and the physics of early stellar evolution \citep{raghavan_survey_2010, moe_mind_2017, offner_origin_2023}.}

\changed{The systematic biases introduced by unresolved binaries have been quantified in several spectroscopic surveys. \citet{el-badry_signatures_2018} predicted that fitting unresolved main-sequence binaries with single-star models in APOGEE introduces systematic offsets of $\sim$330\,K in \teff{} and $\sim$0.1\,dex in [Fe/H] and $\log g$, with the inferred $\log g$ shifted sufficiently low that affected stars scatter above the main sequence. \citet{schlesinger_binary_2010} also found biases introduced to SEGUE spectra: 80K in \teff{} and 0.1 dex in [Fe/H]. \citet{el-badry_discovery_2018} confirmed these predictions by applying a composite binary model directly to APOGEE spectra, recovering $\sim$3000 main-sequence binaries with biases consistent with those predicted. These effects are amplified in APOGEE relative to optical surveys such as GALAH because cooler secondaries contribute a larger fraction of the total flux in the near-infrared.}

\changed{Identifying binary stars within large stellar surveys remains a challenging task, particularly due to the changing binary frequency across different stellar types \citep{price-whelan_close_2020}. In major spectroscopic surveys, identification has been attempted using cross-correlation functions (CCFs) \citep[e.g.][]{fernandez_-sync_2017, merle_gaia-eso_2017}, radial-velocity variability cuts \citep{nidever_data_2015}, and residual-pattern diagnostics \citep{belokurov_unresolved_2020}. In a study of APOGEE spectra \cite{fernandez_-sync_2017} found CCFs to only reliably detect binaries with radial velocity differences exceeding 30\kms, although other studies have shown this to be resolution dependent \citep{merle_gaia-eso_2017, li_massratio_2022}.  These methods predominantly exclude suspect stars from catalogues rather than modelling their composite spectra. More sophisticated approaches overcome this: \citet{el-badry_discovery_2018} bypassed an identification step entirely by comparing a data-driven single-star model to a composite binary model, flagging binaries via a defined improvement fraction ($f_\mathrm{imp}$) to guard against fitting noise. \citet{kovalev_detection_2022, kovalev_detection_2024} identified over 12,000 SB2 candidates in LAMOST using a \mbox{\vsini-based} diagnostic sensitive below the CCF threshold; and \citet{traven_galah_2020} combined CCFs with a machine learning approach (t-Distributed Stochastic Neighbor Embedding) to produce a catalogue of 12,760 spectroscopic binaries in GALAH. The latter study classified a subset of spectra as ``problematic'' objects, which included cluster members, red giant branch stars, and variable stars, that can mimic binary signatures and should be treated with care.}



Given these challenges --- both identifying true binaries and avoiding misclassification of other objects --- it is crucial to not only improve detection methods but also to quantify the impact of unresolved binaries within a specific survey. \citet{el-badry_signatures_2018} made predictions for GALAH specifically, estimating biases on the order of 300\,K in \teff{} and 0.1\,dex in $\log g$ and [Fe/H] for solar-type stars — large enough to place stars in the wrong evolutionary category and to exceed the intrinsic chemical scatter of stellar clusters by a factor of $\sim$2.5. They also demonstrated that combining two single-star models into a composite binary model is preferable to using synthetic ab-initio binary spectra, as incomplete physics in the latter inadvertently compensates for model flaws and leads to unreliable identification. 

\changed{Beyond mitigating systematic biases, composite spectra also represent an opportunity: if the contributions of both stellar components can be disentangled reliably, unresolved binaries may provide access to stellar parameters and chemical abundances for both stars individually. However, before such analyses become possible, it is first necessary to establish robust identification and modelling techniques and to quantify how strongly ignored binarity biases the stellar parameters currently inferred under single-star assumptions. Understanding where such effects occur within GALAH DR4 is essential both for interpreting existing survey measurements and for enabling future component-level analyses of binary systems.}

Building on these insights from previous works, we present an analysis module to de-blend binary spectra and derive the stellar parameters of the component stars for the latest \changed{data release (DR4)} of the GALAH Survey \citep{buder_galah_2018,buder_galah_2021,buder_galah_2025} and present our results as a value-added catalogue to complement these efforts. We address the following key questions:
\begin{enumerate}[leftmargin=2em,labelwidth=0em]
    \item How can we \changed{robustly detect and model spectroscopic} binaries in GALAH DR4 \changed{with a focus on systems whose component spectra can be reliably separated}?
    \item What is the impact of \changed{ignored} binarity on stellar parameter estimations of GALAH DR4?
    \item \changed{Do the binary fraction trends recovered by our approach reproduce the known dependences on primary mass and metallicity seen in the literature?}
\end{enumerate}

The paper is structured as follows: Section~\ref{sec:data} describes the data and where it was sourced. In Section~\ref{sec:analysis} we present our methods and analysis. We then discuss our analysis in the context of our three research questions in Section~\ref{sec:discussion} and conclude in Section~\ref{sec:conclusions}.

\section{Data} \label{sec:data}

Our work makes use of spectroscopic data (Sec.~\ref{sec:data_observations}) from the GALAH survey's data release 4, from which we select a subset of binary candidates (Sec.~\ref{sec:data_selection}). We then use the spectrum model machinery from GALAH DR4 , which we only slightly adjust for our purposes (Sec.~\ref{sec:data_model}), to create  single star model spectra for our analysis. Finally, we also make use of the same set of stellar \textsc{PARSEC} isochrones used for GALAH DR4 (Sec.~\ref{sec:data_parsec})

\subsection{Observed stellar spectra} \label{sec:data_observations}

Spectroscopic data used for this study is taken from the latest release of the GAlactic Archaeology with Hermes (GALAH) survey, GALAH DR4 \citep{buder_galah_2025}. We took each spectrum from the \texttt{allspec} catalogue identified by its \textit{sobject\_id}. This includes multi-epoch spectra for some objects, however allows us to use the maximum available data while avoiding improper co-adding of unidentified binary spectra.  GALAH observes across four distinct spectral bands (blue: 4718 -- 4903\r{A}, green: 5649 -- 5873\r{A}, red: 6481 -- 6739\r{A}, and infrared: 7590 -- 7890\r{A}), with a resolution of approximately R $\approx$ 28,000, and an average signal-to-noise of 30-60 per pixel across the four CCDs. GALAH DR4 offers a catalogue over three times larger than the previous DR3 release \citep{buder_galah_2021} used by \citet{traven_galah_2020} and improvements to its reduction pipeline \citep{kos_galah_2017}. We are using the reduced spectra, which are flat-field corrected, continuum-normalised, have sky and telluric features removed and a barycentric correction applied. As the vast majority of objects in GALAH have also been observed by the \textit{Gaia} survey \citep{gaia_collaboration_gaia_2023}, we have also made use of their reported radial velocities, and the reduced unit weight error (RUWE) in our selection criteria (detailed below). 


\begin{figure}
    \centering
    \includegraphics[width=\columnwidth]{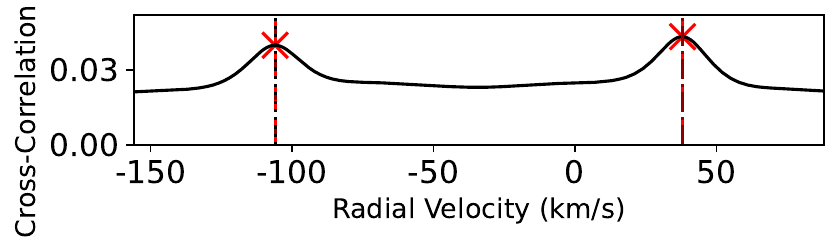}
    \vspace{0.5em} 
    \includegraphics[width=\columnwidth]{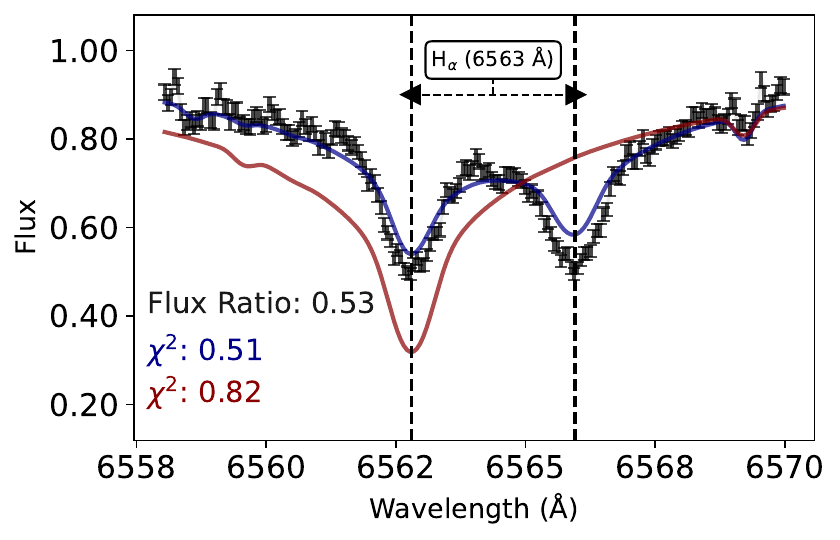}
    \caption{Top: Detection of binarity through the cross-correlation function (CCF) as a function of radial velocity for a candidate spectrum. Bottom: The same binary spectrum with lines fit using our binary model. Dotted lines show the same $H_\alpha$ line shifted due to the relative motion of the two stars. \textit{sobject\_ID}: 170603004601078.}
    \label{fig:example_binary}
\end{figure}

\subsection{Sample selection} \label{sec:data_selection}

\begin{figure}
\centering
\includegraphics[width=\columnwidth]{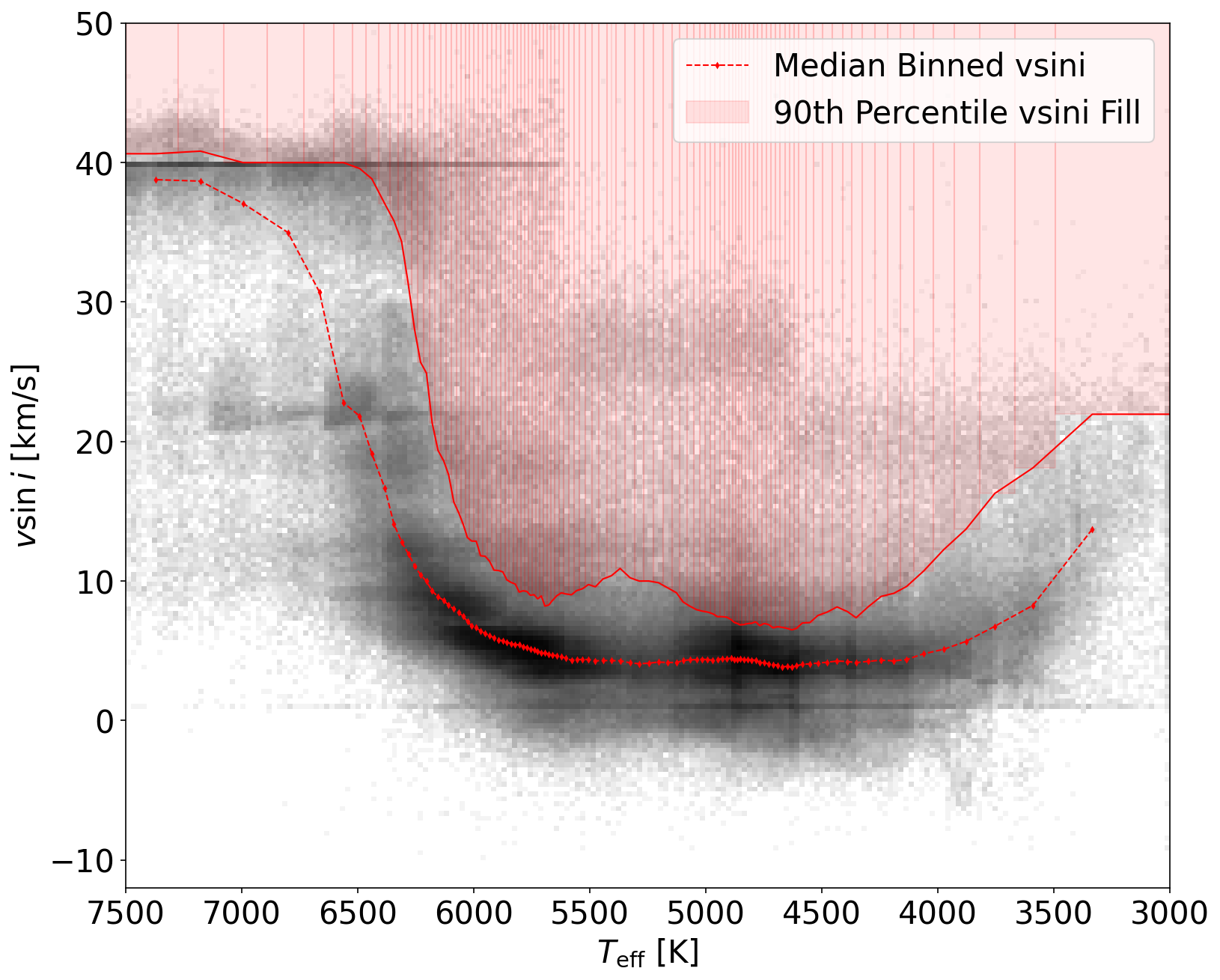}
\caption{Distribution of projected rotational velocity ($v \sin i$) as a function of effective temperature ($T_{\mathrm{eff}}$) for GALAH stars with $T_{\mathrm{eff}} < 7500\,\mathrm{K}$. The background histogram shows the density of all stars in the sample. Red points indicate the median $v \sin i$ in constant bins of 10{,}000 stars, with vertical dashed lines marking the bin edges. Red region highlight stars above the 90th percentile of $v \sin i$ within each bin. We take these to be the high $v \sin i$ outliers described by the 4th criterion in Section \ref{sec:data_selection}. We note that the plateau of points at $40\,\mathrm{km\,s^{-1}}$ arises due to GALAH DR4 neural networks being trained only below this region, causing the model to clip or underestimate higher rotational velocities above this range.}
\label{fig:teff_vs_vsini_binned}
\end{figure}

To determine high-quality possible binary systems, such as the clear candidate from Fig.~\ref{fig:example_binary}, before performing the analysis step we enforced selection criteria on the data:
\begin{enumerate}[leftmargin=2em,labelwidth=0em]
    \item We first included any objects where the number of components as flagged in GALAH DR4 ($rv\_comp\_nr$), was greater than 1. These have been discovered through the use of a CCF identifying blended spectra. As this method is also delivering some of the initial values of our analysis, we describe this in more detail in Sec.~\ref{sec:analysis_fitting}.
    \item Next we included systems where the difference in radial velocities between \textit{Gaia} DR3 and GALAH DR4 were greater than $\sim18\,\mathrm{km\,s^{-1}}$ (representing the 3$\sigma$ interval in the overall distribution of the differences in radial velocities between the surveys). This radial velocity difference is assumed to have occurred due to measuring an the binary system at different epochs, where the component stars were in different  orbital phases at the times of the \textit{Gaia} DR3 and GALAH DR4 observations. 
    \item High error values in the astrometric solution measured by the \textit{Gaia} satellite, with a reduced unit weight error (RUWE) above 1.4 were also used to flag sources with poor \textit{Gaia} astrometric fits, often indicative of unresolved binaries. \citep{belokurov_unresolved_2020,penoyre_astrometric_2022-1,penoyre_astrometric_2022,andrew_binary_2022,castro-ginard_gaia_2024, el-badry_gaias_2024}
    
    
    \item To create our sample selection of high \vsini stars, we took all objects beyond the 90th percentile in 10,000 star bins of \teff. The selection and resulting stars are shown in Fig.~\ref{fig:teff_vs_vsini_binned}, which shows that this effectively selects many objects along the binary MS in addition to a number of objects in the red clump region. Close binaries have been observed to have high rotational velocities (\vsini) \citep{hurley_evolution_2002, el-badry_gaias_2024}.
    Further, binary evolution can result in increased rotational velocity through ``spin-up'' mechanisms such as accretion from one star to the other through transfer of angular momentum \citep{packet_spin-up_1981}. Whilst single-stars can also have increased \vsini as a result of recent formation or advanced evolution, we include all stars with high rotations into our initial sample of \textit{potential} binaries. We note that marginally resolved line splitting can mimic rotational line broadening and address this in Section \ref{reject-criteria}.
    \item We included binary stars detected through other methods or available through other catalogues mutually observed by GALAH in our analysis and validation. These include the selection used by \cite{traven_galah_2020}. We also note that we have some overlap with a study of binary stars in the APOGEE survey \citep{majewski_apache_2017} made by \cite{price-whelan_close_2020}. However we had no mutual stars in our final sample. This is largely due to \cite{price-whelan_close_2020} using multi-epoch data of SB1 binaries whereas we attempt to detect SB2 binaries.
    \item Finally we also include spectra detected as having multiple peaks through the cross-correlation function used in this work. This is distinct from the CCF used in GALAH DR4 and is described in detail in Section\ref{sec:identificiation}.
\end{enumerate}

\begin{figure}
\centering
\includegraphics[width=\columnwidth]{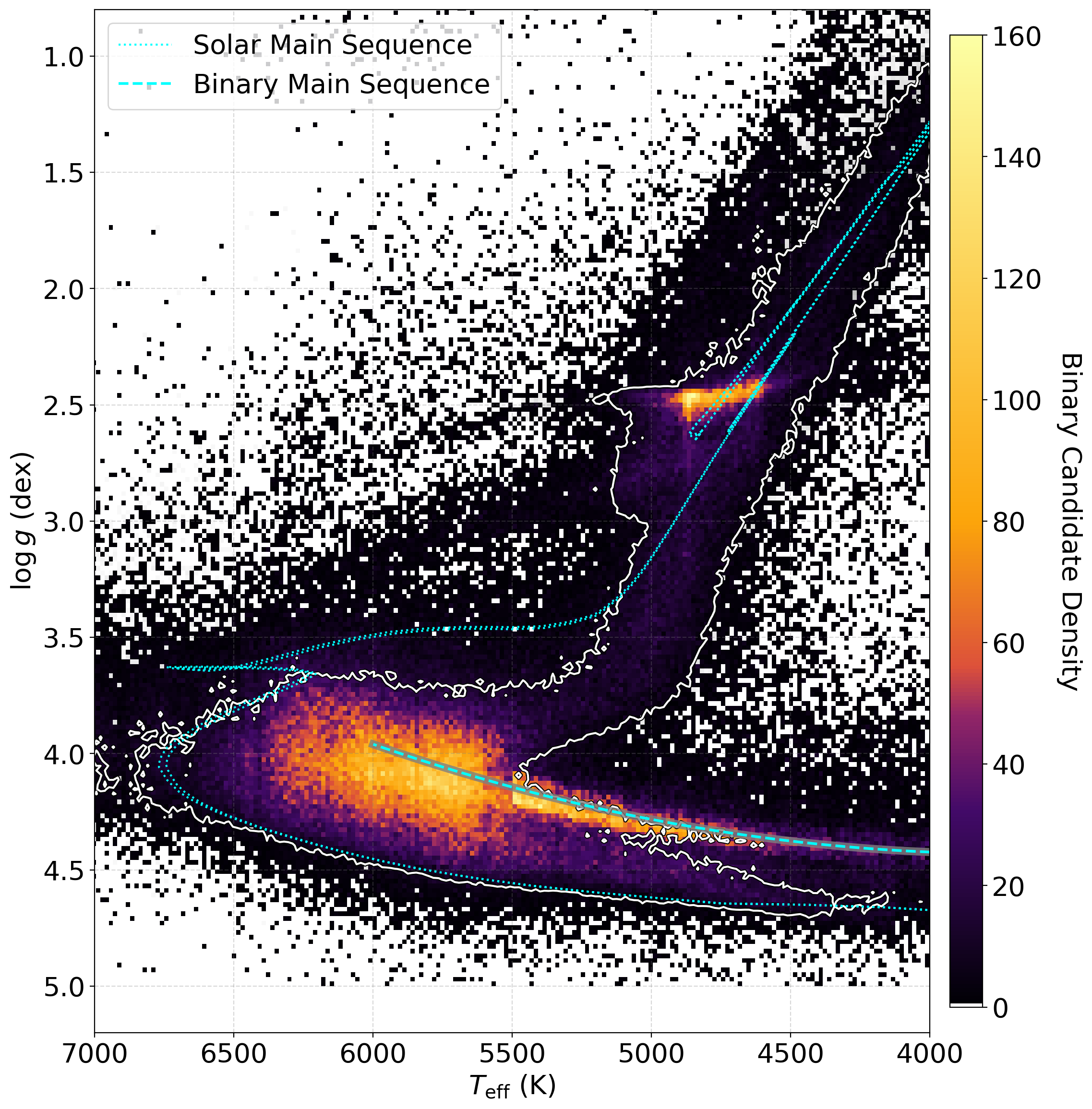}
\caption{Kiel diagram showing the distribution of binary spectra candidates as determined through our selection criteria. Bold cyan line shows the approximate fit to the binary main-sequence which was used to include candidates. Dotted cyan line shows a solar isochrone for reference. White outline is the contour including 90\% of all GALAH DR4 spectra.}
\label{fig:hrd_selection}
\end{figure}

The resulting number of spectra according to each selection criteria is shown in Table~\ref{tab:selection_criteria} and their distribution in the \textit{Gaia} DR3 $T_\mathrm{eff}$-$\log g$ diagram is shown in Fig.~\ref{fig:hrd_selection}. Together these criteria selected \changed{263,785} unique spectra.

Following the selection criteria we also made several quality cuts. These were made to reduce error and prevent the algorithm from perceiving random variation in the spectra or data artifacts as unresolved binarity. First, we required that each spectrum have a minimum S/N per pixel of 15 times the $1 \sigma$ deviation of the noise in each of the GALAH CCDs, thereby excluding spectra with low data quality in a single wavelength region. Second, we imposed a threshold on the combined S/N, calculated in quadrature across all CCDs, requiring $\mathrm{S/N}_\mathrm{combined} > 40$. This hybrid approach balances the need to avoid spectra with unusable regions which could bias parameter estimation. Additionally, we excluded stars with \teff below 4000~K. This cut is not related to noise, but rather reflects the limitations of both the stellar models and the analysis pipeline for very cool stars. For $T_{\text{eff}} < 4000$~K, the reliability of synthetic spectra and derived parameters decreases significantly due to the increasing dominance of molecular absorption features and less-complete molecular line lists. The clean sample thus consisted of 30,186 unique spectra.

\subsection{Modelling single star spectra} \label{sec:data_model}

To model single star spectra, we follow the same methodology of the single star analysis pipeline of GALAH DR4 \citep{buder_galah_2025}. We select the subset of 19\,725 single star synthetic spectra produced by the spectral synthesis code Spectroscopy Made Easy \citep{valenti_spectroscopy_1996,piskunov_spectroscopy_2017} for the GALAH DR4 analysis that sampled a range of global stellar parameters along the grid points of the \textsc{marcs} \citep{gustafsson_grid_2008} stellar atmosphere grids. The sampled parameters included effective temperature $3500 \leq T_\mathrm{eff}~/~\mathrm{K} \leq 7000$, surface gravity $0 \leq \log g \leq 5$, iron abundance $-2.0 \leq \mathrm{[Fe/H]} \leq 0.5$ with adjusted alpha-process elements following the \textsc{marcs} prescription but otherwise scaled-Solar abundances, microturbulence velocity $0.45 \leq v_\mathrm{mic}~/~\mathrm{km\,s^{-2}} < 3.5$, and were later broadened with different rotational velocities $1.5 \leq v \sin i~/~\mathrm{km\,s^{-2}} \leq 24$. For the exact sampling of parameters, we refer the reader to Sec.~3 of \citet{buder_galah_2025}.

We then also use the same neural network architecture and training algorithms as GALAH DR4, which are based on the spectrum interpolation framework of \textit{The Payne} \citep{ting_payne_2019}. The neural network connects $k$ stellar properties/labels $\boldsymbol{\ell}$ with the flux $f$ at each wavelength pixel $\lambda$ via
\begin{equation}
F_\lambda = w \cdot \mathrm{lReLU} \bigg( \tilde{w}_\lambda^i \cdot \mathrm{lReLU}  \Big( w^k_{\lambda i} \ell_k + b_{\lambda i} \Big) + \tilde{b} \bigg) + \bar{f}_\lambda,
\label{eq:neural_network_function}
\end{equation}
which encapsulates the so-called layers of a neural network with $i = 300$ neurons with weights $w$ and biases $b$ as well as a leaky Rectified Linear Unit ($\mathrm{lReLU}$)
\begin{equation}
    \mathrm{lReLU} (x) =  \begin{cases}
        x \qquad &x \geq 0 \\
        0.01 x \qquad &x < 0.
    \end{cases}
\end{equation}

After training (optimized for $10^4$ steps with a mean absolute error loss function and validated against a held-out set), the network can generate a synthetic single-star spectrum for any set of stellar parameters, including a radial velocity shift $v_\mathrm{rad}$:
\begin{equation}
F_\lambda = NN(T_\mathrm{eff}, \log g, \mathrm{[Fe/H]}, v_\mathrm{mic}, v \sin i, v_\mathrm{rad}). \label{eq:neural_network_model}
\end{equation}

To construct a binary spectrum, we simply call this function twice with two different sets of stellar parameters and combine the resulting spectra with a weighting factor.

\subsection{Stellar Isochrones} \label{sec:data_parsec}

We use PAdova and tRieste Stellar Evolutionary Code (PARSEC) isochrones \citep{nguyen_parsec_2022} in our model to restrict exploration of the parameter space to physical regions (see Section \ref{sec:analysis}). We use these isochrones as they are also used by GALAH DR4 and offer a comprehensive, up-to-date, and physically detailed treatment of stellar evolution phases most suited to the parameter space of stars within the catalogue. 

We use the same set of PARSEC isochrones as GALAH DR4 to interpolate over. These cover logarithmic ages in the range $\log(\tau/\mathrm{Gyr}) = 8.00$ to $10.18$ in steps of $0.01$, and metallicities of $[\mathrm{M}/\mathrm{H}] = -2.75$ to $-0.75$ in steps of $0.25$, as well as $[\mathrm{M}/\mathrm{H}] = -0.6$ to $0.7$ in steps of $0.1$. From these, we exclude hot stars with $T_\mathrm{eff} > 10{,}000\,\mathrm{K}$, as well as extremely evolved white dwarfs and highly luminous giant stars (defined as $\log g > 6$ or $J-K_S > 2\,\mathrm{mag}$), as these fall outside the range covered by our spectroscopic pipeline.

To obtain stellar parameters for arbitrary input values of mass, age, and metallicity, we perform a local trilinear interpolation over the PARSEC isochrone tables. Specifically, for each query point $(M_\mathrm{initial}, \log(\tau/\mathrm{Gyr}), [\mathrm{M}/\mathrm{H}])$, we identify the closest grid points in age, mass, and metallicity, and extract all isochrone entries within a narrow window around age (0.1 Gyr), $log$ g (0.01 dex), and mass (0.05 M$_\odot$). We interpolate the initial mass ($M_\mathrm{init}$) rather than the present-day mass ($M$), as it is degenerate with age and does not correspond uniquely to a single row in the isochrone table. This local interpolation approach was adopted, as full isochrone interpolation would otherwise create a major computational bottleneck in our fitting routine. To interface the neural network iron abundance [Fe/H] output with PARSEC isochrones tabulated in [M/H], we apply the GALAH DR3 metallicity conversion \citep{buder_galah_2021} which accounts for empirical $\alpha$-enhancement trends: 
$[\alpha/\mathrm{Fe}] = +0.4$ dex for $[\mathrm{Fe/H}] \leq -1$, 
$[\alpha/\mathrm{Fe}] = 0$ dex for $[\mathrm{Fe/H}] \geq 0$, 
and linearly interpolated in between as $[\alpha/\mathrm{Fe}] = -0.4 \times [\mathrm{Fe/H}]$. 


To ensure robust interpolation, we dynamically adjust the size of the local isochrone window: if the initial selection is too small to construct a valid interpolator, we iteratively expand the age, metallicity, and mass ranges in small increments, always checking that the expanded region remains within the bounds of the isochrone grid. This adaptive process continues until a sufficiently large local sample is obtained, or until a maximum number of expansions is reached. This approach guarantees that the interpolator is neither under-constrained (too few points) nor computationally expensive (too many points), and that it is always constructed from valid regions of the isochrone grid.

We then use a \software{scipy.interpolate} \citep{virtanen_scipy_2020} to interpolate the stellar parameters as a function of logarithmic age, $\log(\tau/\mathrm{Gyr})$, initial mass, $M_\mathrm{init}$ (in $M_\odot$), and metallicity, $[\mathrm{M}/\mathrm{H}]$ (in dex), at the desired location in parameter space. If multiple queries are made within the same local region, we cache the interpolator for computational efficiency\footnote{This decreased the interpolation time by a factor of over $10^3$ compared to creating a new local interpolator.}. The output parameters are thus given by

\begin{align}
    & T_{\mathrm{eff},i},\, \log g_i,\, \log L_i,\, \ldots,\, \mathbf{S}_i
     \nonumber \\
    &\qquad = \mathrm{PARSEC}\left(\log(\tau),\, M_{\mathrm{init},i},\, [\mathrm{M}/\mathrm{H}]\right), \quad i = 1, 2
\label{eq:parsec}
\end{align}

where the function on the right-hand side denotes our trilinear interpolation over the PARSEC isochrone grid for each of the binary components ($i=1,2$). Here, \teff (in K),  $\log g$ (in dex), and $\log(\textrm{L})$ (in $L_\odot$), are the parameters we extract and employ during fitting. $\mathbf{S}$ denotes the full set of 13 parameters interpolated from the isochrone table that we extract once after fitting has completed. These include the M$_{\textrm{present}}$ and magnitude across different bands, which we report as additional outputs requiring no further computation. 

\section{Analysis} \label{sec:analysis}

In this section, we describe the methodology used to identify binary star systems, model their spectra, and evaluate the performance of different optimization techniques. The analysis consists of two main parts: determining the component RVs for the possible systems through cross-correlation (Section~\ref{sec:analysis_fitting}) and constructing and optimizing synthetic binary star models (Section~\ref{sec:modeling}). We refer to binaries identified through our sample selection process as \textit{possible} binary systems, whilst we refer to binaries found through our analysis process as \textit{candidate} binary systems.

\subsection{RV determination through Cross-Correlation} \label{sec:identificiation}


From the initial sample of possible binary systems we attempted to detect radial velocities through the use of a CCF on all GALAH DR4 spectra (see Fig.~\ref{fig:example_binary}). This was initially attempted using the native \software{scipy.signal.correlate} function \citep{virtanen_scipy_2020}, however after some experimentation we developed a bespoke CCF that could process faster and was more suited to analysis of the spectral data. The primary improvements offered by our CCF over the DR4 pipeline include interpolation for mismatched sampling, enhanced peak detection with prominence criteria and normalisation, automated Gaussian peak fitting, and native support for radial velocity shifting. Peak detection in the residuals of the CCF were found using the \software{scipy.signal.find\_peaks} function \citep{virtanen_scipy_2020}. The prominence and height input to the function were set to 10\% of the highest detected peak. These values were found through experimentation and were adjusted to scale to the entire dataset through normalisation of the residuals. The CCF was applied to each spectrum using the best single-star synthetic spectrum offset by the best-fit RV determined by GALAH. We explored RV offsets ranging from -500\kms to 500\kms in steps of 1\kms. This range was selected to account for both the inherent RV of the system and the maximum physical orbital velocities of bound stars.

To identify the secondary RV we also tested a brute-force binary-template RV scan, in which a complete binary model (two DR4 single-star spectra with identical stellar parameters) was constructed and the RV of the secondary stepped from $-500$ to $+500$ km\,s$^{-1}$ in 1 km\,s$^{-1}$ increments, optimizing only the flux ratio at each step. This method was substantially more computationally expensive than the simple CCF-based search and, because the secondary was constrained to share the primary's stellar parameters, did not recover RV separations more reliably. We therefore adopted the simpler CCF implementation for the final analysis.

\subsection{Initial Parameter Assignment} \label{sec:analysis_fitting}

The parameters for each component of the binary model were initialized with the same parameters as the single-fit model from GALAH DR4. We thus assume the second component of the binary to be highly similar, however, the fitting process allows thorough exploration of the parameter space if the secondary is sufficiently different. 


For the radial velocity of the second component (RV$_2$) we used: the radial velocity as determined through our CCF, the RV$_2$ value from GALAH DR4 (if available), or if no RV$_2$ could be obtained, we estimated RV$_2$ by reflecting the primary velocity (RV$_1$) about the systemic velocity as measured by \textit{Gaia} ($\mathrm{RV}_{\text{Gaia}}$), i.e., $\mathrm{RV}_2 = \mathrm{RV}_1 + 2\times(\mathrm{RV}_{\text{Gaia}} - \mathrm{RV}_1)$. This approach leverages the \textit{Gaia} DR3 radial velocity measurement as the approximate average measurement of the system's centre-of-mass velocity. As we discuss below, this assumption does not usually hold for double-lined binaries --- single-star fits to binary spectra do not typically represent a ``middle-ground'' between both stars, rather they align with one component --- meaning the \textit{Gaia} RV or GALAH RV is from one of the components and not the mean system velocity (Section \ref{sec:discussion_parameter_changes_stellar_parameters}). Finally, if a \textit{Gaia} RV was unavailable ($\sim110,000$ spectra) we simply expanded the bounds on RV$_2$ to $\pm150\mathrm{km\,s}^{-1}$ about RV$_1$ (the $3\sigma$ limit in the distribution of GALAH DR4 $\Delta \textrm{RV}$s) and omitted the curve\_fit process from the analysis. Instead fitting directly with particle swarm optimization (PSO - described in detail in Section \ref{sec:algo-selection}) to explore all possible RVs and converge on the best fit, as no initial guess is required for parameters using PSO.

\subsection{Modelling} \label{sec:modeling}

To produce a synthetic model for binary stars we combined two singe-star spectra to produce a binary model. 


The choice to model binary systems through the superposition of two single-star models rather than retraining the existing GALAH neural network or training a new network was driven by considerations of dimensionality. The single star model (Eq.~\ref{eq:neural_network_model}) employs six parameters. A neural network capable of modelling a binary spectrum would double the number of parameters to twelve. Using the already available neural networks of GALAH DR4 would have created significant computational overhead, as these were computed for smaller boxes in Teff, logg, and [Fe/H] and would have to be read in whenever one of the components would have been outside the training range of the neural networks. We thus favoured the superposition approach employing one neural network covering the whole parameter space presumed efficiency and practicality.
\subsubsection{Modelling Binary Spectra}

Previous works have modelled binary systems as a combination of two or more single-star synthetic spectra through either a flux, luminosity \citep{traven_galah_2020}, or mass ratio \citep{el-badry_signatures_2018}. We follow this approach with a combination of two spectra through their weighting ratio ($R$): 

\begin{equation}
    F_{\text{binary}}(\lambda) = RF_1(\lambda) + (1 - R) F_2(\lambda).
    \label{eq:binary_flux}
\end{equation}

The fluxes of each component ($F_1, F_2$) are produced using the neural network described by Eq.~\ref{eq:neural_network_function}. In our initial implementation, we fit the binary spectrum by treating the flux ratio ($R$), as well as the stellar parameters $T_\mathrm{eff}$ and $\log g$ of each component, as free parameters. However, we found that this approach often led to unphysical or non-astrophysical solutions: the fitting routine could converge on parameter combinations that do not correspond to any real stars, especially for the secondary component when its contribution was small.

To address this, we transitioned to a physically-motivated approach in which the stellar parameters are constrained to lie on theoretical isochrones. Specifically, we fit for the initial mass, age, and [Fe/H] of each component, and interpolate \teff, \logg, and luminosity from the PARSEC isochrones (Eq.~\ref{eq:parsec}). The luminosities are then used to calculate the flux ratio,
\begin{equation}
    R_L = \frac{L_1}{L_1 + L_2},
\end{equation}
which replaces $R$ in Eq.~\ref{eq:binary_flux}. This ensures that the fitted parameters correspond to physically plausible stars and that the flux ratio is consistent with stellar evolution models. The binary model (Eq.~\ref{eq:binary_flux}) was then optimized with respect to the free parameters and boundary conditions listed in Table~\ref{tab:boundries}, with all parameters initialised to the GALAH DR4 single-star values for both components. Boundary conditions were chosen according to the distribution of these parameters in the GALAH single-star catalogue. We assume that both components share the same age and [Fe/H] due to their mutual formation history.

We initially enforced a boundary on $R_L$, constraining it between 0.2 and 0.8 to avoid fitting highly disparate systems. However, we found that this constraint often led to less interpretable results. Specifically, the $R_L$ distribution became artificially skewed toward the imposed limits, suggesting that the fitting process was being forced to interpret single-star spectra as binaries --- resulting in false positives. When we removed the constraint and allowed $R_L$ to vary freely, the fit quality improved, and although many resulting fits had extreme flux ratios (e.g., > 0.95), visual inspection allowed these to be easily identified as false positives which proved beneficial, as they could be more easily cut from the final catalogue.

\begin{table}
\renewcommand{\arraystretch}{1.35} 
\centering
\caption{Boundary conditions enforced on the parameters during the optimization process. Note that boundaries for $M_\text{[1,2]}$, $\text{[Fe/H]}$, and age are set by the isochrone.}
\normalsize
\begin{tabular}{lll}
\hline
\textbf{Parameter} & \textbf{Boundary} & \textbf{Unit} \\ \hline
$R_\text{L}$ & 0 -- 1 &  \\
$T_\text{eff[1,2]}$ & 4000 -- 8000 & K \\
$\log g_\text{[1,2]}$ & 0.5 -- 5.0 & dex \\
$M_\text{[1,2]}$ & 0.09 -- 5.348 & $\text{M}_\odot$ \\
$\text{[Fe/H]}$ & -2.19174 -- 0.69525 & dex \\
$\text{age}$ & $0.1$ -- $14.79$ & $\text{Gyr}$ \\
$v_{r\text{[2]}}$ & -250 -- 250 &\kms$^{-1}$ \\
$v_\text{mic[1,2]}$ & 0 -- 4 &\kms$^{-1}$ \\
$v$sin$i_\text{[1,2]}$ & 0 -- 30 &\kms$^{-1}$ \\
\hline
\end{tabular}
\label{tab:boundries}
\end{table}

\subsubsection{Optimization Algorithm Selection} \label{sec:algo-selection}

We ultimately found \software{scipy.optimize.curve\_fit} \citep[][scipy]{virtanen_scipy_2020} as implemented in the GALAH DR4 pipeline to inadequately explore the parameter space, often prematurely converging on solutions that were poor fits or failing to retain the best intermediate fit throughout the optimization. This behaviour persisted despite changing input parameters to the fitting function and additional experimentation. While this behaviour could be prevented for a single case, there were no parameters capable of effectively optimizing the broader dataset.

As a result, we tested various other optimization techniques to identify the method that most robustly approached the global minimum, or at least avoided clearly suboptimal local minima, across the dataset. To quantify the goodness of fits, we used the same $\chi^2$ statistic as \cite{buder_galah_2025}, which is calculated as the uncertainty ($\sigma$) weighted absolute difference of observed and modelled flux ($F_\mathrm{obs}$ and $F_\mathrm{model}$) across all $i$ of the $N$ pixels of the spectrum:
\begin{equation}
\chi^2 = \frac{1}{N} \sum_{i=1}^{N} \frac{\left| F^{(i)}_{\mathrm{obs}} - F^{(i)}_{\mathrm{model}} \right|}{\sigma^{(i)}}.
\end{equation}


Based on a comparison of the resulting $\chi^2$ distributions (Fig. \ref{fig:rchi2_comparisons}), PSO was found to most consistently avoid poorly-fitting local minima across the dataset. We therefore adopted PSO as implemented by \software{pyswarms} as our primary modelling technique. The PSO search bounds are set adaptively based on the quality of the GALAH DR4 single-star parameters as an initial guess. Since both components are initialised to the same GALAH DR4 single-star values (Section~\ref{sec:analysis_fitting}), evaluating the binary model at this point tests the fit of an identical-twin configuration — the natural null hypothesis for an unresolved binary. If this configuration already fits the spectrum well ($\tilde{\chi}^2 < 2$), the true binary solution with differing components is likely nearby in parameter space, and the PSO search is restricted to within $\pm20\%$ of those normalised initial values; otherwise the full normalised parameter space is searched. Implementation details, algorithm comparisons, and hyperparameter selection are given in Appendix~\ref{sec:PSO_details}.

\subsection{Rejection criteria}\label{reject-criteria}

To ensure a clean and reliable sample of binary systems, we applied a series of quality cuts to the results. The criteria are summarised in Table~\ref{tab:data_cuts} and described below.

First, we required the $\chi2$ value of the fit to be less than 3, in order to exclude poorly fitting models and reduce contamination from spurious solutions. This is also sufficiently broad to allow for cases where the spectra are poorly fit in particular regions. This value was chosen as it contained 95\% of all fits with the remaining being fit extremely poorly.
We also applied a luminosity ratio cut (${R_L} < 0.95$) to exclude cases where the system is effectively single-lined, either because only one component is present or because one star overwhelmingly dominates the flux. In such cases, binary decomposition becomes highly degenerate and unreliable and is effectively indistinguishable from a single-star. Examples of both of these cases are shown in Fig.~\ref{fig:binary_false_detections}. In these cases allowing a second component can trivially reduce $\chi^2$ for some spectra by absorbing residuals or noise into an unphysical secondary; our rejection criteria (Table \ref{tab:data_cuts}) explicitly remove such cases despite their lower $\chi^2$. \changedtwo{Systems where the two components are recovered with near-identical radial velocities and comparable flux contributions are also rejected due to failing the \vsini threshold.} \changed{Conversely, we checked if binary fits could produce a higher $\chi^2$ than the GALAH single-star fit due to the isochrone and abundance constraints, making our binary model effectively less flexible in certain parts of parameter space than the original DR4 fitting scheme. Visual inspection of these cases revealed no obvious false-negatives. However, we would suggest some caution when using results where the single-fit $\chi^2$ and binary-fit $\chi^2$ are close. As noted by \cite{el-badry_discovery_2018}\footnote{See their appendix for details.}, a binary model constructed through superposition is always capable of producing an equally good or better fit than the single-star model, since the single-star case is simply a special case of the binary model. In a future study, we would suggest applying an information criterion such as the Bayesian information criterion to more rigorously penalise the additional free parameters of the binary model, allowing a statistically principled comparison between single- and double-component fits. Here however, we adopt a slightly modified version of the improvement fraction $f_\text{imp}$, first used by \citet{el-badry_discovery_2018} and also by \citet{kovalev_detection_2022,kovalev_detection_2024}, to ensure the binary model provides a genuinely better description of the spectrum rather than merely absorbing noise.}

\begin{equation}
    f_{\rm imp} = \frac{\displaystyle\sum \frac{|s_{\rm mod} - s_{\rm ob}| - |b_{\rm mod} - s_{\rm ob}|}{u_{\rm ob}}}{\displaystyle\sum \frac{|s_{\rm mod} - s_{\rm ob}|}{u_{\rm ob}}}
\label{eq:fimp}
\end{equation}

\changed{where $s_\text{mod}$, $b_\text{mod}$, and $s_\text{ob}$ are the single-star model, binary model, and observed spectrum respectively, and $u_\text{ob}$ are the per-pixel uncertainties. This ensures $f_\text{imp}$ expresses the fractional improvement over the single-star fit, and avoids spuriously large values when both models fit the data poorly (see Appendix~\ref{sec:appendix-stats} for further discussion). We require $f_\text{imp}$ to exceed a linear threshold that scales with the improvement in reduced $\chi^2$:}

\begin{equation}
f_{\rm imp} > 0.24 - 0.7\,\Delta\tilde{\chi}^2 ~~~ \text{with} ~~~ \Delta\tilde{\chi}^2 = \tilde{\chi}^2_\text{single} - \tilde{\chi}^2_\text{binary} \geq 0.03.
\label{eq:fimp_threshold}
\end{equation}

\changed{When the binary model already yields a substantially better $\chi^2$, the threshold drops and even a modest $f_\text{imp}$ suffices; when the $\chi^2$ improvement is marginal, a higher $f_\text{imp}$ is required. This formulation was guided through conservative visual inspection of spectra near the $\Delta\chi^2-f_{imp}$ boundary and the piecewise formulation outlined in \citet{el-badry_discovery_2018}.}



\begin{figure}
\centering
\includegraphics[width=\columnwidth]{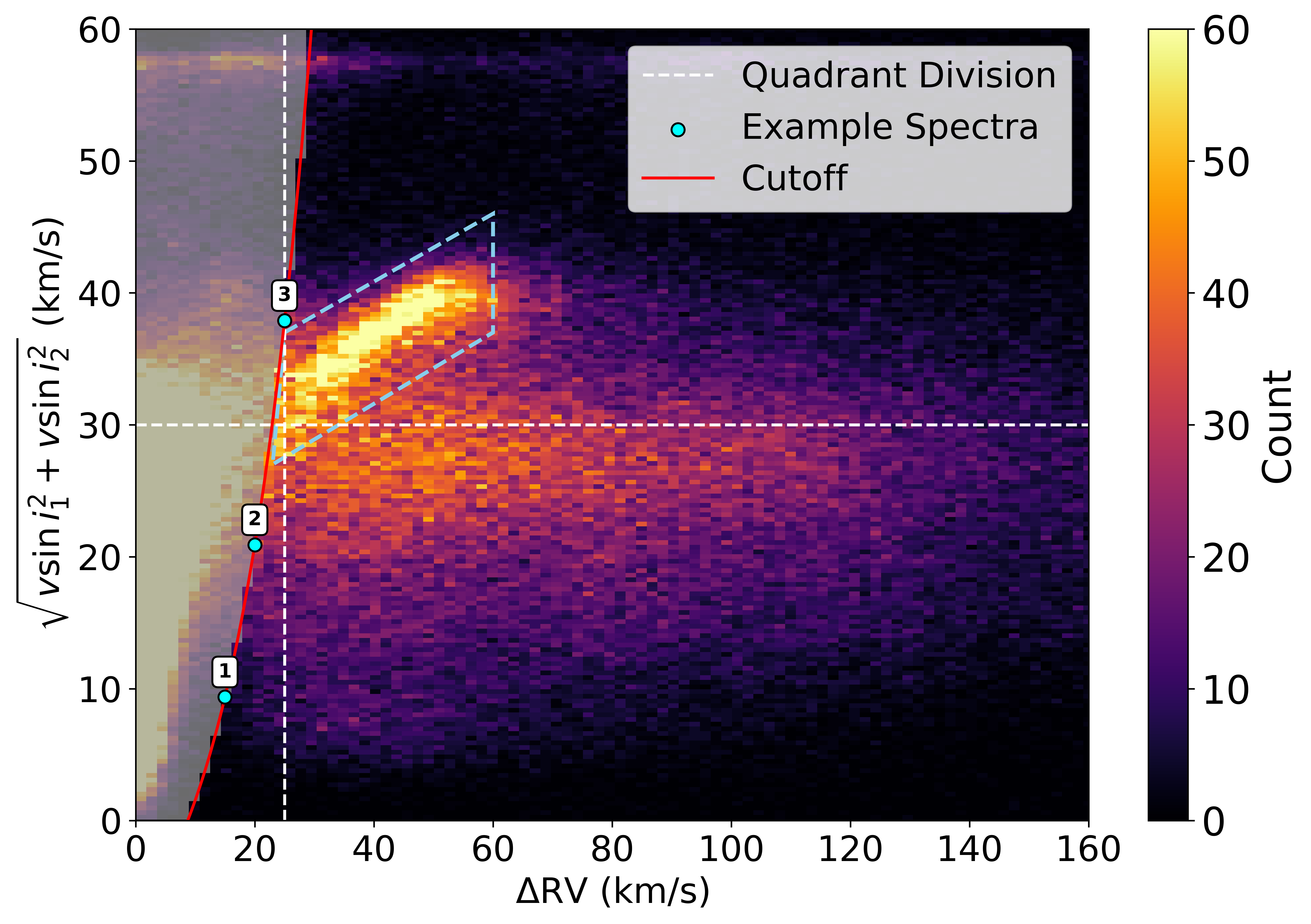}
\caption{Projected rotational velocity ($v\sin i$) as a function of velocity separation ($\Delta$RV) between binary components. The region to the right of the vertical dashed line indicates where spectral lines are sufficiently resolved to be separated in our analysis. The white dashed lines divide the data into four quadrants which we discuss in detail in Section \ref{reject-criteria}. While larger $\Delta$RV values increase line separation, higher $v\sin i$ values broaden the lines, increasing overlap. The empirical cut-off, we define through examination of the spectra is $f(\Delta \text{RV}) = 7.62\,e^{0.0775x} - 15$ (show in red), and marks the boundary above which individual spectral lines become distinguishable in our procedure. Light blue dashed region indicates where a large number ($\sim47\%$) of spectra with \teff > 6000 K reside, we discuss this in more detail in Section \ref{sec:discussion_binary_fraction_completeness} and show how this is due to each component ramping up against the individual limit in Sec. \ref{sec:example_spectra}. Note that this cut is applied in combination with the remaining cuts from Table \ref{tab:data_cuts}. We show example spectra close to the cyan points and discuss the cyan feature in further detail in Sec. \ref{sec:example_spectra}.}
\label{fig:resolution_cut}
\end{figure}

The \teff of both components were constrained to lie within the range $4000-7000$ K, ensuring that we remain within the regime where our spectral models (Sec.~\ref{sec:data_model}) and the GALAH survey are well-calibrated, and to exclude extreme outliers that could skew statistical trends. Additionally, we required a spectral line separation determined from Fig.~\ref{fig:resolution_cut} and given by
\begin{equation}
    f(\Delta\mathrm{RV}) = 7.62 \times e^{0.0775 \Delta\mathrm{RV}} - 15
    \label{eq:rv_cut}
\end{equation}

between the two components, ensuring that the binary nature of the system manifests clearly in the spectrum and is resolvable by our fitting approach. 
Fig. \ref{fig:resolution_cut} shows the projected rotational velocity ($v\sin i$) of the primary star as a function of the velocity separation ($\Delta\mathrm{RV}$) between the two components of the binary system. The underlying 2D histogram represents the density of spectra from our sample in this parameter space. The bright region highlights the parameter space where spectral lines are sufficiently resolved for reliable binary detection and characterisation in our analysis. The dashed curve, defined by eq. \ref{eq:rv_cut}, marks the boundary below which individual spectral lines become distinguishable. This functional form was determined empirically by visually inspecting spectra across the $\Delta\mathrm{RV}$–$v\sin i$ parameter space and identifying where line blending transitions to resolvability. The exponential dependence reflects the non-linear relationship between line broadening ($v\sin i$) and separation ($\Delta\mathrm{RV}$): larger separations allow broader lines to remain distinguishable. Each quadrant can be broken down to better understand this interaction:

\begin{itemize}[leftmargin=2em,labelwidth=0em]
    \item Bottom-left quadrant (low $\Delta\mathrm{RV}$, low $v\sin i$): Spectra here exhibit minimal line blending, but the velocity separation is too small to confidently distinguish components. These systems are challenging to identify as binaries, and their parameters are often unreliable.
    \item Bottom-right quadrant (high $\Delta\mathrm{RV}$, low $v\sin i$): Ideal for binary detection: lines are well-separated and narrow, allowing robust measurement of both components’ RVs and parameters.
    \item Top-left quadrant (low $\Delta\mathrm{RV}$, high $v\sin i$): Severe line blending occurs due to rotational broadening, making binaries nearly indistinguishable from single stars. These systems can be misidentified as single-stars with broad absorption lines.
    \item Top-right quadrant (high $\Delta\mathrm{RV}$, high $v\sin i$): While lines are separated, rotational broadening introduces uncertainty in parameter estimation. We found in the majority of cases, spectral lines exhibited significant broadening beyond the rotational component, such as thermal and pressure broadening, and remained identifiable. 
\end{itemize}

We note the presence of two distinct stellar populations in Figure~\ref{fig:resolution_cut}, broadly divided at $T_\mathrm{eff} = 6000$~K. The hotter stars ($T_\mathrm{eff} > 6000$~K), highlighted by the dotted blue region (containing $\sim47\%$ of these stars), exhibit significantly higher projected rotational velocities, with a median (quadrature summed) $v\sin i$ of 38~km$\,$s$^{-1}$. While this is consistent with previous findings that hot stars tend to rotate more rapidly \citep{barnes_dependence_2005}, this clustering of spectra is not solely explained by this phenomenon. For these high-temperature stars, we observe a positive correlation between $\Delta$RV and $v\sin i$, indicating that greater velocity separation is necessary to resolve binary components as rotational broadening increases. While we believe these stars to be true binaries, we note that the individual rotational velocities are intersecting the upper boundaries of our fit optimisation and neural network training set and their fitting results should thus be treated with caution.




In combination, all of our cuts reduced the initial pool of \changed{$263,785$} possible systems to \changed{$12{,}911$} candidate systems. While this represents a significant reduction, the remaining spectra represent only those we are most confident are true binary systems.

\begin{table*}
\centering
\caption{Summary of data cuts applied to the sample.}
\label{tab:data_cuts}
\begin{tabular}{ll}
\hline
\textbf{Criterion} & \textbf{Description} \\
\hline
$\chi^2 < \chi^2_{\text{GALAH}}$ & Remove spectra where the single-star fit from GALAH is better than the binary-star fit from our analysis\\
$\chi^2 < 3$ & Limit $\chi^2$ to exclude poor fits \\
$R_L < 0.95$ & Exclude systems where one star dominates the luminosity entirely and remove false positives \\
$4000\,\mathrm{K} < T_\mathrm{eff,\,1},T_\mathrm{eff,\,2} < 7000\,\mathrm{K}$ & Restrict primary and secondary temperature to regions in which noise is sufficiently low for fitting. \\
$v \sin i > 7.62\,e^{0.0775\,\Delta \mathrm{RV}} - 15$ & Ensure line separation sufficient to distinguish components and avoid degeneracy (see Figure~\ref{fig:resolution_cut}) \\
$\text{SNR}_i > 15, \quad i = 1,2,3,4$ & Sufficient S/N in each CCD \\
$\sqrt{\sum_{i=1}^4 \text{SNR}_i^2} > 40$ & Sufficient S/N across all CCDs\\
\changed{$f_{\rm imp} > 0.24 - 0.7\,\Delta\tilde{\chi}^2 \quad \text{with} \quad \Delta\tilde{\chi}^2 \geq 0.03$} & \changed{Binary fit must improve on single-star fit by a threshold that scales with $\Delta\tilde{\chi}^2$ (Eq.~\ref{eq:fimp_threshold})} \\
\hline
\end{tabular}
\end{table*}

\begin{table}
\centering
\caption{Summary of columns in the binary star parameter table. Index
$i \in \{1,2\}$ denotes the primary and secondary component
respectively. Columns prefixed \texttt{e\_} are the corresponding
1$\sigma$ uncertainties and are not listed separately.}
\label{tab:table_schema}
\begin{tabular}{lll}
\hline
\textbf{Column} & \textbf{Description} & \textbf{Units} \\
\hline
\multicolumn{3}{l}{\textit{Fit quality}} \\
\texttt{sobject\_id}      & Unique source identifier              & --  \\
\texttt{residual}         & Residual from model fit               & --  \\
\texttt{single\_chi2}     & $\chi^2$, single-star model           & --  \\
\texttt{curvefit\_chi2}   & $\chi^2$, best-fit params             & --  \\
\hline
\multicolumn{3}{l}{\textit{System-level parameters}} \\
\texttt{f\_contr}         & Flux ratio of primary                 & --  \\
\texttt{FeH}              & System [Fe/H]                         & dex \\
\texttt{age}              & System age                            & Gyr \\
\hline
\multicolumn{3}{l}{\textit{Per-component parameters} ($i = 1, 2$)} \\
\texttt{mass\_}$i$        & Stellar mass                          & $\mathrm{M}_\odot$ \\
\texttt{rv\_}$i$          & Radial velocity                       & km\,s$^{-1}$ \\
\texttt{teff\_}$i$        & Effective temperature                 & K   \\
\texttt{logg\_}$i$        & Surface gravity                       & dex \\
\texttt{logl\_}$i$        & Luminosity ($\log$)                   & $L_\odot$ \\
\texttt{vmic\_}$i$        & Microturbulent velocity               & km\,s$^{-1}$ \\
\texttt{vsini\_}$i$       & Projected rotational velocity         & km\,s$^{-1}$ \\
\hline
\multicolumn{3}{l}{\textit{PARSEC isochrone output} ($i = 1, 2$)} \\
\texttt{parsec\_logT\_}$i$   & $\log T_\mathrm{eff}$              & K   \\
\texttt{parsec\_logg\_}$i$   & $\log g$                           & dex \\
\texttt{parsec\_logL\_}$i$   & $\log L$                           & dex \\
\texttt{parsec\_Zini\_}$i$   & Initial metallicity $Z$            & dex \\
\texttt{parsec\_Z\_}$i$      & Current metallicity $Z$            & dex \\
\texttt{parsec\_mass\_}$i$   & Stellar mass                       & $\mathrm{M}_\odot$ \\
\texttt{parsec\_label\_}$i$  & Evolutionary label                 & --  \\
\texttt{parsec\_Gmag\_}$i$   & \textit{Gaia} $G$                  & mag \\
\texttt{parsec\_GBPmag\_}$i$ & \textit{Gaia} $G_\mathrm{BP}$     & mag \\
\texttt{parsec\_GRPmag\_}$i$ & \textit{Gaia} $G_\mathrm{RP}$     & mag \\
\texttt{parsec\_Jmag\_}$i$   & 2MASS $J$                          & mag \\
\texttt{parsec\_Hmag\_}$i$   & 2MASS $H$                          & mag \\
\texttt{parsec\_Ksmag\_}$i$  & 2MASS $K_s$                        & mag \\
\hline
\multicolumn{3}{l}{\textit{Metadata}} \\
\texttt{processing\_time} & Time to analyse spectrum              & s   \\
\hline
\end{tabular}
\end{table}

\subsection{Fitting uncertainties} \label{sec:fitting-uncs}
\changed{To determine uncertainties the PSO chains were analysed for convergence; uncertainties reflect the 16th--84th percentile spread of stable samples. We note some caveats of non-probabilistic sampling apply. We also ran Markov Chain Monte Carlo (MCMC) on the final catalogue to validate and determine our uncertainties probabilistically and compare our PSO uncertainty derivations. This also enables us to compare more directly with the results of \cite{traven_galah_2020} which we present in Fig. \ref{fig:uncertainty_comparisons}. Ultimately we use the uncertainties determined via MCMC in our final catalogue. See section \ref{sec:appendix-unc} for a thorough discussion of our uncertainty analysis.}

\begin{figure*}
\centering
\includegraphics[width=\textwidth]{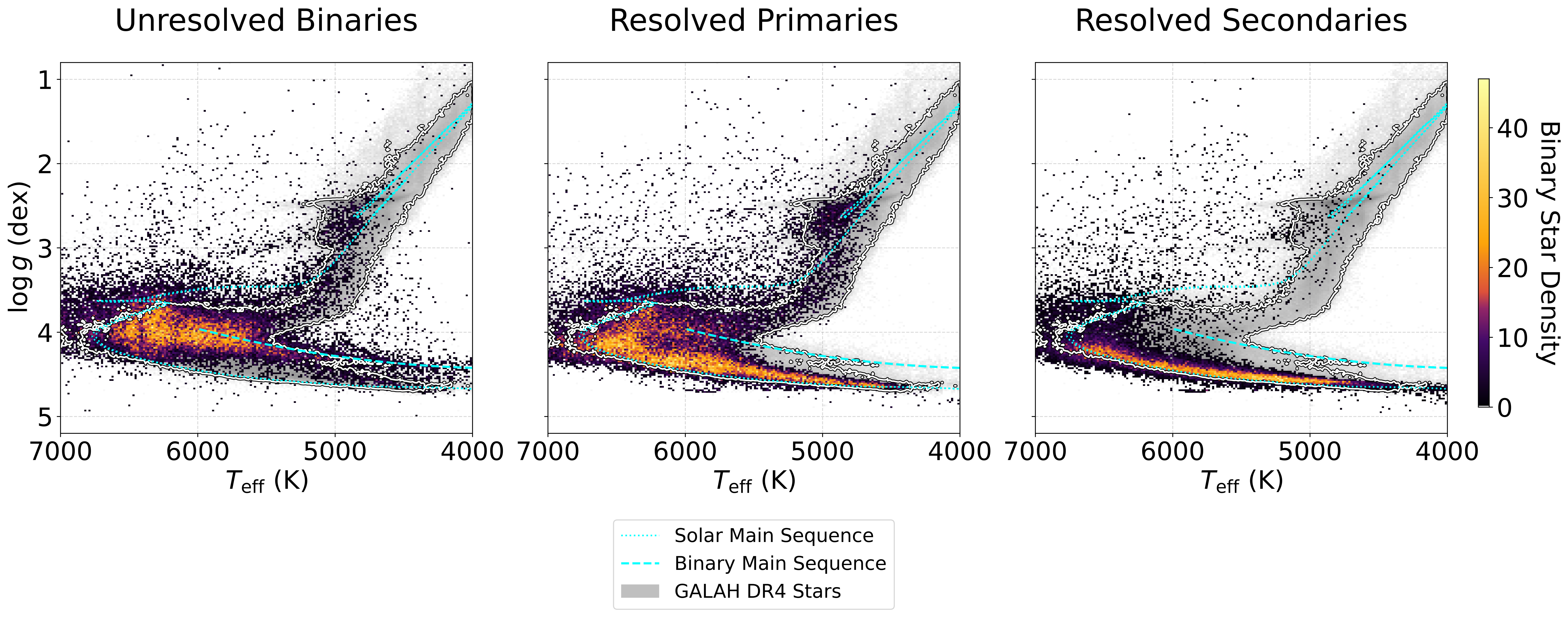}
\caption{Kiel diagrams showing the selected unresolved binaries (left), resolved primaries (centre), and resolved companions (right). Gray region shows the distribution of all stars in GALAH DR4. Stroked white line shows the contour encapsulating the binned distribution of stars where the number of stars per bin exceeds 10 stars. A PARSEC isochrone for a solar-type star ([Fe/H] = 0, M$_\odot$ = 1) is shown by the dotted blue line whilst the approximate position of the binary main sequence is shown by the dashed blue line.}
\label{fig:binary_kiel_optimized}
\end{figure*}

Amongst the final dataset we had 1971 spectra that had repeat observations of which there were 872 unique objects. The majority of these had been observed twice (631), whilst the remaining had been observed three times (70) and more than four times (36). We examined the consistency of the parameters determined for the same objects across these repeat observations and their associated uncertainties to confirm that their difference stays within the expected standard deviation\footnote{For completeness, we attach histograms of these differences in Fig.~S4 of the online supplementary material.}.

\subsection{Catalogue of measured binary properties}

We combine the results of our analysis in a catalogue of best-fit results and their uncertainties, indicators of fit quality and additional predictions of stellar properties, such as magnitudes, from the PARSEC isochrone interpolation. The catalogue schema with column names, descriptions, and units is presented in Table~\ref{tab:table_schema} and includes \changed{12,911} rows for our resolved binaries. For a comparison of the distribution of stars in the $T_\mathrm{eff}$ vs. $\log g$ diagram between unresolved and each of the resolved components, we refer to Fig.~\ref{fig:binary_kiel_optimized}.

\section{Discussion} \label{sec:discussion}

In this section, we discuss our initial research objectives (see Sec.~\ref{sec:introduction}), that is, how we can best detect and model binaries (Sec.~\ref{sec:discussion_detection}), how unresolved binarity impacts the estimation of stellar parameters (Sec.~\ref{sec:discussion_parameter_changes}), and how our binary fractions compare with the literature (Sec.~\ref{sec:discussion_binary_fraction}). Finally, in Sec.~\ref{sec:discussion_binary_fraction_completeness}, we also discuss the influence of possible selection biases in our analysis on these questions.

\subsection{How can we best detect and model binaries?} \label{sec:discussion_detection}
We evaluate our strategies for detecting and modelling binary star systems by first validating our pipeline against the GALAH binary sample of \cite{traven_galah_2020}, then weighing the strengths and limitations of the identification, modelling, and fitting choices that follow from it, and finally setting out the principal limitations of our methodology.

\subsubsection{Algorithm validation through comparison to other samples} \label{sec:comparison_to_other_studies}

\begin{figure}
\centering
\includegraphics[width=\columnwidth]{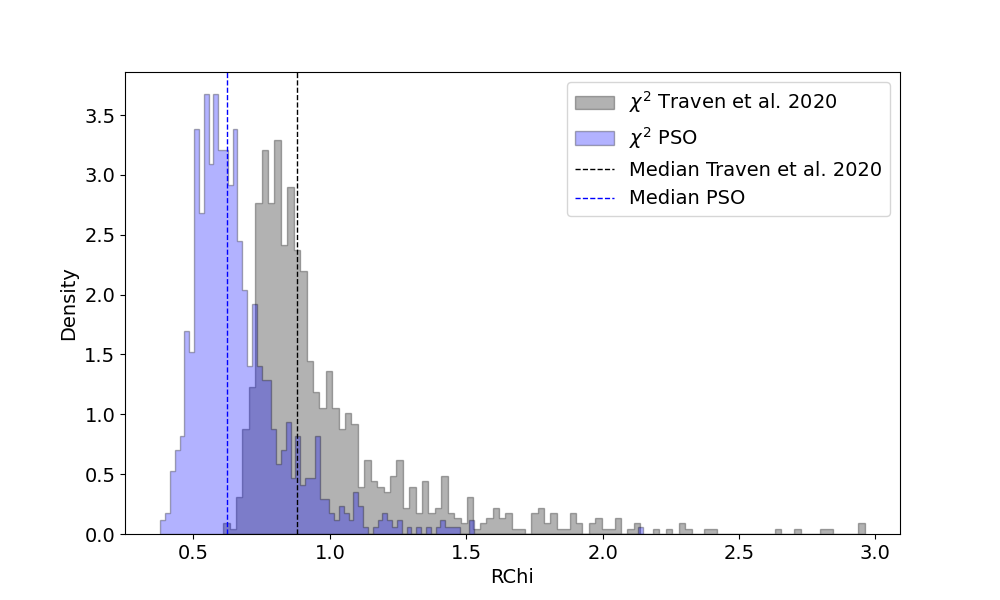}
\caption{
Comparison of $\chi^2$ values from fits to the same spectra, as obtained using our PSO model and the parameters from \citet{traven_galah_2020}. For the latter, $\chi^2$ values are computed by evaluating their published parameters with our forward model. Our fits yield lower $\chi^2$ values, though this likely reflects differences in modelling approach rather than improved agreement with ground truth. As noted by \citet{buder_galah_2025}, we find $\chi^2$ values are consistently less than 1 due to overestimated uncertainties leading to smaller normalized residuals in the fit.
}
\label{fig:delta_rchi}
\end{figure}

\begin{figure}
\centering
\includegraphics[width=\linewidth]{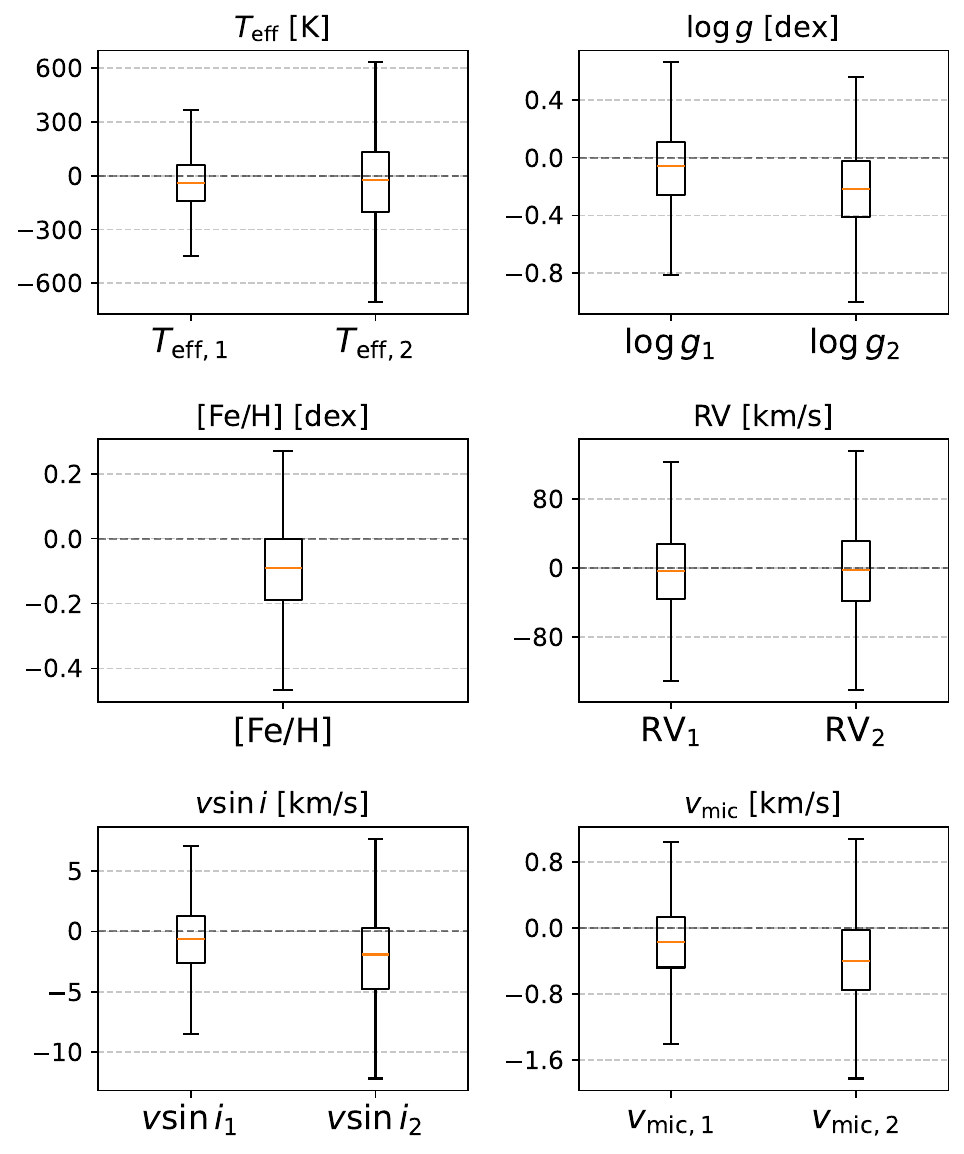}
\caption{A comparison of parameters between this study and \citet{traven_galah_2020}. We use the parameters fitted by \citet{traven_galah_2017} for mutually analysed spectra as inputs to our binary model, enabling a direct comparison of performance and validation of our method. The most discrepant parameters are [Fe/H], $\log g$ (for the secondary component), and an apparent systematic offset of $-100$K in \teff. When restricting the comparison to spectra that our model fits better, we find that \teff is the primary driver of improved fit quality, with differences of around $-300$K in these cases. These differences observed between parameter values may be due to differences in the underlying modelling approaches rather than reflecting true discrepancies in the physical properties of the stars. However we note that \teff is a particularly influential parameter in synthetic spectra fitting.}
\label{fig:box_plot_params}
\end{figure}

To assess the accuracy of the stellar parameters determined by our binary fitting pipeline we compared the outputs to those of mutual stars from other studies. \cite{traven_galah_2020} determined stellar parameters by fitting GALAH binary spectra with MCMC to sample their posterior distributions and also incorporated parallax data and photometry from \textit{Gaia}. We took their median posterior values for each parameter as inputs in our binary model to directly compare the fits of the resulting spectra (Fig.~\ref{fig:delta_rchi}).
This  allowed us to determine which specific parameters were worse fit in our analysis for spectra that are well fit by \cite{traven_galah_2020}. This guided revisions to our pipeline and revealed that \teff is the most significant parameter when achieving a good fit. A comparison showing the difference between parameters across the entire dataset is shown in Fig.~\ref{fig:box_plot_params}. 

\subsubsection{Identification} \label{sec:discussion_detection_identification}

As outlined in Section~\ref{sec:analysis_fitting}, the cross-correlation function did prove to be a reliable method to discern binarity and flag candidate spectra; however, our results indicate that it should not be treated as a single comprehensive tool, rather as one that complements other efforts. We found the input parameters to the cross-correlation function (\software{scipy.find\_peaks}) have significant impact. Slight variation in the input parameters such as the height, distance, prominence, and threshold can result in false positives or failed detections. Fitting multiple Gaussian functions through superposition also worked to improve the detection ability of the CCF; however, we still noted it failed to detect systems identified in our final high-confidence sample. Such a spectrum is shown in Fig.~\ref{fig:binary_CCF_comparison_fig}.

CCF-based binary detection is therefore best used as a fast, complementary tool alongside more flexible approaches such as the PSO fitting employed here or the t-SNE clustering implemented by \citet{traven_galah_2017}, who also identified many candidates with t-SNE that were not flagged by CCF alone. Another complementary strategy is that of \citet{merle_gaia-eso_2017}, who detect multi-lined spectroscopic binaries in \textit{Gaia}-ESO data by computing CCFs between observed spectra and template line masks and then analysing the CCF and its derivatives to identify even strongly blended peaks. Alternatively, recent studies by \cite{kovalev_detection_2022, kovalev_detection_2024} made significant progress toward this by leveraging the strong correlation between the underlying RV separation and the artificially inflated projected rotational velocity \vsini recovered from single-star spectral fits. This diagnostic enables the detection of SB2 candidates with separations as low as $\sim$20\kms. While theoretically sensitive to separations as low as 8\kms, recent updates even allow for the recovery of systems at zero radial velocity separation if the component spectra are sufficiently distinct. This is significantly more sensitive than CCF-based methods, which typically require $\Delta\textrm{RV}$ of $\sim$50\kms to resolve distinct peaks in LAMOST \citep{li_massratio_2022}, $\sim$30\kms in APOGEE \citep{fernandez_-sync_2017}, or the median separation of $\sim$30\kms found in this study\footnote{Median $\Delta\textrm{RV}$ of confirmed binaries in the final catalogue that were originally selected via CCF peak detection. See Table~\ref{tab:selection-survival}}. We would encourage exploration of either of these in any future work.

\subsubsection{\changed{Catalogue Statistics and Comparison to Previous Works}}
\changedtwo{Our final catalogue contains 12,911 candidate binary systems. Of these, 6,683 correspond to objects already flagged as having multiple components in GALAH~DR4, while the remaining 6,228 are newly identified systems that were not previously flagged. Given that GALAH~DR4 contains 25,191 objects flagged as having multiple components, the inclusion of 6,228 additional unflagged systems represents a substantial expansion of the binary population accessible in GALAH. It should be noted that the flagging procedure in DR4 was rudimentary, intended primarily to indicate the need for follow-up rather than to constitute a rigorous binary classification. Of the 25,191 flagged objects, 18,508 did not survive our quality cuts, leaving the 6,683 systems in common between both catalogues. The additional objects recovered here arise primarily from our supplementary selection criteria, in particular the inclusion of objects with high RUWE and high rotational velocity (Table~\ref{tab:selection-survival}).}

Our results confirm that one of the strongest indicators of binarity is a large discrepancy in radial velocities between temporally separated surveys (GALAH DR4 and \textit{Gaia}). This is unsurprising: observations taken at different epochs will sample different orbital phases of a true binary, yielding distinct radial velocities. The CCF criterion demonstrated the highest purity of any selection criterion, with a survival rate of 68.9\%, though it identified few objects exclusively (236) due to substantial overlap with other criteria. Similarly, the large RV difference criterion achieved a survival rate of 18.9\%, but it likewise showed considerable overlap with the remainder of the sample, contributing only 210 unique objects. Notably, the GALAH DR4 CCF selected a larger parent sample (25,191 objects) than our CCF criterion (15,581), yet fewer objects from the GALAH selection survived to the final catalogue (6,683 versus 10,730). This inverse relationship between selection size and survival --- a factor of $\sim$2.6 difference in purity --- suggests that the GALAH DR4 flagging procedure, whilst more inclusive, introduced substantial contamination from non-binary interlopers. This shows a well tuned CCF makes a large difference in detection quality.

Conversely, the high RUWE and high rotational velocity criteria exhibited lower survival rates of 2.4\% and 3.1\% respectively, indicating substantial contamination from non-binary interlopers. Nevertheless, these criteria contribute the largest numbers of exclusively unique objects to the final catalogue, 632 and 1,313 respectively, objects that would otherwise have been excluded entirely. Their inclusion is therefore justified despite the elevated contamination fraction.

Finally, we compare our binary candidates with those of \citet{traven_galah_2020} and \textit{Gaia}. Of the $\sim$263,000 objects in our initial sample, 12,171 are in common with \citet{traven_galah_2020} and 3,190 with \textit{Gaia}. Following our analysis and data cuts, \changed{4,733} ($\sim$\changed{39}\%) of the Traven cross-matched objects and \changed{465} ($\sim$\changed{15}\%) of the \textit{Gaia} cross-matched objects survived to the final catalogue. For the Traven subsample, the majority of removed objects ($\sim$40\%) were excluded either by the \vsini--$\Delta$RV criterion or because the single-star fit was preferred on the basis of a lower reduced $\chi^2$. For the \textit{Gaia} subsample, the dominant sources of attrition were objects exceeding the \teff cut-off and, similarly, cases where the single-star fit was found to be superior. The objects lost from these cross-matched subsamples likely include systems whose binarity is detectable through astrometric or photometric signatures, as exploited by \textit{Gaia} and \citet{traven_galah_2020}, but which do not exhibit sufficient spectroscopic signatures to satisfy our selection criteria. This represents an inherent limitation of a purely spectroscopic approach, which is insensitive to binaries with unfavourable inclinations, long orbital periods, or large flux ratio contrasts between components. A detailed breakdown of the statistics including where all of the cuts impacted both the cross-matched objects and those from the initial selection criteria is available in Appendix \ref{sec:appendix-stats}.

\begin{table*}
\centering
\caption{Summary of selection criteria applied to identify binary candidates, showing the number of spectra selected by each criterion, the number surviving to the final catalogue, the survival rate, and the number of spectra unique to each criterion in the final catalogue. "Unique to criterion" means: stars in the final catalogue that were selected exclusively by that one criterion --- i.e. they would have been excluded from the sample entirely if that criterion were removed, because no other criterion selects for them.}
\label{tab:selection_criteria}
\renewcommand{\arraystretch}{1}
\begin{tabular}{lrrrr}
\hline
\textbf{Selection} & \textbf{N in Selection} & \textbf{N in Final Catalogue} & \textbf{Survival Rate (\%)} & \textbf{N Unique to Criterion} \\
\hline
Multiple RV Components   & 25{,}191  & 6{,}683  & 26.5 & 341         \\
Large RV Difference      & 31{,}240  & 5{,}912  & 18.9 & 210         \\
High RUWE                & 160{,}761 & 3{,}891  & 2.4  & 632         \\
High Rotational Velocity & 204{,}303 & 6{,}341  & 3.1  & 1{,}313     \\
Traven Sample            & 12{,}171  & 4{,}733  & 38.9 & 354         \\
CCF                      & 15{,}581  & 10,730 & 68.9 & 236         \\
\hline
\end{tabular}
\label{tab:selection-survival}
\end{table*}

\subsubsection{Modelling and Fitting} \label{sec:discussion_detection_modelling}

The selection of the fitting algorithm in our analysis evolved as it became clear that alternatives were inadequate to accurately fit the data. Our initial attempt employed the same fitting approach as \cite{buder_galah_2025}, using \software{scipy.curvefit} to perform a simple least-squares fit. This is fast, however, can suffer from inadequately exploring the full parameter space. This became especially relevant in this case where the number of parameters required for modelling is effectively doubled for binary systems.

When inputting the parameters found in \cite{traven_galah_2020} into our model, the resulting fit is typically poorer. This is not unexpected, as our analysis is designed to find the best-fitting parameters within our own modelling framework. However, this exercise highlights where our model produces systematically different results from \cite{traven_galah_2020}, most notably in \teff, log g, and [Fe/H]. We expect these differences arise because \cite{traven_galah_2020} also includes photometric and astrometric data, leading to solutions that differ from those obtained with our purely spectroscopic model. We also note that the quadratic Cannon model used by \citet{traven_galah_2020} is intrinsically less flexible than our current generative neural-network model, and is therefore more prone to systematic modelling errors in parts of parameter space where the spectral dependence on the labels is strongly non‑linear.

Specifically, we obtain lower \teff for the primary stars, with a median difference of approximately -100 K, and slightly higher surface gravities log g by about + 0.1 dex. For the secondary components, the discrepancies are more pronounced, with our \teff values being lower by around -300 K and log g values lower by -0.25 dex. Similarly to \cite{traven_galah_2017}, we find unresolved binaries in GALAH DR4 to be measured with systematically lower [Fe/H] than comparable single stars, largely because the single-star pipeline compensates for the extra flux by increasing \teff{}, which through the known \teff{}–[Fe/H] degeneracy—drives [Fe/H] down.

One possible explanation for these differences lies in the constraints used during the fitting process. Our approach assumes that both stars in a binary system share the same age and [Fe/H], this may not always be a valid assumption, particularly for systems undergoing mass transfer or significant evolution. In contrast, \cite{traven_galah_2020} rely on photometry and astrometry, which could provide additional constraints on the spectral energy distribution (SED) but may also introduce biases due to uncertain extinction corrections and potential contamination from poor astrometry. Their method may be more robust in some cases where colour information provides a better handle on the overall SED, but it could also be susceptible to systematic errors if extinction and distance estimates are unreliable. \changed{Both our method and that of \citet{traven_galah_2020} fit all four GALAH CCDs simultaneously; the H$\alpha$ and H$\beta$ exclusion in \citet{traven_galah_2020} applies only to their CCF-based detection step, not to the spectral fitting itself. Nevertheless, H$\alpha$ is both highly temperature-sensitive and systematically double-lined in SB2 systems. Future work should evaluate whether masking H$\alpha$ during fitting --- or adopting a per-CCD fitting strategy to avoid cross-CCD continuum normalisation artefacts --- could reduce the $\sim$100--300\,K \teff offset we observe.}

\subsubsection{Model Limitations} \label{sec:discussion_detection_limitations}

While our method for identifying and characterizing binary systems is robust, several inherent limitations must be acknowledged. First, the ability to detect binaries is strongly dependent on the contrast between the two components. Systems where the secondary star is significantly fainter than the primary may not produce sufficiently distinct spectral features for reliable detection. Consequently, binaries with low-mass companions or extreme flux ratios are more likely to remain unidentified in our analysis. We also note that we removed such systems when we applied quality cuts to prevent false positives from being included. 

Secondly, even in cases where binary companions exist, the system may appear single if the stars share similar radial velocities at the time of observation. Such systems, despite potentially showing significant radial velocity variations over longer timescales, can evade detection when observed at specific orbital phases. As a result, some binaries with unresolved secondaries are likely to be misclassified, emphasising the need for multi-epoch observations to fully characterise the binary population or the development of new methods capable of disentangling binary spectra. We assume that when the difference in RVs is very small (according to eq. \ref{eq:rv_cut}), it is impossible to spectroscopically distinguish between very close binaries and single star systems.

In addition to these primary limitations, selection biases inherent to the observed sample, assumptions in the spectral modelling, and the influence of stellar activity may also impact the completeness and reliability of binary detection. We also make no attempt to identify interacting binaries, which would show distinct spectral features and skew the reported multiplicity fraction. All of these factors should be considered when interpreting the completeness and characteristics of the binary population derived from this study.

\subsubsection{\changed{Recommendations for Binaries Flagged in GALAH DR4 and Future Surveys}}
\changed{Based on our analysis, we offer the following recommendations for binary classification in future GALAH data releases. First, the CCF-based flagging used in DR4 should be treated as an indicative screen rather than a reliable classification --- our results show that 69\% of DR4-flagged objects do not survive rigorous quality cuts, suggesting substantial contamination from non-binary objects. We recommend that future releases supplement CCF-based detection with additional selection criteria, in particular high RUWE and high rotational velocity, which together recover a significant fraction of genuine binaries missed by the CCF alone (Table~\ref{tab:selection-survival}). Second, where computational resources permit, a composite spectral model is preferred over single-star fitting for flagged objects, as it provides a principled basis for both confirming binarity and extracting per-component stellar parameters. Third, future binary fitting pipelines for GALAH should consider adopting a per-CCD fitting strategy and masking H$\alpha$ during spectral fitting: fitting each CCD independently avoids cross-CCD continuum normalisation artefacts, and masking H$\alpha$ removes a region that is both highly temperature-sensitive and systematically double-lined in SB2 systems (Section~\ref{sec:discussion_detection_modelling}).}

\changed{Ultimately, the ideal approach would be to fit every object in the parent catalogue with both a single-star and a binary model, retaining the binary solution where it yields a statistically significant improvement in fit quality. This removes the need for pre-selection criteria entirely and avoids the incompleteness introduced by any finite set of binary indicators. At the rate reported in our analysis ($\sim$40 wall-clock hours per 100,000 spectra on a 50-core machine; Fig.~\ref{fig:rchi2_comparisons}), processing a survey of one million objects would require approximately 15 days — comparable in overhead to the GALAH DR4 stellar parameter step alone, which requires $\sim$700--1,000 hours per 100,000 spectra on equivalent hardware. Running a binary model in parallel therefore introduces only modest additional cost relative to what is already invested in single-star analysis. With access to larger high-performance computing infrastructure or optimisations to the fitting routine, this could be reduced further, making blind binary modelling of full survey catalogues, such as those expected from 4MOST or future GALAH releases, a realistic prospect.}

\subsection{The Impact of Unresolved Binarity on Stellar Parameter Estimation} \label{sec:discussion_parameter_changes}

In the following sections, we examine how each of the key stellar parameters is affected by successful binary detection. An overview of the change in \teff, $\log g$, [Fe/H], and age is shown for the \teff-logg parameter space in Fig.~\ref{fig:teff_logg_smoothed_coolwarm}. For each quantity, we have also calculated the 16th and 84th percentiles of change from the unresolved to resolved primary properties and list them in Tab.~\ref{tab:percentiles}. Before the discussion of individual parameters, we subsequently discuss how changes manifest across different regions of the parameter space.


\subsubsection{Change of Stellar Parameters - General Observations} \label{sec:discussion_parameter_changes_stellar_parameters}

\begin{figure}
\centering
\includegraphics[width=8cm]{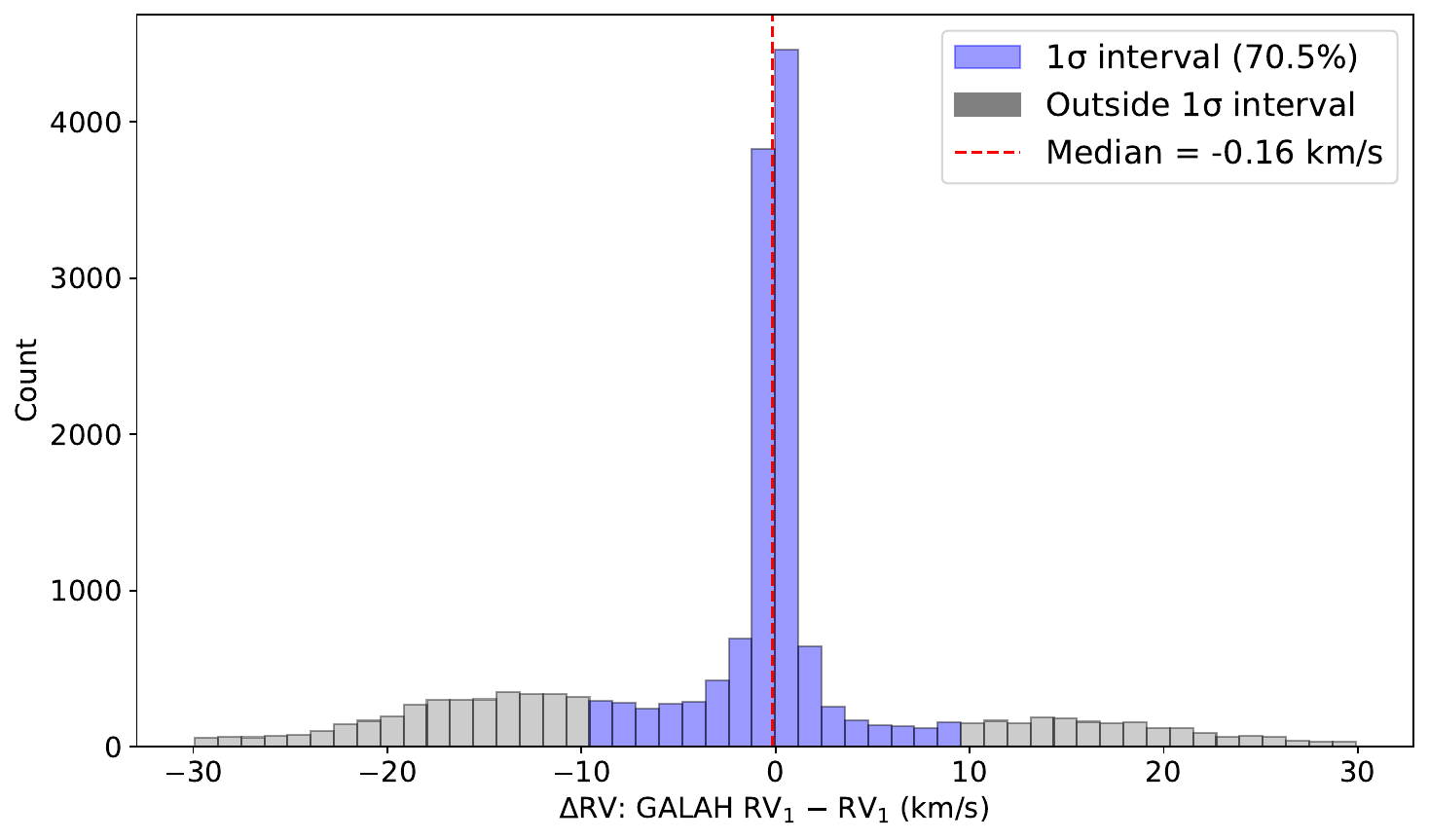}
\caption{The difference between the RV fit by the GALAH single-star model and by our analysis (RV$_\text{GALAH}$ - RV$_1$). The distribution is sharply peaked at zero, with the vast majority of values falling within $\pm10\,\mathrm{km\,s^{-1}}$ (the $1\sigma$ interval), indicating that in most cases the single-star fit adopts the RV of one of the binary components.
}
\label{fig:rv_diffs}
\end{figure}

Even though we exclude systems with radial velocity separations (Fig.~\ref{fig:resolution_cut}) which reduces the effects of line blending, unresolved spectra are still influenced by line dilution. When absorption features are detected in only one component, the combined flux from both stars elevates the continuum level, causing the lines to appear shallower after normalisation. This dilution distorts the spectral features used to derive stellar parameters, introducing systematic biases when a single-star model is applied to a composite spectrum. The consequences of these distortions vary depending on the spectral types and flux contributions of the two components. It is also important to note when interpreting these results that the classification of the unresolved star is not necessarily reflective of the classification of the resolved components, i.e. an unresolved ``giant'' star may in fact be comprised of two MS stars. The broad trend for stars across the binary main sequence is to be resolved as hotter, denser stars (increased \logg), which relocates them to the single-star MS.
\paragraph{Single-star fits almost always align with one component.}
Across the vast majority of single-star fits, we observe that the single-star model typically aligns with one component of the binary system, rather than representing an average or "middle-ground" between the two stars. Fig.~\ref{fig:rv_diffs} shows the difference between the single-star fit RV and that of the primary component from the binary-star fit. The clear distribution around zero demonstrates that in most cases the single-star fit adopts the RV of one of the binary components, strongly supporting this interpretation. While this behaviour is expected when the flux ratio is significantly skewed (e.g., $>0.6$), it can also occur when the two stars have similar flux contributions. In such cases, which star the single-star model aligns with may be determined by subtle differences in their spectra.
\paragraph{Single-star fits produce deeper lines than binary star fits.}
Most individual spectral lines in the single-star fits are modelled as deeper than those found by our analysis. This occurs because the single-star fit attempts to minimize residuals by aligning with one component’s spectral features while partially accounting for the second component’s Doppler-shifted lines. By fitting a deeper line to the primary component, the model effectively overlaps with the same line from the secondary component, which is shifted in wavelength due to the RV difference. The  maximum line displacement we find is between 3–6 Å depending on wavelength, and corresponds to a maximum $\Delta$RV difference ($\sim 200\,\mathrm{km\,s^{-1}}$) in our sample. The median RV difference we find is approximately $50\,\mathrm{km\,s^{-1}}$, corresponding to a wavelength shift of about 0.6–1.0 Å across the four GALAH bands, with the exact shift depending on the central wavelength of each band. Since the widths of the most prominent spectral lines range from 0.4-3.0 Å, they are almost always close together if not partially blended. Fitting deeper lines that can encompass both components becomes an effective way for the single-star model to minimize residuals and improve the fit. 
\paragraph{Single-star models prioritise fitting dominant spectral features at the expense of weaker metal lines.}
In contrast to the above, elements with shallower, narrower, or fewer spectral lines --- such as O or Ti --- exhibit the opposite trend. In these regions, the single-star fits produce lines that are noticeably shallower than what is observed. This occurs because the fitting process prioritizes matching the deepest and most dominant features that drive the $\chi^2$ fit at the expense of the weaker metal lines, as the relationship between \teff and line depth differs for hydrogen and metals. Increasing \teff decreases the line depth for metals but increases it for hydrogen lines. The single-star fit ultimately settles on fitting overly deep lines for H features to best capture the neighbouring lines resulting from binarity and also fitting the lower continuum. We note that this is of course dependent on the number of metal lines being fit and their profiles. We emphasise, however, that the bulk of the systematic offsets in \teff and $\log g$ across our sample are driven primarily by metal-line blending rather than H$\alpha$ fitting: stars with $T_{\mathrm{eff}} < 5000\,\mathrm{K}$, where H$\alpha$ carries minimal temperature information, show the largest median \teff offset ($+227\,\mathrm{K}$), while stars with $T_{\mathrm{eff}} > 6000\,\mathrm{K}$, where H$\alpha$ is most temperature-sensitive, show little systematic offset ($-24\,\mathrm{K}$).
\paragraph{Single-star fits systematically underestimate both $T_{\mathrm{eff}}$ and $\log g$.}
The single-star model consistently finds lower \teff and lower $\log g$ than the binary solution assigns to the resolved primary. The \logg underestimate arises from continuum dilution: the secondary's flux raises the overall continuum level, so that after normalisation all absorption lines appear shallower than they truly are for the primary alone. A single-star model interprets shallower lines as lower \logg (less pressure broadening), and consequently assigns a lower \logg than the true primary value. This effect operates across all temperature regimes, consistent with the relatively uniform median $\Delta\log g$ we observe: $+0.21\,\mathrm{dex}$ for $T_{\mathrm{eff}} < 5000\,\mathrm{K}$, $+0.22\,\mathrm{dex}$ for $5000$--$6000\,\mathrm{K}$, and $+0.13\,\mathrm{dex}$ for $T_{\mathrm{eff}} > 6000\,\mathrm{K}$. The \teff underestimate, by contrast, is strongly temperature-dependent --- largest for cool stars where metal-line blending dominates, and negligible for hot stars --- as demonstrated by the H$\alpha$ analysis above.

\begin{table}
\centering
\caption{16th and 84th
percentiles of residuals between resolved and unresolved stellar
parameters..}
\label{tab:percentiles}
\begin{tabular}{lcc}
    \hline
    Parameter & 16th & 84th \\
    \hline
    $\Delta T_{\rm eff}$ (K)  & $-142$   & $200$   \\
    $\Delta\log g$ (dex)      & $-0.070$ & $0.346$  \\
    $\Delta$[Fe/H] (dex)      & $-0.055$ & $0.129$ \\
    $\Delta$Age (Gyr)         & $-0.015$ & $0.018$ \\
    \hline
\end{tabular}
\end{table}

\begin{figure*}
\centering
\includegraphics[width=\textwidth]{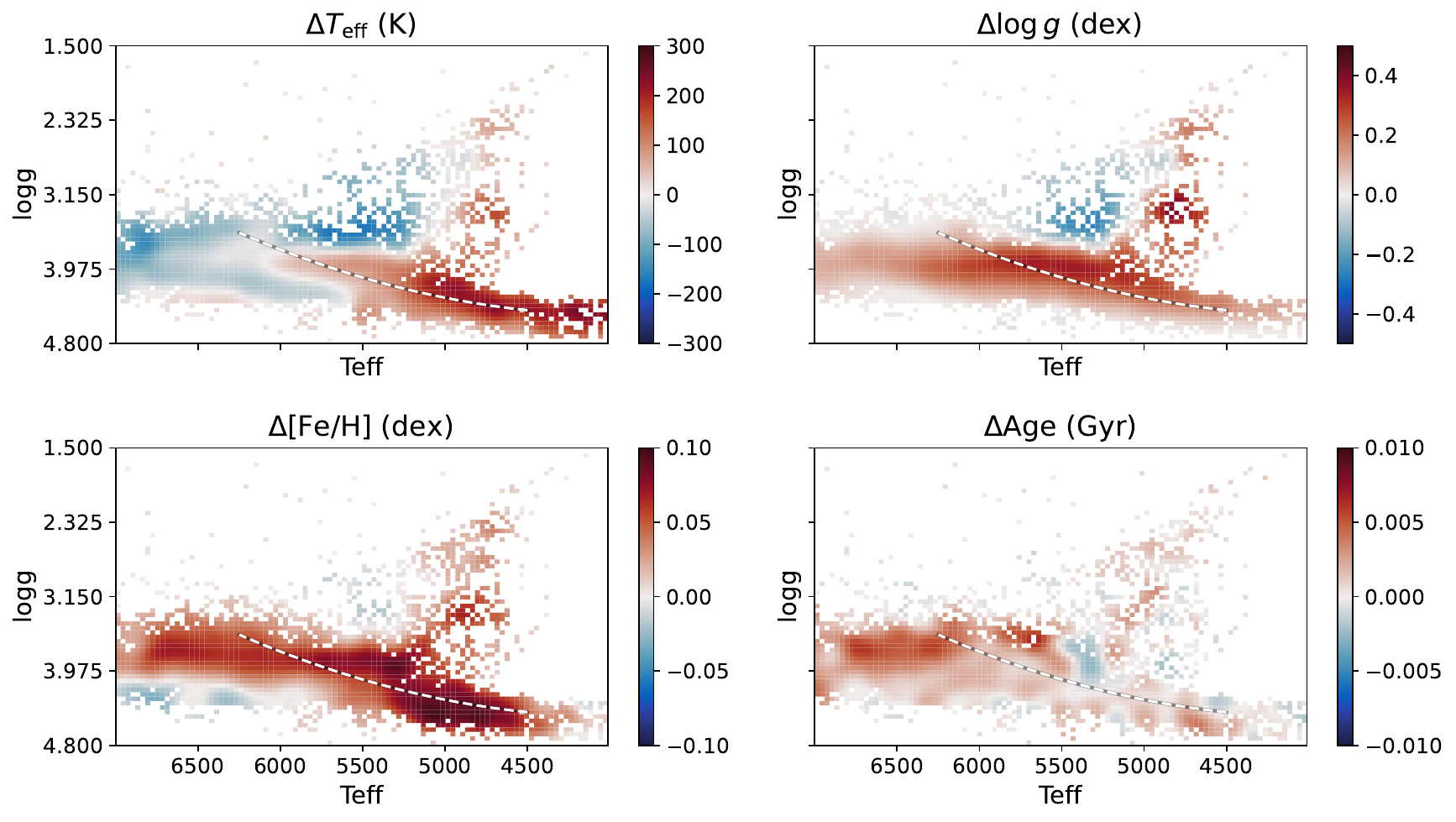}
\caption{Smoothed differences between resolved and
unresolved stellar parameters in the unresolved $T_{\mathrm{eff}}$--$\log
g$ space. Panels show differences for $T_\mathrm{eff}$ (a), $\log g$
(b), [Fe/H] (c), and age (d). Solar-like isochrones (Age\,=\,4.5\,Gyr,
[Fe/H]\,=\,0) are shown as dashed lines.}
\label{fig:teff_logg_smoothed_coolwarm}
\end{figure*}

\begin{figure}
  \centering
  \includegraphics[width=\linewidth]{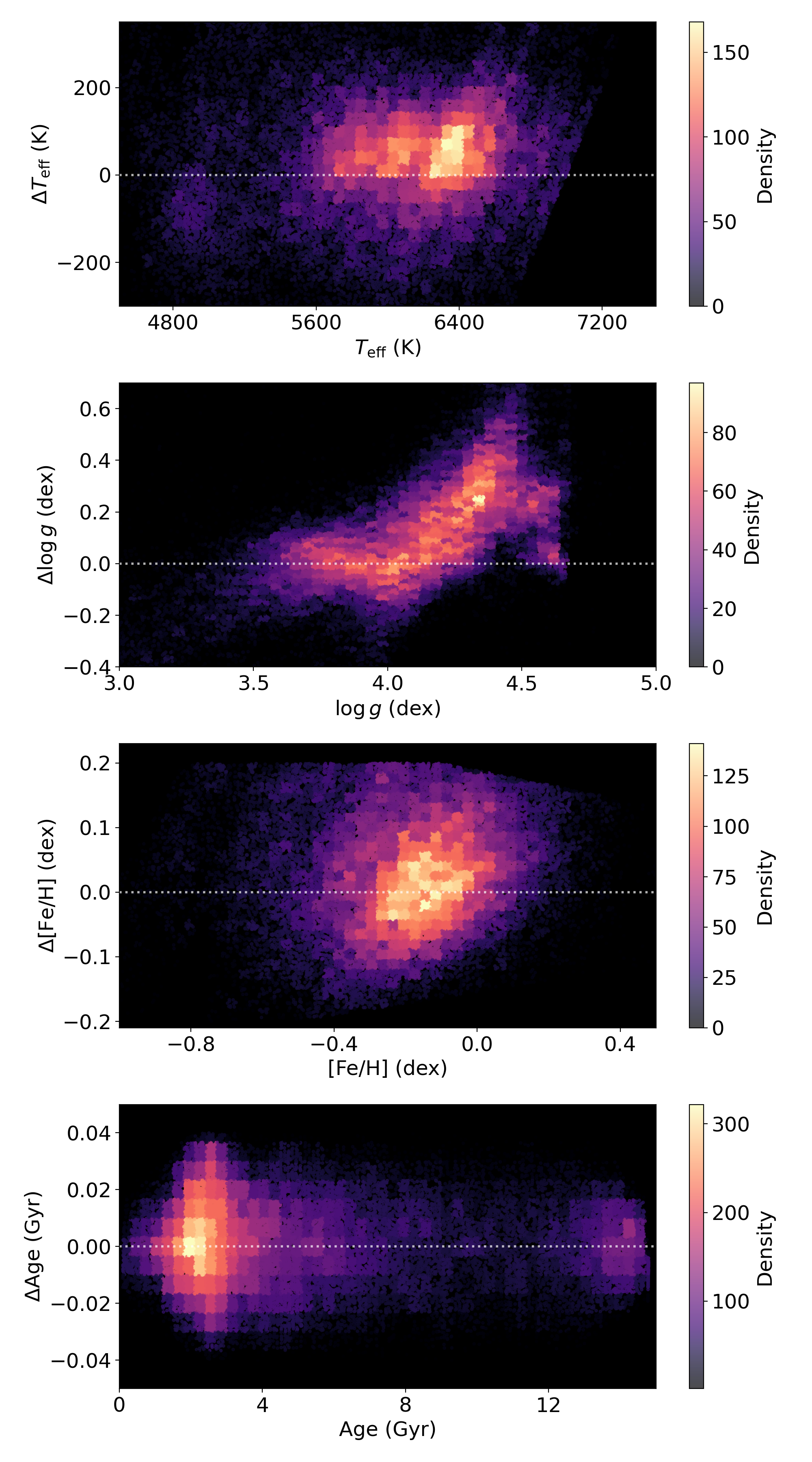}
  \caption{Differences in stellar parameters between unresolved and resolved binary model fits, shown as a function of the primary component's value from the binary fit. Each panel displays the difference (unresolved-star $-$ resolved binary star) for [Fe/H], $T_{\mathrm{eff}}$, and $\log g$, respectively, plotted against the corresponding parameter value for the primary star. Point density is indicated by colour.}
  \label{fig:delta_columns}
\end{figure}

\subsubsection{Effective Temperature $T_\mathrm{eff}$} \label{sec:discussion_parameter_changes_temperature}

When examining the change in \teff of stars specifically across the binary main sequence (Fig.~\ref{fig:teff_logg_smoothed_coolwarm}), we note that when they are resolved and appropriately fitted with a binary model the relative temperature of the primary systematically increases by around 200 K across the MS and increases by around 100 K everywhere else (see also Fig.~\ref{fig:delta_columns}). Upon further inspecting these systems we find that in the unresolved case, the fitting procedure had simply modelled the dominant component --- the primary. The contribution from the cooler secondary is thus neglected, albeit contributes to reddening the overall SED and can result in producing a synthetic spectra with a lower \teff. By contrast, hot stars (\teff > 6000 K) outside the binary MS show the opposite trend. These are found to have primaries with \teff $50 - 100$ K cooler than their unresolved values. This may be due to the cooler secondary being captured by the fitting process rather than the primary, due to the increased number of lines at cooler temperatures. 


The largest differences in temperature are observed in the populations of stars with \teff < 4500 K where \teff is observed to be systematically higher by $\sim200$ K. This low temperature region includes both dwarf and giant stars, where spectra can become noisier due to increased accretion activity and convective processes. These factors combined with a dearth of stars in this region result in higher uncertainties and many false-positives due to spectra exhibiting a higher line density, from both atomic and molecular lines. It is unlikely that any single factor fully accounts for the observed changes in \teff, and more plausibly, these shifts arise from a confluence of effects --- including spectral degeneracies, line blending, temperature-sensitive line features --- all of which are difficult to disentangle in practice.

\subsubsection{Surface Gravity $\log g$} \label{sec:discussion_parameter_changes_logg}

In our comparison of unresolved and resolved stellar parameters, we find a consistent trend of increased $\log g$ in the resolved spectra (see Figs.~\ref{fig:teff_logg_smoothed_coolwarm} and \ref{fig:delta_columns}). This effect is observed across a broad range of unresolved stars and is likely a direct consequence of how \logg is inferred in single-star models. In the case of binaries, the combined light from both stars results in a total luminosity that is higher than that of either component alone. When this combined flux is interpreted as originating from a single star, the inferred radius is correspondingly overestimated. Assuming a roughly constant mass, this inflated radius leads to an underestimated \logg. Thus, single-star models fitted to unresolved binaries systematically yield \logg values that are too low.

When the same systems are fit with our binary model, the two components are no longer forced to mimic a single, over-luminous star. Instead, the primary and secondary are placed self-consistently on PARSEC isochrones, which constrains their combinations of $T_{\mathrm{eff}}$, $\log g$, and luminosity to correspond to physically plausible single stars. Each component is assigned a lower luminosity, resulting in a smaller inferred radius and hence a higher \logg. This correction is reflected in the systematically higher \logg values observed for the resolved stars. 

The most significant change in $\log g$ is observed for unresolved stars that appear in the red clump region. This area has previously been found to show mismatches in GALAH data analyses \citep[see for example the discussion in Sec.~6.5.1 of][]{buder_galah_2021}. Similarly, several of these actual red clump stars have been found to have unreliable astrometric solutions which has led to the DR4 fitting routine to placing them in a less luminous region below the red clump \citep{buder_galah_2025}. Our detection of a secondary component to several of these stars is thus in agreement with an unreliable astrometric solution and the primary component truly being more luminous.

\subsubsection{Metallicity} \label{sec:discussion_parameter_changes_stellar_metallicity}

\begin{figure}
\centering
\includegraphics[width=\columnwidth]{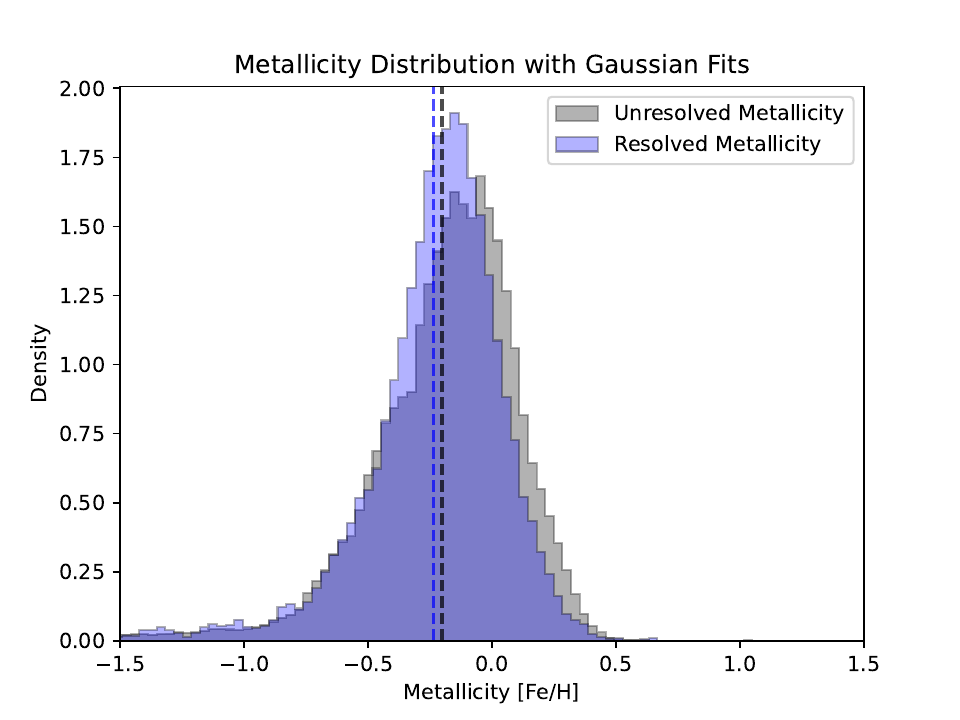}
\caption{[Fe/H] distribution for the binary sample and the remaining stars from GALAH DR4. We find a difference of 0.16\,dex in the mean [Fe/H] of the binaries and the remaining stars (mean [Fe/H]$_{\text{binaries}}$ = -0.37, mean [Fe/H]$_{\text{all}}$ = -0.21). We note that the remaining stars likely contain a large number of undetected binary systems which would alter the [Fe/H] distributions toward a larger difference. }
\label{fig:feh_hist}
\end{figure}

Figs.~\ref{fig:teff_logg_smoothed_coolwarm} and \ref{fig:delta_columns} show the change in unresolved and resolved [Fe/H]. In Fig.~\ref{fig:teff_logg_smoothed_coolwarm}, there appears to be a clear trend differentiating stars below and above the main-sequence turn-off around $T_\mathrm{eff} \sim 5500\,\mathrm{K}$. Specifically, below $5500\,\text{K}$ we find that stars with lower unresolved [Fe/H] show an increase in resolved [Fe/H], suggesting that the [Fe/H] values for these stars are often underestimated when a single-star model is incorrectly applied. 
The mean [Fe/H] of the binary spectra compared to that of the remaining stars shows a difference of 0.16 dex (Fig.~\ref{fig:feh_hist}).

Inspecting the stars with the highest changes in [Fe/H], \mbox{|$\Delta$ [Fe/H]| > 0.1|}, reveals that these are systems where the primary dominates; producing a significant fraction of the total flux. In contrast to cooler stars, in hotter binaries the resolved [Fe/H] is often lower than that inferred from single-star fits --- indicating that single-star models tend to overestimate [Fe/H] in these systems. One possible explanation is that hotter stars exhibit fewer and broader spectral lines, and the introduction of a secondary adds additional features. In a single-star model, the fitting algorithm may misinterpret this added structure as evidence of higher [Fe/H], since it lacks the flexibility to attribute the lines to a second, cooler component. Consequently, the [Fe/H] value is artificially inflated. When the binary is correctly modelled, the excess structure is properly attributed to the secondary, and the inferred [Fe/H] decreases to better reflect the true photospheric values.

This behaviour differs from that observed in cooler binaries, where even in systems with near-equal flux contributions, the fitting routine often converges on a single component and effectively ignores the secondary. In these cases, the additional lines from the secondary are not misinterpreted as higher [Fe/H] but instead partially fill in the primary's absorption lines, making them appear weaker in the combined spectrum than they would for the primary alone, leading to an underestimation of [Fe/H]. This dilution effect contrasts with the misidentification seen in hotter stars, suggesting that the temperature of the primary plays a critical role: in hot stars, where general line density is low, any added structure is more easily mistaken for higher [Fe/H]; whereas in cooler stars, fitting algorithms often converge on one dominant component and treat additional features as noise or dilution, rather than enrichment.

\begin{figure}
\centering
\includegraphics[width=\columnwidth]{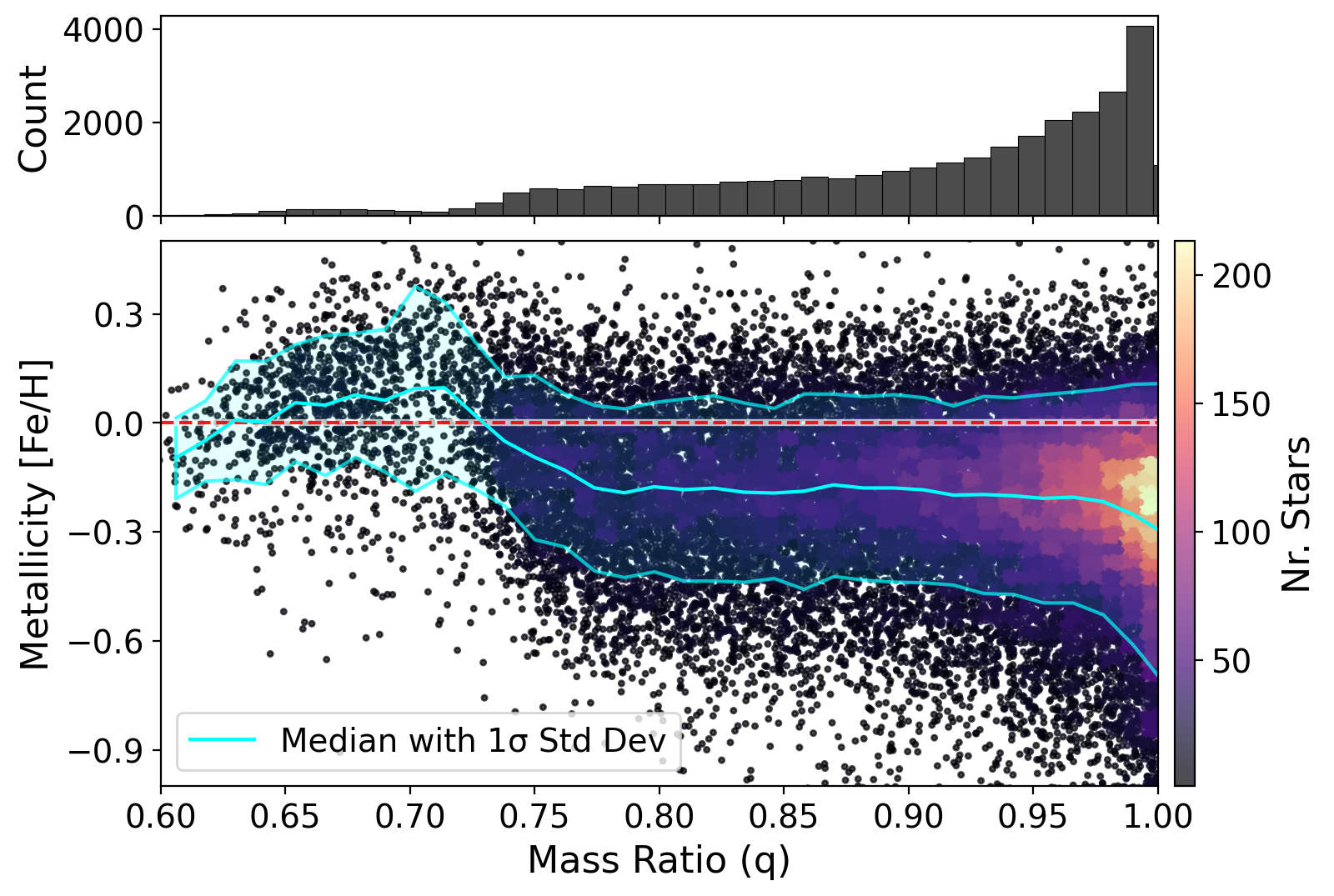}
\caption{Distribution of [Fe/H] as a function of mass ratio $\frac{M_1}{M_2}$. A linear fit (red dashed line) shows that equal-mass systems tend to have lower [Fe/H] values, or systems with lower [Fe/H] values preferentially have equal masses.}
\label{fig:fe_h_mass}
\end{figure}

To further probe formation pathways, we examined the relationship between the mass ratio and [Fe/H]. If metal-rich systems show a higher fraction of equal-mass twins, this could indicate that enhanced cooling and fragmentation in metal-rich disks influences binary formation as generally accepted by literature \cite{batten_binary_1973, bonnell_formation_1994, tokovinin_formation_2020}. In Fig.~\ref{fig:fe_h_mass} we examine resolved MS binary systems with ages < 10 Gyr and find the opposite trend albeit insignificant; equal mass systems have lower [Fe/H]. The difference in mean [Fe/H] between equal mass systems and the most disparate systems was found to be 0.02 dex.

\subsubsection{Age} \label{sec:discussion_parameter_changes_age}

\changed{While the ages of most unresolved stars remain broadly consistent with their resolved counterparts, we observe the largest discrepancies near $T_{\mathrm{eff}} \sim 6700$\,K and $\log g \sim 3.6$ (Fig.~\ref{fig:teff_logg_smoothed_coolwarm}). This region corresponds to a transitional evolutionary phase where stars are turning off the main sequence onto the subgiant branch; stellar parameters change rapidly with age here, producing heightened sensitivity to small variations in $T_{\mathrm{eff}}$ or $\log g$ \citep{soderblom_ages_2010}.}

\changed{This sensitivity is further compounded for unresolved binaries composed of similar-mass hot stars. Such systems are mischaracterised as more massive single stars, biasing age estimates toward younger values. Once the components are resolved, the inferred masses are lower and the estimated ages increase accordingly. The relatively modest observed shifts ($\sim$0.01\,Gyr) reflect the fact that our conservative quality cuts preferentially retain systems where both components are well-constrained, limiting the sample to cases where the binary bias on the inferred parameters — and hence on age — is comparatively small and our age uncertainties are much smaller than is typical in the literature.}

\changed{More broadly, the accuracy of isochronal ages depends on both the quality of the input spectroscopic parameters and on the stellar models themselves. For field stars, isochronal ages carry typical uncertainties of $\sim$20–50\%. These errors are dominated by uncertainties in \teff, typically 80–100 K, which alone can produce age offsets of $\sim1$ Gyr for stars near the turn-off older than 5 Gyr. Our PARSEC-based masses are accurate at the $\sim$3--5\% level for well-constrained main-sequence stars, degrading for evolved and cool stars \citep{serenelli_weighing_2021}. Independent benchmarks from cluster membership and asteroseismology \citep{soderblom_ages_2010, serenelli_weighing_2021} offer higher-fidelity age estimates; the latter is particularly valuable as it constrains $\log g$ directly from oscillation frequencies, bypassing the spectroscopic biases that affect our single-star fits. Our recovered $T_{\rm eff}$ and $\log g$ biases lie within the regime where asteroseismic ages have been shown to remain robust \citep{bellinger_stellar_2019}, and a future cross-match with \textit{Kepler} or TESS asteroseismic samples would provide a direct validation of our resolved component ages.}

\subsection{Putting our binary fraction estimates into context} \label{sec:discussion_binary_fraction}

In this section, we are comparing our GALAH-based estimates of binary fractions for close, line-splitting spectroscopic binaries with the literature. We discuss our completeness and then define a binary fraction in Sec.~\ref{sec:discussion_binary_fraction_completeness}, before looking at the fraction as a function of (primary) mass (Sec.~\ref{sec:discussion_binary_fraction_mass}), metallicity (Sec.~\ref{sec:discussion_binary_fraction_metallicity}), and 
stellar types (Sec.~\ref{sec:discussion_binary_fraction_stellar_types}).

\subsubsection{A note on completeness and our definition of binary fraction} \label{sec:discussion_binary_fraction_completeness}

\begin{figure}
\centering
\includegraphics[width=\columnwidth]{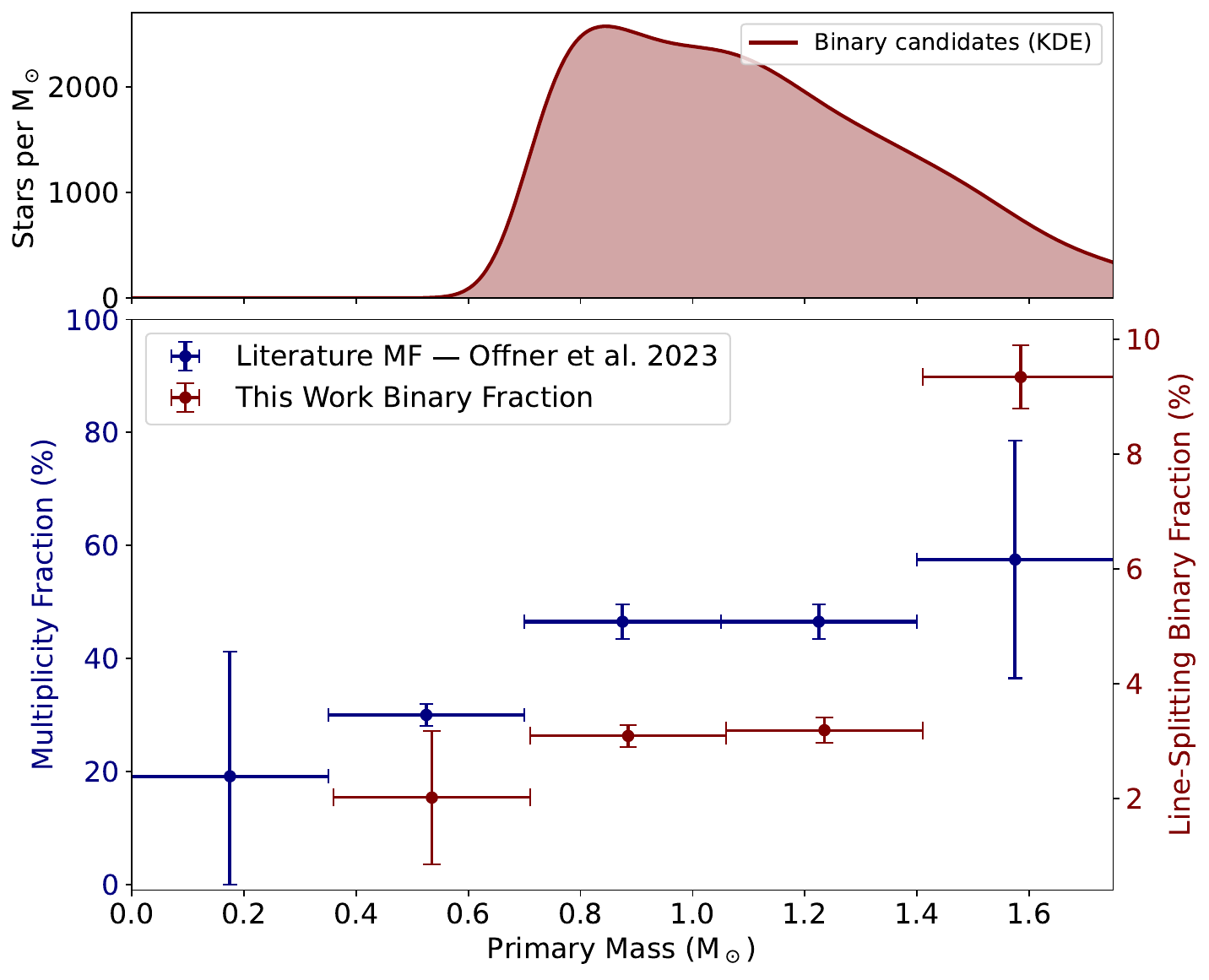}
\caption{Multiplicity fraction as a function of primary mass for the literature estimates by \citet[][blue]{offner_origin_2023} and our estimates (red). Note that we show two different scales for the multiplicity fraction, as our estimates only include line-splitting binaries while \citet{offner_origin_2023} show total multiplicity fraction. Our analysis includes no data for the first bin from Offner ($0<M_\odot<0.4$). A KDE showing the number of stars per $M_\odot$ is shown above.}
\label{fig:bf_v_mass}
\end{figure}

A crude estimate of the expected number of binaries in the GALAH survey can be obtained by applying the mass distribution of GALAH stars to the most recent, bias-corrected multiplicity fraction as a function of stellar mass. \citet{offner_origin_2023} provide a comprehensive review of multiplicity surveys, and their compiled distribution in five different primary mass bins is shown in Fig.~\ref{fig:bf_v_mass}. By weighting the GALAH DR4 stellar mass distribution (with detected binaries from this study removed) with this multiplicity function, we find that approximately 48\% of stars in the survey are predicted to be binaries. However, there are important and significant limitations to this estimate:

Assessing the completeness of our binary detection in GALAH is inherently limited by several factors. First, the survey design typically provides only a single observation per star, restricting our sensitivity to binaries with sufficiently large instantaneous velocity separations ($\Delta$RV) and favourable flux ratios. Second, the GALAH pipeline explicitly excludes objects that appear as extended sources rather than point sources from the main parameter analysis, further biasing our sample against the detection of binaries, particularly double-lined systems (SB2s). As a result, our search focussed only on SB2 binaries with large $\Delta$RV and comparable component luminosities at the time of observation, and is insensitive to SB1s, close SB2s with blended lines, and systems with large flux ratios. We thus expect to only identify a subset of the overall binaries. Nonetheless, we can study the similarity of SB2 fractions across the same properties as previous more complete studies.

\changed{We also note that hotter and faster rotating systems are more common in our sample as seen in Figure \ref{fig:resolution_cut}. This effect has been found in the literature, with binaries typically being more prevalent with increasing mass and temperature \citep{offner_origin_2023}, and increasing temperature also correlating with increased rotation \citep{barnes_dependence_2005}. As a result we are more likely to find stars approaching the rotation limit of our model above 30\kms. In addition, our selection is biased toward brighter objects because we impose S/N thresholds and GALAH is itself magnitude limited, so intrinsically more luminous dwarf stars enter the parent sample more readily.} 

For the subsequent comparisons, we define the binary fraction as the ratio of those stars that have passed our rejection criteria (see Tab.~\ref{tab:data_cuts}) and consider all remaining spectra (assumed to be single-star systems) that passed a basic quality selection cut to be considered a reasonable parent sample. For the remaining sample, we only included spectra with $\log g > 3.5$, $4000\,\text{K} < $\teff$ < 7000\,\text{K}$, and the same two S/N cuts as described in Tab. \ref{tab:data_cuts}.
\subsubsection{Binary fraction and mass ratios across (primary) mass} \label{sec:discussion_binary_fraction_mass}

In qualitative agreement with \citet{offner_origin_2023}, we find a significant relative increase of binary fraction with increasing stellar primary mass. In particular, we find an increase by more than a factor of 2 between sub-solar and super-solar primary masses. The lower absolute fractions can be explained by the study setups. While \citet{offner_origin_2023} looked at the overall bias-corrected multiplicity fraction, our study focusses on only the subset of line-splitting SB2 binaries and neglect wide binaries or SB1s.

Similar to previous studies \citep[e.g.][]{el-badry_gaias_2024}, we can also study the mass ratios $q=\frac{m_2}{m_1}$ of our binary sample (see Fig.~\ref{fig:fe_h_mass}). Similarly to previous studies, we also find an over-representation of equal mass ($q \sim 1$) binaries or ``twin binaries'' where the mass ratio is inconsistent with random sampling of pairs from the initial mass function. We find 37\% of binaries in our sample to belong to this population of twins with mass ratios $q > 0.9$. This has been particularly noted to be the case for close binaries at small separations ($\approx 0.1 \text{AU}$) due to the theorised formation mechanisms that serve to equalise the primordial stellar masses \citep{tokovinin_architecture_2021, el-badry_gaias_2024}. 

\subsubsection{Binary fraction across metallicity} \label{sec:discussion_binary_fraction_metallicity}

\begin{figure}
\centering
\includegraphics[width=8cm]{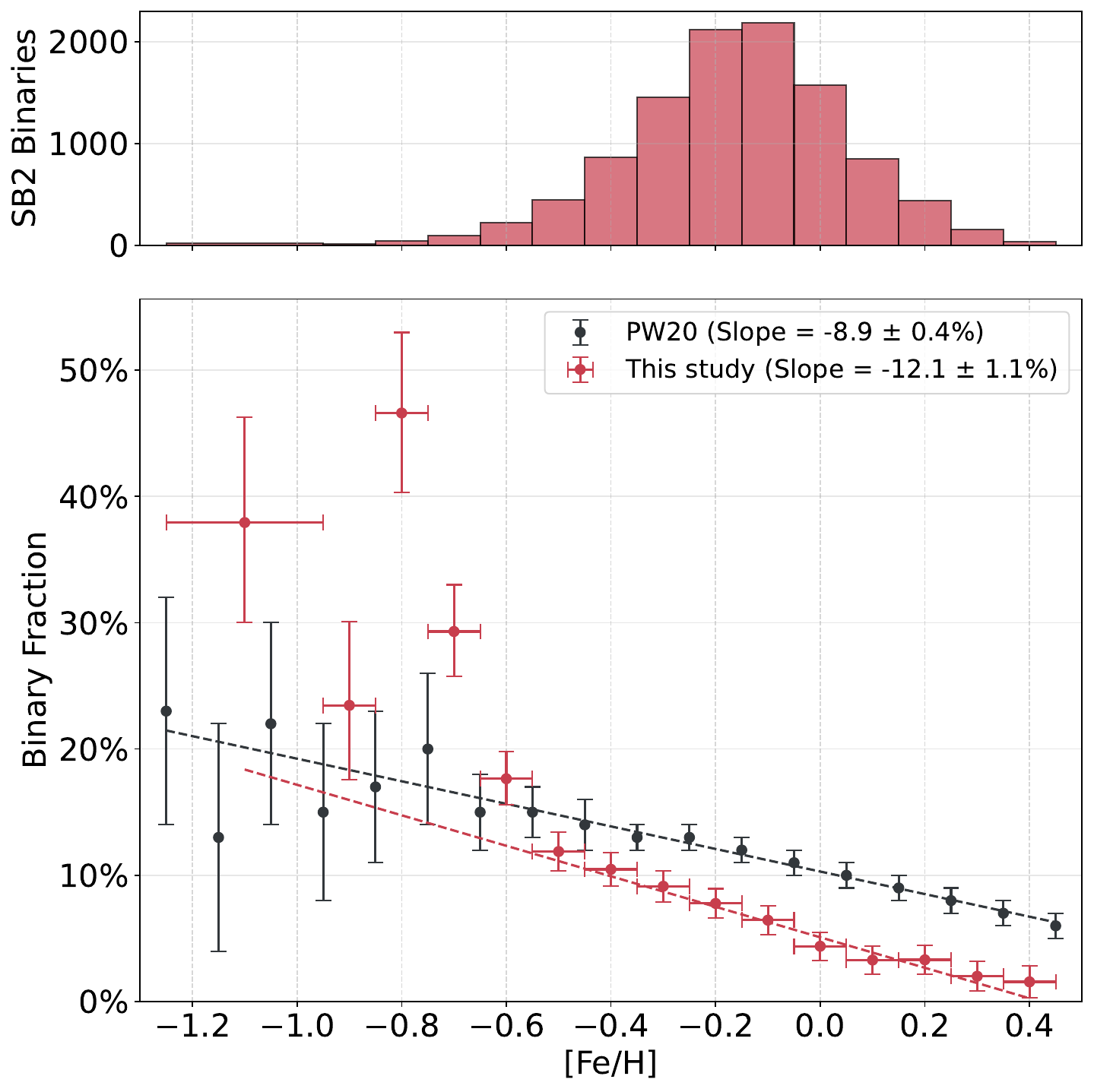}
\caption{Metallicity vs. number of binaries (top panel) and binary fraction (bottom panel). In the bottom panel, we show both our estimates as well as those by \citet{price-whelan_close_2020} binned by metallicity. Linear fits (dashed lines) are shown for each data set. We approximate our error as the $ 1\sigma $ binomial uncertainty on the binary fraction in each metallicity bin (smallest SB2 bin contains 75 objects). Note that we do not attempt to include our detection efficiency here. Slopes are consistent however we note we find a reduced binary fraction across most bins. We found the binary fraction to be highly dependent on which cuts are made to the data, which spectral types are used and how they are classified --- e.g. classifying main sequence stars according to photometry such as in \citet{raghavan_survey_2010} vs spectroscopically by \citet{price-whelan_close_2020}.}
\label{fig:binary_fraction_vs_metallicity_all}
\end{figure}

The binary fraction as a function of [Fe/H] is shown in Fig.~\ref{fig:binary_fraction_vs_metallicity_all}. As reported by previous studies \citep{moe_close_2019, price-whelan_close_2020}, we find the fraction of binary systems decreases with increasing metallicity. In particular, we find an increase of SB2 fraction from \changed{$\sim$5\,\%} for super-solar metallicities to roughly \changed{40\,\%} for stars around $\mathrm{[Fe/H]} \sim -1$. We note that our trend is in qualitative agreement with \citet{price-whelan_close_2020}. We note that we do not account for our detection efficiency or the selection function of GALAH DR4 and as such find smaller binary fractions for at higher metallicities than in \cite{price-whelan_close_2020} or \cite{raghavan_survey_2010}. Although not shown here, we also find a good agreement with the bias-corrected close binary fractions collected by \citet{moe_close_2019}, who found roughly 20\,\% around $\mathrm{[Fe/H]} \sim 0$. They find, however a higher binary fraction of 40\,\% for $\mathrm{[Fe/H]} \sim -1$ and even higher fractions below $\mathrm{[Fe/H]} < -1$ that rise slightly above 50\% (on average) at $\mathrm{[Fe/H]} \sim -3$.

\subsubsection{Binary fraction across stellar types} \label{sec:discussion_binary_fraction_stellar_types}

\begin{figure}
  \centering 
  \includegraphics[width=\columnwidth]{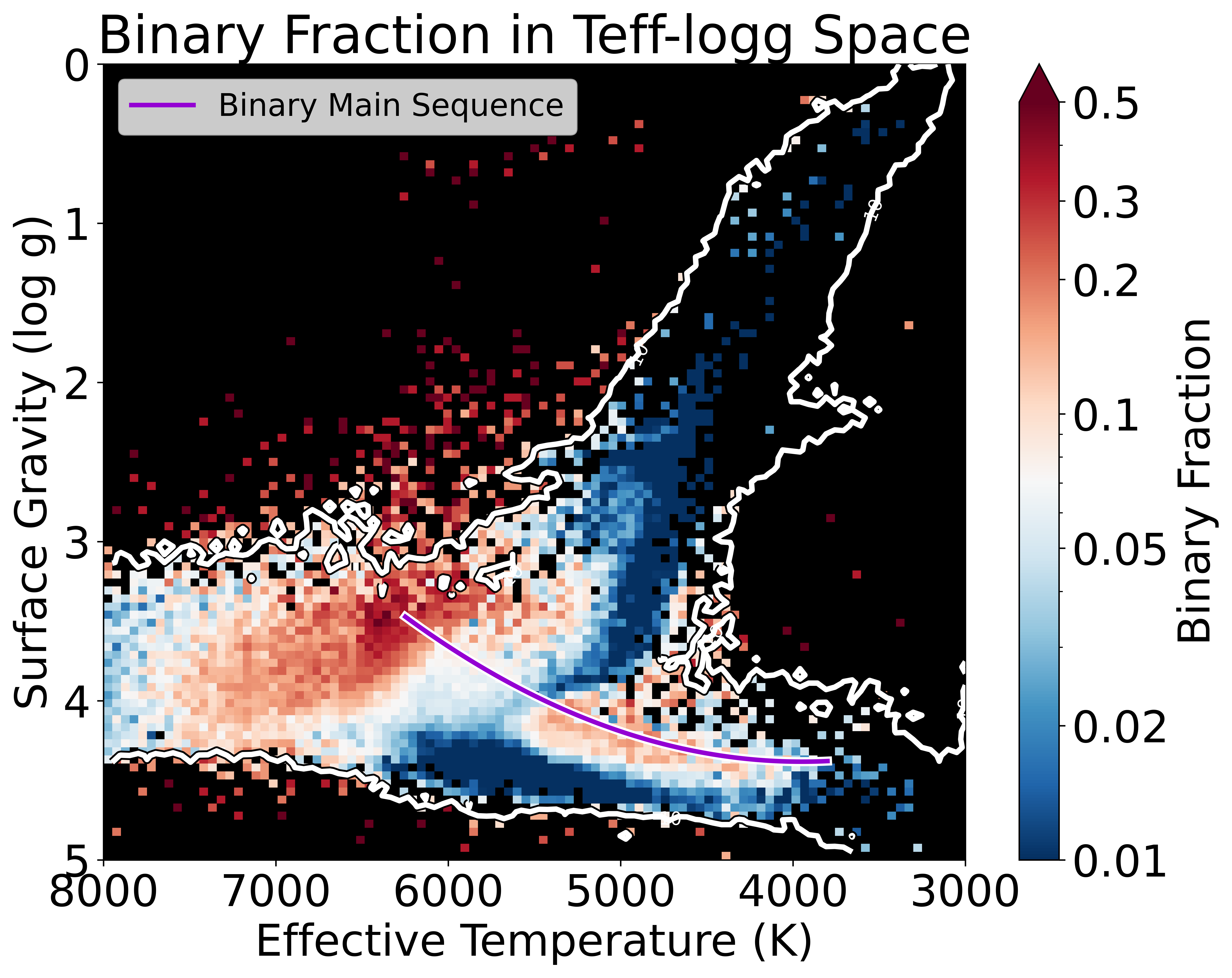}
  \caption{Binary fraction across the $T_\mathrm{eff}$-$\log g$ diagram, showing significantly increased fractions for hotter main-sequence and turn-off stars as well as the binary main sequence (indicated with a purple line).}
  \label{fig:binary_fraction_teff_logg}
\end{figure}

To better understand the binary fraction across stellar types, we estimate the binary fraction for individual bins in $T_\mathrm{eff}$-$\log g$ space as shown in Fig.~\ref{fig:binary_fraction_teff_logg}. Here we note the particularly high binary fractions above 10\,\% for hotter main-sequence turn-off stars (reaching up to 50\,\%) as well as the binary main sequence (around $\log g \sim 4$ for $T_\mathrm{eff} < 5500\,\mathrm{K}$). This finding is again in very good qualitative agreement with the close binary fraction as mapped by \citet{price-whelan_close_2020} for APOGEE DR16 (see their Fig.~5). In agreement with our estimates, they only find smaller percentages of binaries along the cooler main sequences and for evolved stars along the red giant branch.

\section{Conclusions} \label{sec:conclusions}

\subsection{Take-away}
\changed{In this study, we developed a bespoke extension to the GALAH DR4 spectrum analysis pipeline to fit binary spectra and extract stellar parameters of both components.} Our final catalogue contains \changed{12,911} candidate binary systems, of which \changed{6,683} overlap with the 25,191 objects flagged in GALAH DR4. \changed{The remaining DR4-flagged objects did not survive quality cuts following our analysis, suggesting they are likely unreliable detections, which was expected due to its rudimentary flagging routine. Whilst our data cuts also reduced the number of objects from cross-matched catalogues including \citet{traven_galah_2020} ($\sim40\%$ remaining) and \textit{Gaia} ($\sim15\%$ remaining), this attrition reflects the result of our purely spectroscopic approach. Our pipeline extracts stellar parameters for both components by fitting the composite spectrum, adopting the primary's chemical abundances, as derived from the single-star GALAH DR4 analysis, as a fixed prior when modelling the secondary. While this assumes chemical homogeneity between components, it provides a more self-consistent characterisation of each system than approaches that fit a single-star model to the blended spectrum.}

\changed{Our study yields several practical learnings for detecting and modelling spectroscopic binaries at scale. The CCF proved a fast and reliable tool for identifying clear double-lined systems, but is sensitive to its peak-detection thresholds and routinely misses binaries that are unambiguous in a full spectral fit. Among the optimization algorithms tested, PSO outperformed BFGS and \textsc{scipy.optimize.curve\_fit} in spectral fit quality across the full sample, and we adopt a hybrid PSO+BFGS strategy to recover poorly-converged cases. However, our bespoke PSO-derived uncertainties are systematically 5--10 times larger than MCMC estimates, indicating that an adaptively decaying inertia weight or extended run is needed for reliable posterior characterisation. Our conservative quality cuts achieve high sample purity at the cost of completeness, as evidenced by the substantial attrition from external catalogues. While this reduces the total number of recovered systems, it preferentially retains binaries whose component signatures can be reliably separated and modelled. Such systems are particularly valuable because they provide not only cleaner constraints on binary-induced parameter biases but also a foundation for future component-level analyses.}


\changed{We also quantified the impact of unresolved binarity on the determination of key stellar parameters in the sample of high-confidence binary systems from GALAH DR4. We find that \teff can be affected by as much as \(-142\) K to \(+200\) K, \logg by \(-0.070\) to \(+0.346\) dex, [Fe/H] by \(-0.055\) to \(+0.129\) dex, and age by \(-0.015\) to \(0.018\) Gyr.} These biases arise from the confluence of multiple factors, including the blending of spectral lines, which lead to degeneracies of stellar parameters. The influence of binarity is particularly significant for systems where the secondary \changed{contributes insufficiently distinct signatures to be separated robustly,} leading to systematic errors in parameter estimations.

While we caution against the completeness of our sample, we find similar trends to previous literature: increasing binary fraction with increasing primary mass \citep{offner_origin_2023}, and increasing binary fractions with decreasing [Fe/H] \citep{moe_close_2019, price-whelan_close_2020}. These fractions vary across stellar types, with the main-sequence turn-off region showing the highest binary fractions in addition to the binary main-sequence region, again in good agreement with previous studies \citep{price-whelan_close_2020}.

\subsection{Future Work}

One important avenue for future work is improving the modularity of the analysis pipeline. While the current method is designed around a specific dataset and resolution, future versions should aim to generalize to a wider range of spectra, including those at different spectral resolutions and signal-to-noise levels. \changed{Adopting per-CCD fitting and masking the H$\alpha$ region during spectral fitting are targeted improvements that could reduce systematic \teff biases and should be evaluated in future work (Section~\ref{sec:discussion_detection_modelling}).} Expanding the pipeline's flexibility would also allow it to handle more complex systems such as triple or higher-order multiples, and potentially fainter binaries where the secondary is harder to detect. Building this modularity will be key for applying the method to diverse spectroscopic surveys.

\changed{Another priority is the study of equal-mass spectroscopic binaries and systems with small or no $\Delta$RV separation, which our purely CCF-based detection systematically misses. These systems can be particularly challenging because their spectral features overlap significantly, making detection and characterisation more difficult. For general zero-$\Delta$RV systems, the improvement fraction ($f_\mathrm{imp}$) introduced by \citet{el-badry_discovery_2018} offers a complementary identification route: even when both components share the same radial velocity, a composite binary model can still yield a statistically better fit than a single-star model. Techniques leveraging the \vsini-inflation diagnostic \citep{kovalev_detection_2022, kovalev_detection_2024} provide a further route, enabling detection at separations as low as $\sim$8\kms and potentially extending to zero separation where component spectra are sufficiently distinct. For equal-mass zero-$\Delta$RV systems in particular, their doubled luminosity relative to a single star of the same spectral type provides an additional photometric diagnostic when parallax measurements are available --- making them well suited to a dedicated future study combining spectroscopic and astrometric data.}

Finally, extending the current methodology to derive accurate elemental abundances for both components in binary systems is a \changed{natural} next step. \changed{The present work intentionally prioritises systems whose component spectra can be robustly separated, creating a sample well suited for such analyses. In favourable systems with sufficiently separated spectral features, including resolved absorption signatures from both components, abundance information may be extracted individually for each star. This would allow unresolved binaries to transition from a source of systematic uncertainty into an additional astrophysical laboratory and enable studies of abundance differences within binary systems, tests of stellar evolution, and improved applications of chemical tagging and Galactic Archaeology.}

\section*{Acknowledgments}

SB acknowledges support from the Australian Research Council under grant numbers DE240100150 and CE170100013.
GT acknowledges financial support of the Slovenian Research Agency (research core funding No. P1-0188) and the European Space Agency (Prodex Experiment Arrangement No. 4000143450).

This research is supported by an Australian Government Research Training Program (RTP) Scholarship \url{doi.org/10.82133/C42F-K220}.

\section*{Data Availability}

All code to reproduce the analysis and figures can be publicly accessed via \url{https://github.com/YaniL12/binaries-GALAH}. Access to data from GALAH DR4 is available via \citep{buder_galah_2025}.

\section*{Supporting Information}

Supplementary data are available at \textit{MNRAS} online.

\begin{description}[leftmargin=0pt,labelindent=0pt,itemsep=0pt,parsep=0pt]
\item \textbf{Figure S1.} The improvement fraction $f_{\rm imp}$ cut-off overlaid on all data, and after all other cuts excluding $f_{\rm imp}$.
\item \textbf{Table S1.} Sequential data cuts for each selection criterion.
\item \textbf{Figure S2.} \vsini vs $\Delta$RV distribution of items in the binary catalogue before cuts for the primary.
\item \textbf{Figure S3.} Distributions of \vsini vs \teff for all spectra in the final catalogue.
\item \textbf{Figure S4.} Comparison of parameters for spectra across repeat observations.
\end{description}

Please note: Oxford University Press is not responsible for the content or functionality of any supporting materials supplied by the authors. Any queries (other than missing material) should be directed to the corresponding author for the article.

\bibliographystyle{mnras}
\bibliography{zot_260113,bib}

@article{kovalev_detection_2022,
    title = {Detection of 2460 double-lined spectroscopic binary candidates in the {LAMOST}-{MRS} using projected rotational velocities and a binary spectral model},
    volume = {517},
    issn = {0035-8711},
    url = {https://ui.adsabs.harvard.edu/abs/2022MNRAS.517..356K},
    doi = {10.1093/mnras/stac2513},
    urldate = {2026-04-29},
    journal = {Monthly Notices of the Royal Astronomical Society},
    publisher = {OUP},
    author = {Kovalev, Mikhail and Chen, Xuefei and Han, Zhanwen},
    month = nov,
    year = {2022},
    note = {ADS Bibcode: 2022MNRAS.517..356K},
    pages = {356--373},
}

@article{kovalev_detection_2024,
    title = {Detection of 12 426 {SB2} candidates in the {LAMOST}-{MRS}, using a binary spectral model},
    volume = {527},
    issn = {0035-8711},
    url = {https://doi.org/10.1093/mnras/stad3222},
    doi = {10.1093/mnras/stad3222},
    number = {1},
    urldate = {2026-04-29},
    journal = {Monthly Notices of the Royal Astronomical Society},
    author = {Kovalev, Mikhail and Zhou, Zenghua and Chen, Xuefei and Han, Zhanwen},
    month = jan,
    year = {2024},
    pages = {521--530},
}

@article{serenelli_weighing_2021,
    title = {Weighing stars from birth to death: mass determination methods across the {HRD}},
    volume = {29},
    issn = {0935-4956},
    shorttitle = {Weighing stars from birth to death},
    url = {https://ui.adsabs.harvard.edu/abs/2021A&ARv..29....4S},
    doi = {10.1007/s00159-021-00132-9},
    urldate = {2026-05-06},
    journal = {Astronomy and Astrophysics Review},
    publisher = {Springer},
    author = {Serenelli, Aldo and Weiss, Achim and Aerts, Conny and Angelou, George C. and Baroch, David and Bastian, Nate and Beck, Paul G. and Bergemann, Maria and Bestenlehner, Joachim M. and Czekala, Ian and Elias-Rosa, Nancy and Escorza, Ana and Van Eylen, Vincent and Feuillet, Diane K. and Gandolfi, Davide and Gieles, Mark and Girardi, Léo and Lebreton, Yveline and Lodieu, Nicolas and Martig, Marie and Miller Bertolami, Marcelo M. and Mombarg, Joey S. G. and Morales, Juan Carlos and Moya, Andrés and Nsamba, Benard and Pavlovski, Krešimir and Pedersen, May G. and Ribas, Ignasi and Schneider, Fabian R. N. and Silva Aguirre, Victor and Stassun, Keivan G. and Tolstoy, Eline and Tremblay, Pier-Emmanuel and Zwintz, Konstanze},
    month = dec,
    year = {2021},
    note = {ADS Bibcode: 2021A\&ARv..29....4S},
    pages = {4},
}

@article{bellinger_stellar_2019,
    title = {Stellar ages, masses, and radii from asteroseismic modeling are robust to systematic errors in spectroscopy},
    volume = {622},
    copyright = {© ESO 2019},
    issn = {0004-6361, 1432-0746},
    url = {https://www.aanda.org/articles/aa/abs/2019/02/aa34461-18/aa34461-18.html},
    doi = {10.1051/0004-6361/201834461},
    language = {en},
    urldate = {2026-05-06},
    journal = {Astronomy \& Astrophysics},
    publisher = {EDP Sciences},
    author = {Bellinger, E. P. and Hekker, S. and Angelou, G. C. and Stokholm, A. and Basu, S.},
    month = feb,
    year = {2019},
    pages = {A130},
}

@article{li_massratio_2022,
    title = {Mass-ratio {Distribution} of {Binaries} from the {LAMOST}-{MRS} {Survey}},
    volume = {933},
    issn = {0004-637X},
    url = {https://doi.org/10.3847/1538-4357/ac731d},
    doi = {10.3847/1538-4357/ac731d},
    language = {en},
    number = {2},
    urldate = {2026-05-06},
    journal = {The Astrophysical Journal},
    publisher = {The American Astronomical Society},
    author = {Li, Jiangdan and Li, Jiao and Liu, Chao and Li, Chunqian and Guo, Yanjun and Wang, Luqian and Chen, Xuefei and Xing, Lifeng and Hou, Yonghui and Han, Zhanwen},
    month = jul,
    year = {2022},
    pages = {119},
}

@article{schlesinger_binary_2010,
    author = {Schlesinger, Katharine J. and Johnson, Jennifer A. and Lee, Young Sun and Masseron, Thomas and Yanny, Brian and Rockosi, Constance M. and Gaudi, B. Scott and Beers, Timothy C.},
    title = {Binary Contamination in the {SEGUE} Sample: Effects on {SSPP} Determinations of Stellar Atmospheric Parameters},
    journal = {The Astrophysical Journal},
    volume = {719},
    pages = {996--1020},
    year = {2010},
    doi = {10.1088/0004-637X/719/2/996},
}

@article{soderblom_ages_2010,
    author = {Soderblom, David R.},
    title = {The Ages of Stars},
    journal = {Annual Review of Astronomy and Astrophysics},
    volume = {48},
    pages = {581--629},
    year = {2010},
    doi = {10.1146/annurev-astro-081309-130806},
}

@article{nidever_data_2015,
	title = {The {Data} {Reduction} {Pipeline} for the {Apache} {Point} {Observatory} {Galactic} {Evolution} {Experiment}},
	volume = {150},
	issn = {0004-6256},
	url = {https://ui.adsabs.harvard.edu/abs/2015AJ....150..173N},
	doi = {10.1088/0004-6256/150/6/173},
	urldate = {2026-04-02},
	journal = {The Astronomical Journal},
	publisher = {IOP},
	author = {Nidever, David L. and Holtzman, Jon A. and Allende Prieto, Carlos and Beland, Stephane and Bender, Chad and Bizyaev, Dmitry and Burton, Adam and Desphande, Rohit and Fleming, Scott W. and García Pérez, Ana E. and Hearty, Fred R. and Majewski, Steven R. and Mészáros, Szabolcs and Muna, Demitri and Nguyen, Duy and Schiavon, Ricardo P. and Shetrone, Matthew and Skrutskie, Michael F. and Sobeck, Jennifer S. and Wilson, John C.},
	month = dec,
	year = {2015},
	note = {ADS Bibcode: 2015AJ....150..173N},
	pages = {173},
}

@article{el-badry_wide_2019,
	adsurl = {https://ui.adsabs.harvard.edu/abs/2019MNRAS.482L.139E},
	archiveprefix = {arXiv},
	author = {{El-Badry}, Kareem and {Rix}, Hans-Walter},
	doi = {10.1093/mnrasl/sly206},
	eprint = {1809.06860},
	journal = {\mnras},
	month = jan,
	number = {1},
	pages = {L139-L144},
	primaryclass = {astro-ph.SR},
	title = {{The wide binary fraction of solar-type stars: emergence of metallicity dependence at a < 200 au}},
	volume = {482},
	year = 2019}

@article{el-badry_gaias_2024,
	adsurl = {https://ui.adsabs.harvard.edu/abs/2024NewAR..9801694E},
	archiveprefix = {arXiv},
	author = {{El-Badry}, Kareem},
	doi = {10.1016/j.newar.2024.101694},
	eid = {101694},
	eprint = {2403.12146},
	journal = {\nar},
	month = jun,
	pages = {101694},
	primaryclass = {astro-ph.SR},
	title = {{Gaia's binary star renaissance}},
	volume = {98},
	year = 2024}

@article{price-whelan_close_2020,
	adsurl = {https://ui.adsabs.harvard.edu/abs/2020ApJ...895....2P},
	archiveprefix = {arXiv},
	author = {{Price-Whelan}, Adrian M. and {Hogg}, David W. and {Rix}, Hans-Walter and {Beaton}, Rachael L. and {Lewis}, Hannah M. and {Nidever}, David L. and {Almeida}, Andr{\'e}s and {Badenes}, Carles and {Barba}, Rodolfo and {Beers}, Timothy C. and {Carlberg}, Joleen K. and {De Lee}, Nathan and {Fern{\'a}ndez-Trincado}, Jos{\'e} G. and {Frinchaboy}, Peter M. and {Garc{\'\i}a-Hern{\'a}ndez}, D.~A. and {Green}, Paul J. and {Hasselquist}, Sten and {Longa-Pe{\~n}a}, Pen{\'e}lope and {Majewski}, Steven R. and {Nitschelm}, Christian and {Sobeck}, Jennifer and {Stassun}, Keivan G. and {Stringfellow}, Guy S. and {Troup}, Nicholas W.},
	doi = {10.3847/1538-4357/ab8acc},
	eid = {2},
	eprint = {2002.00014},
	journal = {\apj},
	month = may,
	number = {1},
	pages = {2},
	primaryclass = {astro-ph.SR},
	title = {{Close Binary Companions to APOGEE DR16 Stars: 20,000 Binary-star Systems Across the Color-Magnitude Diagram}},
	volume = {895},
	year = 2020}

@inproceedings{offner_origin_2023,
	adsurl = {https://ui.adsabs.harvard.edu/abs/2023ASPC..534..275O},
	archiveprefix = {arXiv},
	author = {{Offner}, S.~S.~R. and {Moe}, M. and {Kratter}, K.~M. and {Sadavoy}, S.~I. and {Jensen}, E.~L.~N. and {Tobin}, J.~J.},
	booktitle = {Protostars and Planets VII},
	doi = {10.48550/arXiv.2203.10066},
	editor = {{Inutsuka}, S. and {Aikawa}, Y. and {Muto}, T. and {Tomida}, K. and {Tamura}, M.},
	eprint = {2203.10066},
	month = jul,
	pages = {275},
	primaryclass = {astro-ph.SR},
	series = {Astronomical Society of the Pacific Conference Series},
	title = {{The Origin and Evolution of Multiple Star Systems}},
	volume = {534},
	year = 2023}

@article{el-badry_discovery_2018,
	author = {El-Badry, Kareem and Ting, Yuan-Sen and Rix, Hans-Walter and Quataert, Eliot and Weisz, Daniel R. and Cargile, Phillip and Conroy, Charlie and Hogg, David W. and Bergemann, Maria and Liu, Chao},
	doi = {10.1093/mnras/sty240},
	issn = {0035-8711},
	journal = {\mnras},
	month = may,
	note = {Publisher: OUP ADS Bibcode: 2018MNRAS.476..528E},
	pages = {528--553},
	title = {Discovery and characterization of 3000+ main-sequence binaries from {APOGEE} spectra},
	url = {https://ui.adsabs.harvard.edu/abs/2018MNRAS.476..528E},
	urldate = {2025-12-07},
	volume = {476},
	year = {2018}}

@article{foreman-mackey_emcee_2013,
	author = {Foreman-Mackey, Daniel and Hogg, David W. and Lang, Dustin and Goodman, Jonathan},
	doi = {10.1086/670067},
	issn = {0004-6280},
	journal = {\pasp},
	month = mar,
	note = {Publisher: IOP ADS Bibcode: 2013PASP..125..306F},
	pages = {306},
	shorttitle = {emcee},
	title = {emcee: {The} {MCMC} {Hammer}},
	url = {https://ui.adsabs.harvard.edu/abs/2013PASP..125..306F},
	urldate = {2025-09-29},
	volume = {125},
	year = {2013}}

@article{virtanen_scipy_2020,
	author = {Virtanen, Pauli and Gommers, Ralf and Oliphant, Travis E. and Haberland, Matt and Reddy, Tyler and Cournapeau, David and Burovski, Evgeni and Peterson, Pearu and Weckesser, Warren and Bright, Jonathan and van der Walt, St{\'e}fan J. and Brett, Matthew and Wilson, Joshua and Millman, K. Jarrod and Mayorov, Nikolay and Nelson, Andrew R. J. and Jones, Eric and Kern, Robert and Larson, Eric and Carey, C. J. and Polat, {\.I}lhan and Feng, Yu and Moore, Eric W. and VanderPlas, Jake and Laxalde, Denis and Perktold, Josef and Cimrman, Robert and Henriksen, Ian and Quintero, E. A. and Harris, Charles R. and Archibald, Anne M. and Ribeiro, Ant{\^o}nio H. and Pedregosa, Fabian and van Mulbregt, Paul and {SciPy 1. 0 Contributors}},
	doi = {10.1038/s41592-019-0686-2},
	journal = {Nature Methods},
	month = feb,
	note = {ADS Bibcode: 2020NatMe..17..261V},
	pages = {261--272},
	shorttitle = {{SciPy} 1.0},
	title = {{SciPy} 1.0: fundamental algorithms for scientific computing in {Python}},
	url = {https://ui.adsabs.harvard.edu/abs/2020NatMe..17..261V},
	urldate = {2025-09-26},
	volume = {17},
	year = {2020}}

@article{buder_galah_2025,
	adsurl = {https://ui.adsabs.harvard.edu/abs/2025PASA...42...51B},
	archiveprefix = {arXiv},
	author = {{Buder}, Sven and {Kos}, Janez and {Wang}, Xi Ella and {McKenzie}, Madeleine and {Howell}, Madeleine and {Martell}, Sarah and {Hayden}, Michael R. and {Zucker}, Daniel B. and {Nordlander}, Thomas and {Montet}, Benjamin and {Traven}, Gregor and {Bland-Hawthorn}, Joss and {de Silva}, Gayandhi M. and {Freeman}, Kenneth and {Lewis}, Geraint and {Lind}, Karin and {Sharma}, Sanjib and {Simpson}, Jeffrey D. and {Stello}, Dennis and {Zwitter}, Tomaz and {Amarsi}, Anish M. and {Armstrong}, Joseph J. and {Banks}, Kirsten and {Beavis}, Mark and {Beeson}, Kevin-Luke and {Chen}, Boquan and {Ciuc{\u{a}}}, Ioana and {da Costa}, Gary S. and {de Grijs}, Richard and {Martin}, Bailey and {Nataf}, David Moise and {Ness}, Melissa and {Rains}, Adam D. and {Scarr}, Tim and {Vogrin{\v{c}}i{\v{c}}}, Rok and {Wang}, Zixian Purmortal and {Wittenmyer}, Rob A. and {Xie}, Yi Anne and {The Galah Collaboration}},
	doi = {10.1017/pasa.2025.26},
	eid = {e051},
	eprint = {2409.19858},
	journal = {\pasa},
	month = may,
	pages = {e051},
	primaryclass = {astro-ph.GA},
	title = {{The GALAH survey: Data release 4}},
	volume = {42},
	year = 2025}

@article{barnes_dependence_2005,
	author = {Barnes, J. R. and Collier Cameron, A. and Donati, J. -F. and James, D. J. and Marsden, S. C. and Petit, P.},
	doi = {10.1111/j.1745-3933.2005.08587.x},
	issn = {0035-8711},
	journal = {\mnras},
	month = feb,
	note = {Publisher: OUP ADS Bibcode: 2005MNRAS.357L...1B},
	pages = {L1--L5},
	title = {The dependence of differential rotation on temperature and rotation},
	url = {https://ui.adsabs.harvard.edu/abs/2005MNRAS.357L...1B},
	urldate = {2025-06-11},
	volume = {357},
	year = {2005}}

@article{hurley_evolution_2002,
       author = {{Hurley}, Jarrod R. and {Tout}, Christopher A. and {Pols}, Onno R.},
        title = "{Evolution of binary stars and the effect of tides on binary populations}",
      journal = {\mnras},
         year = 2002,
        month = feb,
       volume = {329},
       number = {4},
        pages = {897-928},
          doi = {10.1046/j.1365-8711.2002.05038.x},
archivePrefix = {arXiv},
       eprint = {astro-ph/0201220},
 primaryClass = {astro-ph},
       adsurl = {https://ui.adsabs.harvard.edu/abs/2002MNRAS.329..897H}
}

@article{freeman_new_2002,
	author = {Freeman, Ken and Bland-Hawthorn, Joss},
	doi = {10.1146/annurev.astro.40.060401.093840},
	issn = {0066-4146, 1545-4282},
	journal = {\araa},
	month = sep,
	note = {arXiv:astro-ph/0208106},
	number = {1},
	pages = {487--537},
	shorttitle = {The {New} {Galaxy}},
	title = {The {New} {Galaxy}: {Signatures} of its {Formation}},
	url = {http://arxiv.org/abs/astro-ph/0208106},
	urldate = {2025-01-31},
	volume = {40},
	year = {2002}}

@article{james_v_miranda_pyswarms_2018,
	author = {James V. Miranda, Lester},
	doi = {10.21105/joss.00433},
	journal = {The Journal of Open Source Software},
	month = jan,
	note = {ADS Bibcode: 2018JOSS....3..433J},
	pages = {433},
	shorttitle = {{PySwarms}},
	title = {{PySwarms}: a research toolkit for {Particle} {Swarm} {Optimization} in {Python}},
	url = {https://ui.adsabs.harvard.edu/abs/2018JOSS....3..433J},
	urldate = {2025-01-20},
	volume = {3},
	year = {2018}}

@article{andrew_binary_2022,
	author = {Andrew, Shion and Penoyre, Zephyr and Belokurov, Vasily and Evans, N. Wyn and Oh, Semyeong},
	doi = {10.1093/mnras/stac2532},
	issn = {0035-8711, 1365-2966},
	journal = {\mnras},
	month = sep,
	note = {arXiv:2206.04392 [astro-ph]},
	number = {3},
	pages = {3661--3684},
	title = {Binary parameters from astrometric and spectroscopic errors--candidate hierarchical triples and massive dark companions in {Gaia} {DR3}},
	url = {http://arxiv.org/abs/2206.04392},
	urldate = {2025-01-16},
	volume = {516},
	year = {2022}}

@ARTICLE{penoyre_astrometric_2022-1,
       author = {{Penoyre}, Zephyr and {Belokurov}, Vasily and {Evans}, N. Wyn},
        title = "{Astrometric identification of nearby binary stars - I. Predicted astrometric signals}",
      journal = {\mnras},
         year = 2022,
        month = jun,
       volume = {513},
       number = {2},
        pages = {2437-2456},
          doi = {10.1093/mnras/stac959},
archivePrefix = {arXiv},
       eprint = {2111.10380},
 primaryClass = {astro-ph.SR},
       adsurl = {https://ui.adsabs.harvard.edu/abs/2022MNRAS.513.2437P}
}

@article{penoyre_astrometric_2022,
       author = {{Penoyre}, Zephyr and {Belokurov}, Vasily and {Evans}, N. Wyn},
        title = "{Astrometric identification of nearby binary stars - II. Astrometric binaries in the Gaia Catalogue of Nearby Stars}",
      journal = {\mnras},
         year = 2022,
        month = jul,
       volume = {513},
       number = {4},
        pages = {5270-5289},
          doi = {10.1093/mnras/stac1147},
archivePrefix = {arXiv},
       eprint = {2202.06963},
 primaryClass = {astro-ph.SR},
       adsurl = {https://ui.adsabs.harvard.edu/abs/2022MNRAS.513.5270P}
}

@article{belokurov_unresolved_2020,
	author = {Belokurov, Vasily and Penoyre, Zephyr and Oh, Semyeong and Iorio, Giuliano and Hodgkin, Simon and Evans, N Wyn and Everall, Andrew and Koposov, Sergey E and Tout, Christopher A and Izzard, Robert and Clarke, Cathie J and Brown, Anthony G A},
	doi = {10.1093/mnras/staa1522},
	issn = {0035-8711},
	journal = {\mnras},
	month = aug,
	number = {2},
	pages = {1922--1940},
	title = {Unresolved stellar companions with {Gaia} {DR2} astrometry},
	url = {https://doi.org/10.1093/mnras/staa1522},
	urldate = {2025-01-16},
	volume = {496},
	year = {2020}}

@article{merle_gaia-eso_2017,
	author = {Merle, T. and Van Eck, S. and Jorissen, A. and Van der Swaelmen, M. and Masseron, T. and Zwitter, T. and Hatzidimitriou, D. and Klutsch, A. and Pourbaix, D. and Blomme, R. and Worley, C. C. and Sacco, G. and Lewis, J. and Abia, C. and Traven, G. and Sordo, R. and Bragaglia, A. and Smiljanic, R. and Pancino, E. and Damiani, F. and Hourihane, A. and Gilmore, G. and Randich, S. and Koposov, S. and Casey, A. and Morbidelli, L. and Franciosini, E. and Magrini, L. and Jofre, P. and Costado, M. T. and Jeffries, R. D. and Bergemann, M. and Lanzafame, A. C. and Bayo, A. and Carraro, G. and Flaccomio, E. and Monaco, L. and Zaggia, S.},
	doi = {10.1051/0004-6361/201730442},
	issn = {0004-6361},
	journal = {\aap},
	month = dec,
	note = {ADS Bibcode: 2017A\&A...608A..95M},
	pages = {A95},
	shorttitle = {The {Gaia}-{ESO} {Survey}},
	title = {The {Gaia}-{ESO} {Survey}: double-, triple-, and quadruple-line spectroscopic binary candidates},
	url = {https://ui.adsabs.harvard.edu/abs/2017A&A...608A..95M},
	urldate = {2025-01-15},
	volume = {608},
	year = {2017}}

@article{ness_homogeneity_2022,
	author = {Ness, Melissa K. and Wheeler, Adam J. and McKinnon, Kevin and Horta, Danny and Casey, Andrew R. and Cunningham, Emily C. and Price-Whelan, Adrian M.},
	doi = {10.3847/1538-4357/ac4754},
	issn = {0004-637X},
	journal = {\apj},
	month = feb,
	note = {Publisher: IOP ADS Bibcode: 2022ApJ...926..144N},
	pages = {144},
	title = {The {Homogeneity} of the {Star}-forming {Environment} of the {Milky} {Way} {Disk} over {Time}},
	url = {https://ui.adsabs.harvard.edu/abs/2022ApJ...926..144N},
	urldate = {2025-01-15},
	volume = {926},
	year = {2022}}

@article{fernandez_-sync_2017,
	author = {Fernandez, M. A. and Covey, Kevin R. and De Lee, Nathan and Chojnowski, S. Drew and Nidever, David and Ballantyne, Richard and Cottaar, Michiel and Da Rio, Nicola and Foster, Jonathan B. and Majewski, Steven R. and Meyer, Michael R. and Reyna, A. M. and Roberts, G. W. and Skinner, Jacob and Stassun, Keivan and Tan, Jonathan C. and Troup, Nicholas and Zasowski, Gail},
	doi = {10.1088/1538-3873/aa77e0},
	issn = {0004-6280},
	journal = {\pasp},
	month = aug,
	note = {Publisher: IOP ADS Bibcode: 2017PASP..129h4201F},
	pages = {084201},
	title = {{IN}-{SYNC} {VI}. {Identification} and {Radial} {Velocity} {Extraction} for 100+ {Double}-{Lined} {Spectroscopic} {Binaries} in the {APOGEE}/{IN}-{SYNC} {Fields}},
	url = {https://ui.adsabs.harvard.edu/abs/2017PASP..129h4201F},
	urldate = {2024-10-27},
	volume = {129},
	year = {2017}}

@article{nguyen_parsec_2022,
	author = {Nguyen, C. T. and Costa, G. and Girardi, L. and Volpato, G. and Bressan, A. and Chen, Y. and Marigo, P. and Fu, X. and Goudfrooij, P.},
	doi = {10.1051/0004-6361/202244166},
	issn = {0004-6361},
	journal = {\aap},
	month = sep,
	note = {ADS Bibcode: 2022A\&A...665A.126N},
	pages = {A126},
	shorttitle = {{PARSEC} {V2}.0},
	title = {{PARSEC} {V2}.0: {Stellar} tracks and isochrones of low- and intermediate-mass stars with rotation},
	url = {https://ui.adsabs.harvard.edu/abs/2022A&A...665A.126N},
	urldate = {2024-10-22},
	volume = {665},
	year = {2022}}

@article{ting_payne_2019,
	author = {Ting, Yuan-Sen and Conroy, Charlie and Rix, Hans-Walter and Cargile, Phillip},
	doi = {10.3847/1538-4357/ab2331},
	issn = {0004-637X},
	journal = {\apj},
	month = jul,
	note = {Publisher: IOP ADS Bibcode: 2019ApJ...879...69T},
	pages = {69},
	shorttitle = {The {Payne}},
	title = {The {Payne}: {Self}-consistent ab initio {Fitting} of {Stellar} {Spectra}},
	url = {https://ui.adsabs.harvard.edu/abs/2019ApJ...879...69T},
	urldate = {2024-10-22},
	volume = {879},
	year = {2019}}

@article{gaia_collaboration_gaia_2023,
	author = {{Gaia Collaboration} and Vallenari, A. and Brown, A. G. A. and Prusti, T. and de Bruijne, J. H. J. and Arenou, F. and Babusiaux, C. and Biermann, M. and Creevey, O. L. and Ducourant, C. and Evans, D. W. and Eyer, L. and Guerra, R. and Hutton, A. and Jordi, C. and Klioner, S. A. and Lammers, U. L. and Lindegren, L. and Luri, X. and Mignard, F. and Panem, C. and Pourbaix, D. and Randich, S. and Sartoretti, P. and Soubiran, C. and Tanga, P. and Walton, N. A. and Bailer-Jones, C. A. L. and Bastian, U. and Drimmel, R. and Jansen, F. and Katz, D. and Lattanzi, M. G. and van Leeuwen, F. and Bakker, J. and Cacciari, C. and Casta{\~n}eda, J. and De Angeli, F. and Fabricius, C. and Fouesneau, M. and Fr{\'e}mat, Y. and Galluccio, L. and Guerrier, A. and Heiter, U. and Masana, E. and Messineo, R. and Mowlavi, N. and Nicolas, C. and Nienartowicz, K. and Pailler, F. and Panuzzo, P. and Riclet, F. and Roux, W. and Seabroke, G. M. and Sordo, R. and Th{\'e}venin, F. and Gracia-Abril, G. and Portell, J. and Teyssier, D. and Altmann, M. and Andrae, R. and Audard, M. and Bellas-Velidis, I. and Benson, K. and Berthier, J. and Blomme, R. and Burgess, P. W. and Busonero, D. and Busso, G. and C{\'a}novas, H. and Carry, B. and Cellino, A. and Cheek, N. and Clementini, G. and Damerdji, Y. and Davidson, M. and de Teodoro, P. and Nu{\~n}ez Campos, M. and Delchambre, L. and Dell'Oro, A. and Esquej, P. and Fern{\'a}ndez-Hern{\'a}ndez, J. and Fraile, E. and Garabato, D. and Garc{\'\i}a-Lario, P. and Gosset, E. and Haigron, R. and Halbwachs, J. -L. and Hambly, N. C. and Harrison, D. L. and Hern{\'a}ndez, J. and Hestroffer, D. and Hodgkin, S. T. and Holl, B. and Jan{\ss}en, K. and Jevardat de Fombelle, G. and Jordan, S. and Krone-Martins, A. and Lanzafame, A. C. and L{\"o}ffler, W. and Marchal, O. and Marrese, P. M. and Moitinho, A. and Muinonen, K. and Osborne, P. and Pancino, E. and Pauwels, T. and Recio-Blanco, A. and Reyl{\'e}, C. and Riello, M. and Rimoldini, L. and Roegiers, T. and Rybizki, J. and Sarro, L. M. and Siopis, C. and Smith, M. and Sozzetti, A. and Utrilla, E. and van Leeuwen, M. and Abbas, U. and {\'A}brah{\'a}m, P. and Abreu Aramburu, A. and Aerts, C. and Aguado, J. J. and Ajaj, M. and Aldea-Montero, F. and Altavilla, G. and {\'A}lvarez, M. A. and Alves, J. and Anders, F. and Anderson, R. I. and Anglada Varela, E. and Antoja, T. and Baines, D. and Baker, S. G. and Balaguer-N{\'u}{\~n}ez, L. and Balbinot, E. and Balog, Z. and Barache, C. and Barbato, D. and Barros, M. and Barstow, M. A. and Bartolom{\'e}, S. and Bassilana, J. -L. and Bauchet, N. and Becciani, U. and Bellazzini, M. and Berihuete, A. and Bernet, M. and Bertone, S. and Bianchi, L. and Binnenfeld, A. and Blanco-Cuaresma, S. and Blazere, A. and Boch, T. and Bombrun, A. and Bossini, D. and Bouquillon, S. and Bragaglia, A. and Bramante, L. and Breedt, E. and Bressan, A. and Brouillet, N. and Brugaletta, E. and Bucciarelli, B. and Burlacu, A. and Butkevich, A. G. and Buzzi, R. and Caffau, E. and Cancelliere, R. and Cantat-Gaudin, T. and Carballo, R. and Carlucci, T. and Carnerero, M. I. and Carrasco, J. M. and Casamiquela, L. and Castellani, M. and Castro-Ginard, A. and Chaoul, L. and Charlot, P. and Chemin, L. and Chiaramida, V. and Chiavassa, A. and Chornay, N. and Comoretto, G. and Contursi, G. and Cooper, W. J. and Cornez, T. and Cowell, S. and Crifo, F. and Cropper, M. and Crosta, M. and Crowley, C. and Dafonte, C. and Dapergolas, A. and David, M. and David, P. and de Laverny, P. and De Luise, F. and De March, R. and De Ridder, J. and de Souza, R. and de Torres, A. and del Peloso, E. F. and del Pozo, E. and Delbo, M. and Delgado, A. and Delisle, J. -B. and Demouchy, C. and Dharmawardena, T. E. and Di Matteo, P. and Diakite, S. and Diener, C. and Distefano, E. and Dolding, C. and Edvardsson, B. and Enke, H. and Fabre, C. and Fabrizio, M. and Faigler, S. and Fedorets, G. and Fernique, P. and Fienga, A. and Figueras, F. and Fournier, Y. and Fouron, C. and Fragkoudi, F. and Gai, M. and Garcia-Gutierrez, A. and Garcia-Reinaldos, M. and Garc{\'\i}a-Torres, M. and Garofalo, A. and Gavel, A. and Gavras, P. and Gerlach, E. and Geyer, R. and Giacobbe, P. and Gilmore, G. and Girona, S. and Giuffrida, G. and Gomel, R. and Gomez, A. and Gonz{\'a}lez-N{\'u}{\~n}ez, J. and Gonz{\'a}lez-Santamar{\'\i}a, I. and Gonz{\'a}lez-Vidal, J. J. and Granvik, M. and Guillout, P. and Guiraud, J. and Guti{\'e}rrez-S{\'a}nchez, R. and Guy, L. P. and Hatzidimitriou, D. and Hauser, M. and Haywood, M. and Helmer, A. and Helmi, A. and Sarmiento, M. H. and Hidalgo, S. L. and Hilger, T. and H{\l}adczuk, N. and Hobbs, D. and Holland, G. and Huckle, H. E. and Jardine, K. and Jasniewicz, G. and Jean-Antoine Piccolo, A. and Jim{\'e}nez-Arranz, {\'O}. and Jorissen, A. and Juaristi Campillo, J. and Julbe, F. and Karbevska, L. and Kervella, P. and Khanna, S. and Kontizas, M. and Kordopatis, G. and Korn, A. J. and K{\'o}sp{\'a}l, {\'A}. and Kostrzewa-Rutkowska, Z. and Kruszy{\'n}ska, K. and Kun, M. and Laizeau, P. and Lambert, S. and Lanza, A. F. and Lasne, Y. and Le Campion, J. -F. and Lebreton, Y. and Lebzelter, T. and Leccia, S. and Leclerc, N. and Lecoeur-Taibi, I. and Liao, S. and Licata, E. L. and Lindstr{\o}m, H. E. P. and Lister, T. A. and Livanou, E. and Lobel, A. and Lorca, A. and Loup, C. and Madrero Pardo, P. and Magdaleno Romeo, A. and Managau, S. and Mann, R. G. and Manteiga, M. and Marchant, J. M. and Marconi, M. and Marcos, J. and Marcos Santos, M. M. S. and Mar{\'\i}n Pina, D. and Marinoni, S. and Marocco, F. and Marshall, D. J. and Martin Polo, L. and Mart{\'\i}n-Fleitas, J. M. and Marton, G. and Mary, N. and Masip, A. and Massari, D. and Mastrobuono-Battisti, A. and Mazeh, T. and McMillan, P. J. and Messina, S. and Michalik, D. and Millar, N. R. and Mints, A. and Molina, D. and Molinaro, R. and Moln{\'a}r, L. and Monari, G. and Mongui{\'o}, M. and Montegriffo, P. and Montero, A. and Mor, R. and Mora, A. and Morbidelli, R. and Morel, T. and Morris, D. and Muraveva, T. and Murphy, C. P. and Musella, I. and Nagy, Z. and Noval, L. and Oca{\~n}a, F. and Ogden, A. and Ordenovic, C. and Osinde, J. O. and Pagani, C. and Pagano, I. and Palaversa, L. and Palicio, P. A. and Pallas-Quintela, L. and Panahi, A. and Payne-Wardenaar, S. and Pe{\~n}alosa Esteller, X. and Penttil{\"a}, A. and Pichon, B. and Piersimoni, A. M. and Pineau, F. -X. and Plachy, E. and Plum, G. and Poggio, E. and Pr{\v s}a, A. and Pulone, L. and Racero, E. and Ragaini, S. and Rainer, M. and Raiteri, C. M. and Rambaux, N. and Ramos, P. and Ramos-Lerate, M. and Re Fiorentin, P. and Regibo, S. and Richards, P. J. and Rios Diaz, C. and Ripepi, V. and Riva, A. and Rix, H. -W. and Rixon, G. and Robichon, N. and Robin, A. C. and Robin, C. and Roelens, M. and Rogues, H. R. O. and Rohrbasser, L. and Romero-G{\'o}mez, M. and Rowell, N. and Royer, F. and Ruz Mieres, D. and Rybicki, K. A. and Sadowski, G. and S{\'a}ez N{\'u}{\~n}ez, A. and Sagrist{\`a} Sell{\'e}s, A. and Sahlmann, J. and Salguero, E. and Samaras, N. and Sanchez Gimenez, V. and Sanna, N. and Santove{\~n}a, R. and Sarasso, M. and Schultheis, M. and Sciacca, E. and Segol, M. and Segovia, J. C. and S{\'e}gransan, D. and Semeux, D. and Shahaf, S. and Siddiqui, H. I. and Siebert, A. and Siltala, L. and Silvelo, A. and Slezak, E. and Slezak, I. and Smart, R. L. and Snaith, O. N. and Solano, E. and Solitro, F. and Souami, D. and Souchay, J. and Spagna, A. and Spina, L. and Spoto, F. and Steele, I. A. and Steidelm{\"u}ller, H. and Stephenson, C. A. and S{\"u}veges, M. and Surdej, J. and Szabados, L. and Szegedi-Elek, E. and Taris, F. and Taylor, M. B. and Teixeira, R. and Tolomei, L. and Tonello, N. and Torra, F. and Torra, J. and Torralba Elipe, G. and Trabucchi, M. and Tsounis, A. T. and Turon, C. and Ulla, A. and Unger, N. and Vaillant, M. V. and van Dillen, E. and van Reeven, W. and Vanel, O. and Vecchiato, A. and Viala, Y. and Vicente, D. and Voutsinas, S. and Weiler, M. and Wevers, T. and Wyrzykowski, {\L}. and Yoldas, A. and Yvard, P. and Zhao, H. and Zorec, J. and Zucker, S. and Zwitter, T.},
	doi = {10.1051/0004-6361/202243940},
	issn = {0004-6361},
	journal = {\aap},
	month = jun,
	note = {ADS Bibcode: 2023A\&A...674A...1G},
	pages = {A1},
	title = {Gaia {Data} {Release} 3. {Summary} of the content and survey properties},
	url = {https://ui.adsabs.harvard.edu/abs/2023A&A...674A...1G},
	urldate = {2024-10-16},
	volume = {674},
	year = {2023}}

@article{buder_galah_2018,
	author = {Buder, Sven and Asplund, Martin and Duong, Ly and Kos, Janez and Lind, Karin and Ness, Melissa K. and Sharma, Sanjib and Bland-Hawthorn, Joss and Casey, Andrew R. and de Silva, Gayandhi M. and D'Orazi, Valentina and Freeman, Ken C. and Lewis, Geraint F. and Lin, Jane and Martell, Sarah L. and Schlesinger, Katharine J. and Simpson, Jeffrey D. and Zucker, Daniel B. and Zwitter, Toma{\v z} and Amarsi, Anish M. and Anguiano, Borja and Carollo, Daniela and Casagrande, Luca and {\v C}otar, Klemen and Cottrell, Peter L. and da Costa, Gary and Gao, Xudong D. and Hayden, Michael R. and Horner, Jonathan and Ireland, Michael J. and Kafle, Prajwal R. and Munari, Ulisse and Nataf, David M. and Nordlander, Thomas and Stello, Dennis and Ting, Yuan-Sen and Traven, Gregor and Watson, Fred and Wittenmyer, Robert A. and Wyse, Rosemary F. G. and Yong, David and Zinn, Joel C. and {\v Z}erjal, Maru{\v s}a and {Galah Collaboration}},
	doi = {10.1093/mnras/sty1281},
	issn = {0035-8711},
	journal = {\mnras},
	month = aug,
	note = {Publisher: OUP ADS Bibcode: 2018MNRAS.478.4513B},
	pages = {4513--4552},
	shorttitle = {The {GALAH} {Survey}},
	title = {The {GALAH} {Survey}: second data release},
	url = {https://ui.adsabs.harvard.edu/abs/2018MNRAS.478.4513B},
	urldate = {2024-10-15},
	volume = {478},
	year = {2018}}

@misc{badenes_stellar_2018,
	author = {Badenes, Carles and Mazzola, Christine and Thompson, Todd A. and Covey, Kevin and Freeman, Peter E. and Walker, Matthew G. and Moe, Maxwell and Troup, Nicholas and Nidever, David and Prieto, Carlos Allende and Andrews, Brett and Barb{\'a}, Rodolfo H. and Beers, Timothy C. and Bovy, Jo and Carlberg, Joleen K. and Lee, Nathan De and Johnson, Jennifer and Lewis, Hannah and Majewski, Steven R. and Pinsonneault, Marc and Sobeck, Jennifer and Stassun, Keivan G. and Stringfellow, Guy and Zasowski, Gail},
	doi = {10.48550/arXiv.1711.00660},
	month = jan,
	note = {arXiv:1711.00660},
	publisher = {arXiv},
	shorttitle = {Stellar {Multiplicity} {Meets} {Stellar} {Evolution} {And} {Metallicity}},
	title = {Stellar {Multiplicity} {Meets} {Stellar} {Evolution} {And} {Metallicity}: {The} {APOGEE} {View}},
	url = {http://arxiv.org/abs/1711.00660},
	urldate = {2024-10-11},
	year = {2018}}

@article{packet_spin-up_1981,
	author = {Packet, W.},
	issn = {0004-6361},
	journal = {\aap},
	month = sep,
	note = {ADS Bibcode: 1981A\&A...102...17P},
	pages = {17--19},
	title = {On the spin-up of the mass accreting component in a close binary system},
	url = {https://ui.adsabs.harvard.edu/abs/1981A&A...102...17P},
	urldate = {2024-10-10},
	volume = {102},
	year = {1981}}

@article{tokovinin_formation_2020,
	author = {Tokovinin, Andrei and Moe, Maxwell},
	doi = {10.1093/mnras/stz3299},
	issn = {0035-8711},
	journal = {\mnras},
	month = feb,
	number = {4},
	pages = {5158--5171},
	title = {Formation of close binaries by disc fragmentation and migration, and its statistical modelling},
	url = {https://doi.org/10.1093/mnras/stz3299},
	urldate = {2024-10-01},
	volume = {491},
	year = {2020}}

@book{batten_binary_1973,
	author = {Batten, Alan Henry},
	publisher = {Pergamon Press Ltd., Oxford},
	title = {Binary and multiple systems of stars},
	year = {1973}}

@article{traven_galah_2020,
	author = {Traven, G. and Feltzing, S. and Merle, T. and Van der Swaelmen, M. and {\v C}otar, K. and Church, R. and Zwitter, T. and Ting, Y.-S. and Sahlholdt, C. and Asplund, M. and Bland-Hawthorn, J. and De Silva, G. and Freeman, K. and Martell, S. and Sharma, S. and Zucker, D. and Buder, S. and Casey, A. and D'Orazi, V. and Kos, J. and Lewis, G. and Lin, J. and Lind, K. and Simpson, J. and Stello, D. and Munari, U. and Wittenmyer, R. A.},
	doi = {10.1051/0004-6361/202037484},
	issn = {0004-6361, 1432-0746},
	journal = {\aap},
	month = jun,
	note = {arXiv:2005.00014 [astro-ph]},
	pages = {A145},
	shorttitle = {The {GALAH} survey},
	title = {The {GALAH} survey: {Multiple} stars and our {Galaxy}. {I}. {A} comprehensive method for deriving properties of {FGK} binary stars},
	url = {http://arxiv.org/abs/2005.00014},
	urldate = {2024-09-30},
	volume = {638},
	year = {2020}}

@article{castro-ginard_gaia_2024,
	adsurl = {https://ui.adsabs.harvard.edu/abs/2024A&A...688A...1C},
	archiveprefix = {arXiv},
	author = {{Castro-Ginard}, Alfred and {Penoyre}, Zephyr and {Casey}, Andrew R. and {Brown}, Anthony G.~A. and {Belokurov}, Vasily and {Cantat-Gaudin}, Tristan and {Drimmel}, Ronald and {Fouesneau}, Morgan and {Khanna}, Shourya and {Kurbatov}, Evgeny P. and {Price-Whelan}, Adrian M. and {Rix}, Hans-Walter and {Smart}, Richard L.},
	doi = {10.1051/0004-6361/202450172},
	eid = {A1},
	eprint = {2404.14127},
	journal = {\aap},
	month = aug,
	pages = {A1},
	primaryclass = {astro-ph.GA},
	title = {{Gaia DR3 detectability of unresolved binary systems}},
	volume = {688},
	year = 2024}

@article{tokovinin_architecture_2021,
	author = {Tokovinin, Andrei},
	doi = {10.3390/universe7090352},
	journal = {Universe},
	month = sep,
	pages = {352},
	title = {Architecture of {Hierarchical} {Stellar} {Systems} and {Their} {Formation}},
	volume = {7},
	year = {2021}}

@article{moe_mind_2017,
	author = {Moe, Maxwell and Stefano, Rosanne Di},
	doi = {10.3847/1538-4365/aa6fb6},
	issn = {0067-0049},
	journal = {\apjs},
	language = {en},
	month = jun,
	note = {Publisher: The American Astronomical Society},
	number = {2},
	pages = {15},
	shorttitle = {Mind {Your} {Ps} and {Qs}},
	title = {Mind {Your} {Ps} and {Qs}: {The} {Interrelation} between {Period} ({P}) and {Mass}-ratio ({Q}) {Distributions} of {Binary} {Stars}},
	url = {https://dx.doi.org/10.3847/1538-4365/aa6fb6},
	urldate = {2024-04-22},
	volume = {230},
	year = {2017}}

@article{bonnell_formation_1994,
	author = {Bonnell, I. A. and Bate, M. R.},
	doi = {10.1093/mnras/271.4.999},
	issn = {0035-8711},
	journal = {\mnras},
	month = dec,
	note = {Publisher: OUP ADS Bibcode: 1994MNRAS.271..999B},
	pages = {999--1004},
	title = {The {Formation} of {Close} {Binary} {Systems}},
	url = {https://ui.adsabs.harvard.edu/abs/1994MNRAS.271..999B},
	urldate = {2024-04-17},
	volume = {271},
	year = {1994}}

@article{traven_galah_2017,
	author = {Traven, G. and Matijevi{\v c}, G. and Zwitter, T. and {\v Z}erjal, M. and Kos, J. and Asplund, M. and Bland-Hawthorn, J. and Casey, A. R. and De Silva, G. and Freeman, K. and Lin, J. and Martell, S. L. and Schlesinger, K. J. and Sharma, S. and Simpson, J. D. and Zucker, D. B. and Anguiano, B. and Da Costa, G. and Duong, L. and Horner, J. and Hyde, E. A. and Kafle, P. R. and Munari, U. and Nataf, D. and Navin, C. A. and Reid, W. and Ting, Y. -S.},
	doi = {10.3847/1538-4365/228/2/24},
	issn = {0067-0049},
	journal = {\apjs},
	month = feb,
	note = {ADS Bibcode: 2017ApJS..228...24T},
	pages = {24},
	shorttitle = {The {Galah} {Survey}},
	title = {The {Galah} {Survey}: {Classification} and {Diagnostics} with t-{SNE} {Reduction} of {Spectral} {Information}},
	url = {https://ui.adsabs.harvard.edu/abs/2017ApJS..228...24T},
	urldate = {2024-03-21},
	volume = {228},
	year = {2017}}

@article{el-badry_signatures_2018,
	author = {El-Badry, Kareem and Rix, Hans-Walter and Ting, Yuan-Sen and Weisz, Daniel R. and Bergemann, Maria and Cargile, Phillip and Conroy, Charlie and Eilers, Anna-Christina},
	doi = {10.1093/mnras/stx2758},
	issn = {0035-8711},
	journal = {\mnras},
	month = feb,
	note = {ADS Bibcode: 2018MNRAS.473.5043E},
	pages = {5043--5049},
	shorttitle = {Signatures of unresolved binaries in stellar spectra},
	title = {Signatures of unresolved binaries in stellar spectra: implications for spectral fitting},
	url = {https://ui.adsabs.harvard.edu/abs/2018MNRAS.473.5043E},
	urldate = {2024-03-20},
	volume = {473},
	year = {2018}}

@article{kos_galah_2017,
	author = {Kos, Janez and Lin, Jane and Zwitter, Toma{\v z} and {\v Z}erjal, Maru{\v s}ka and Sharma, Sanjib and Bland-Hawthorn, Joss and Asplund, Martin and Casey, Andrew R. and De Silva, Gayandhi M. and Freeman, Ken C. and Martell, Sarah L. and Simpson, Jeffrey D. and Schlesinger, Katharine J. and Zucker, Daniel and Anguiano, Borja and Bacigalupo, Carlos and Bedding, Timothy R. and Betters, Christopher and Da Costa, Gary and Duong, Ly and Hyde, Elaina and Ireland, Michael and Kafle, Prajwal R. and Leon-Saval, Sergio and Lewis, Geraint F. and Munari, Ulisse and Nataf, David and Stello, Dennis and Tinney, C. G. and Traven, Gregor and Watson, Fred and Wittenmyer, Robert A.},
	doi = {10.1093/mnras/stw2064},
	issn = {0035-8711},
	journal = {\mnras},
	month = jan,
	note = {ADS Bibcode: 2017MNRAS.464.1259K},
	pages = {1259--1281},
	shorttitle = {The {GALAH} survey},
	title = {The {GALAH} survey: the data reduction pipeline},
	url = {https://ui.adsabs.harvard.edu/abs/2017MNRAS.464.1259K},
	urldate = {2024-03-20},
	volume = {464},
	year = {2017}}

@article{duchene_stellar_2013,
	author = {Duch{\^e}ne, Gaspard and Kraus, Adam},
	doi = {10.1146/annurev-astro-081710-102602},
	issn = {0066-4146},
	journal = {\araa},
	month = aug,
	note = {ADS Bibcode: 2013ARA\&A..51..269D},
	pages = {269--310},
	title = {Stellar {Multiplicity}},
	url = {https://ui.adsabs.harvard.edu/abs/2013ARA&A..51..269D},
	urldate = {2024-03-19},
	volume = {51},
	year = {2013}}

@article{raghavan_survey_2010,
	author = {Raghavan, Deepak and McAlister, Harold A. and Henry, Todd J. and Latham, David W. and Marcy, Geoffrey W. and Mason, Brian D. and Gies, Douglas R. and White, Russel J. and ten Brummelaar, Theo A.},
	doi = {10.1088/0067-0049/190/1/1},
	issn = {0067-0049},
	journal = {\apjs},
	month = sep,
	note = {ADS Bibcode: 2010ApJS..190....1R},
	pages = {1--42},
	shorttitle = {A {Survey} of {Stellar} {Families}},
	title = {A {Survey} of {Stellar} {Families}: {Multiplicity} of {Solar}-type {Stars}},
	url = {https://ui.adsabs.harvard.edu/abs/2010ApJS..190....1R},
	urldate = {2024-03-19},
	volume = {190},
	year = {2010}}

@article{lada_stellar_2006,
	author = {Lada, Charles J.},
	doi = {10.1086/503158},
	issn = {0004-637X},
	journal = {\apj},
	month = mar,
	note = {ADS Bibcode: 2006ApJ...640L..63L},
	pages = {L63--L66},
	shorttitle = {Stellar {Multiplicity} and the {Initial} {Mass} {Function}},
	title = {Stellar {Multiplicity} and the {Initial} {Mass} {Function}: {Most} {Stars} {Are} {Single}},
	url = {https://ui.adsabs.harvard.edu/abs/2006ApJ...640L..63L},
	urldate = {2024-03-15},
	volume = {640},
	year = {2006}}

@article{moe_close_2019,
	author = {Moe, Maxwell and Kratter, Kaitlin M. and Badenes, Carles},
	doi = {10.3847/1538-4357/ab0d88},
	issn = {0004-637X},
	journal = {\apj},
	month = apr,
	note = {ADS Bibcode: 2019ApJ...875...61M},
	pages = {61},
	title = {The {Close} {Binary} {Fraction} of {Solar}-type {Stars} {Is} {Strongly} {Anticorrelated} with {Metallicity}},
	url = {https://ui.adsabs.harvard.edu/abs/2019ApJ...875...61M},
	urldate = {2024-03-14},
	volume = {875},
	year = {2019}}

@article{buder_galah_2021,
	author = {Buder, Sven and Sharma, Sanjib and Kos, Janez and Amarsi, Anish M and Nordlander, Thomas and Lind, Karin and Martell, Sarah L and Asplund, Martin and Bland-Hawthorn, Joss and Casey, Andrew R and De Silva, Gayandhi M and D'Orazi, Valentina and Freeman, Ken C and Hayden, Michael R and Lewis, Geraint F and Lin, Jane and Schlesinger, Katharine J and Simpson, Jeffrey D and Stello, Dennis and Zucker, Daniel B and Zwitter, Toma{\v z} and Beeson, Kevin L and Buck, Tobias and Casagrande, Luca and Clark, Jake T and {\v C}otar, Klemen and Da Costa, Gary S and de Grijs, Richard and Feuillet, Diane and Horner, Jonathan and Kafle, Prajwal R and Khanna, Shourya and Kobayashi, Chiaki and Liu, Fan and Montet, Benjamin T and Nandakumar, Govind and Nataf, David M and Ness, Melissa K and Spina, Lorenzo and Tepper-Garc{\'\i}a, Thor and Ting(丁源森), Yuan-Sen and Traven, Gregor and Vogrin{\v c}i{\v c}, Rok and Wittenmyer, Robert A and Wyse, Rosemary F G and {\v Z}erjal, Maru{\v s}a and {GALAH Collaboration}},
	doi = {10.1093/mnras/stab1242},
	issn = {0035-8711},
	journal = {\mnras},
	month = sep,
	number = {1},
	pages = {150--201},
	shorttitle = {The {GALAH}+ survey},
	title = {The {GALAH}+ survey: {Third} data release},
	url = {https://doi.org/10.1093/mnras/stab1242},
	urldate = {2023-09-20},
	volume = {506},
	year = {2021}}

@article{majewski_apache_2017,
	author = {Majewski, Steven R. and Schiavon, Ricardo P. and Frinchaboy, Peter M. and Allende Prieto, Carlos and Barkhouser, Robert and Bizyaev, Dmitry and Blank, Basil and Brunner, Sophia and Burton, Adam and Carrera, Ricardo and Chojnowski, S. Drew and Cunha, K{\'a}tia and Epstein, Courtney and Fitzgerald, Greg and Garc{\'\i}a P{\'e}rez, Ana E. and Hearty, Fred R. and Henderson, Chuck and Holtzman, Jon A. and Johnson, Jennifer A. and Lam, Charles R. and Lawler, James E. and Maseman, Paul and M{\'e}sz{\'a}ros, Szabolcs and Nelson, Matthew and Nguyen, Duy Coung and Nidever, David L. and Pinsonneault, Marc and Shetrone, Matthew and Smee, Stephen and Smith, Verne V. and Stolberg, Todd and Skrutskie, Michael F. and Walker, Eric and Wilson, John C. and Zasowski, Gail and Anders, Friedrich and Basu, Sarbani and Beland, Stephane and Blanton, Michael R. and Bovy, Jo and Brownstein, Joel R. and Carlberg, Joleen and Chaplin, William and Chiappini, Cristina and Eisenstein, Daniel J. and Elsworth, Yvonne and Feuillet, Diane and Fleming, Scott W. and Galbraith-Frew, Jessica and Garc{\'\i}a, Rafael A. and Garc{\'\i}a-Hern{\'a}ndez, D. An{\'\i}bal and Gillespie, Bruce A. and Girardi, L{\'e}o and Gunn, James E. and Hasselquist, Sten and Hayden, Michael R. and Hekker, Saskia and Ivans, Inese and Kinemuchi, Karen and Klaene, Mark and Mahadevan, Suvrath and Mathur, Savita and Mosser, Beno{\^\i}t and Muna, Demitri and Munn, Jeffrey A. and Nichol, Robert C. and O'Connell, Robert W. and Parejko, John K. and Robin, A. C. and Rocha-Pinto, Helio and Schultheis, Matthias and Serenelli, Aldo M. and Shane, Neville and Silva Aguirre, Victor and Sobeck, Jennifer S. and Thompson, Benjamin and Troup, Nicholas W. and Weinberg, David H. and Zamora, Olga},
	doi = {10.3847/1538-3881/aa784d},
	issn = {0004-6256},
	journal = {\aj},
	month = sep,
	note = {ADS Bibcode: 2017AJ....154...94M},
	pages = {94},
	title = {The {Apache} {Point} {Observatory} {Galactic} {Evolution} {Experiment} ({APOGEE})},
	url = {https://ui.adsabs.harvard.edu/abs/2017AJ....154...94M},
	urldate = {2023-06-06},
	volume = {154},
	year = {2017}}

@article{gustafsson_grid_2008,
	author = {Gustafsson, B. and Edvardsson, B. and Eriksson, K. and J{\o}rgensen, U. G. and Nordlund, {\AA}. and Plez, B.},
	doi = {10.1051/0004-6361:200809724},
	issn = {0004-6361},
	journal = {\aap},
	month = aug,
	note = {ADS Bibcode: 2008A\&A...486..951G},
	pages = {951--970},
	title = {A grid of {MARCS} model atmospheres for late-type stars. {I}. {Methods} and general properties},
	url = {https://ui.adsabs.harvard.edu/abs/2008A&A...486..951G},
	urldate = {2023-06-04},
	volume = {486},
	year = {2008}}

@article{valenti_spectroscopy_1996,
       author = {{Valenti}, J.~A. and {Piskunov}, N.},
        title = "{Spectroscopy made easy: A new tool for fitting observations with synthetic spectra.}",
      journal = {\aaps},
         year = 1996,
        month = sep,
       volume = {118},
        pages = {595-603},
       adsurl = {https://ui.adsabs.harvard.edu/abs/1996A&AS..118..595V}
}

@article{piskunov_spectroscopy_2017,
	author = {Piskunov, Nikolai and Valenti, Jeff A.},
	doi = {10.1051/0004-6361/201629124},
	issn = {0004-6361},
	journal = {\aap},
	month = jan,
	note = {ADS Bibcode: 2017A\&A...597A..16P},
	pages = {A16},
	shorttitle = {Spectroscopy {Made} {Easy}},
	title = {Spectroscopy {Made} {Easy}: {Evolution}},
	url = {https://ui.adsabs.harvard.edu/abs/2017A&A...597A..16P},
	urldate = {2023-05-30},
	volume = {597},
	year = {2017}}

@article{jofre_accuracy_2019,
	author = {Jofr{\'e}, Paula and Heiter, Ulrike and Soubiran, Caroline},
	doi = {10.1146/annurev-astro-091918-104509},
	issn = {0066-4146},
	journal = {\araa},
	month = aug,
	note = {ADS Bibcode: 2019ARA\&A..57..571J},
	pages = {571--616},
	title = {Accuracy and {Precision} of {Industrial} {Stellar} {Abundances}},
	url = {https://ui.adsabs.harvard.edu/abs/2019ARA&A..57..571J},
	urldate = {2023-04-26},
	volume = {57},
	year = {2019}}

@article{shu_star_1987,
	author = {Shu, Frank H. and Adams, Fred C. and Lizano, Susana},
	doi = {10.1146/annurev.aa.25.090187.000323},
	journal = {\araa},
	note = {\_eprint: https://doi.org/10.1146/annurev.aa.25.090187.000323},
	number = {1},
	pages = {23--81},
	shorttitle = {Star {Formation} in {Molecular} {Clouds}},
	title = {Star {Formation} in {Molecular} {Clouds}: {Observation} and {Theory}},
	url = {https://doi.org/10.1146/annurev.aa.25.090187.000323},
	urldate = {2023-03-30},
	volume = {25},
	year = {1987}}
\FloatBarrier
\begin{appendix}


\section{Algorithm Comparisons and Selection} \label{sec:PSO_details}
Beyond \textsc{scipy.optimize.curve\_fit}, we test the commonly used methods BFGS and Nelder-Mead of the \software{scipy.optimize.minimize} package, Bayesian Markov Chain Monte Carlo (MCMC) using \software{emcee} \citep[][emcee]{foreman-mackey_emcee_2013}, and also test Particle Swarm Optimization (PSO) with \software{pyswarms} \citep[][pyswarms]{james_v_miranda_pyswarms_2018}. Additionally, we experimented with a gridding approach to parameter optimization, providing an alternative baseline comparison to the other iterative methods.

\begin{figure}
  \centering
  \includegraphics[width=\columnwidth]{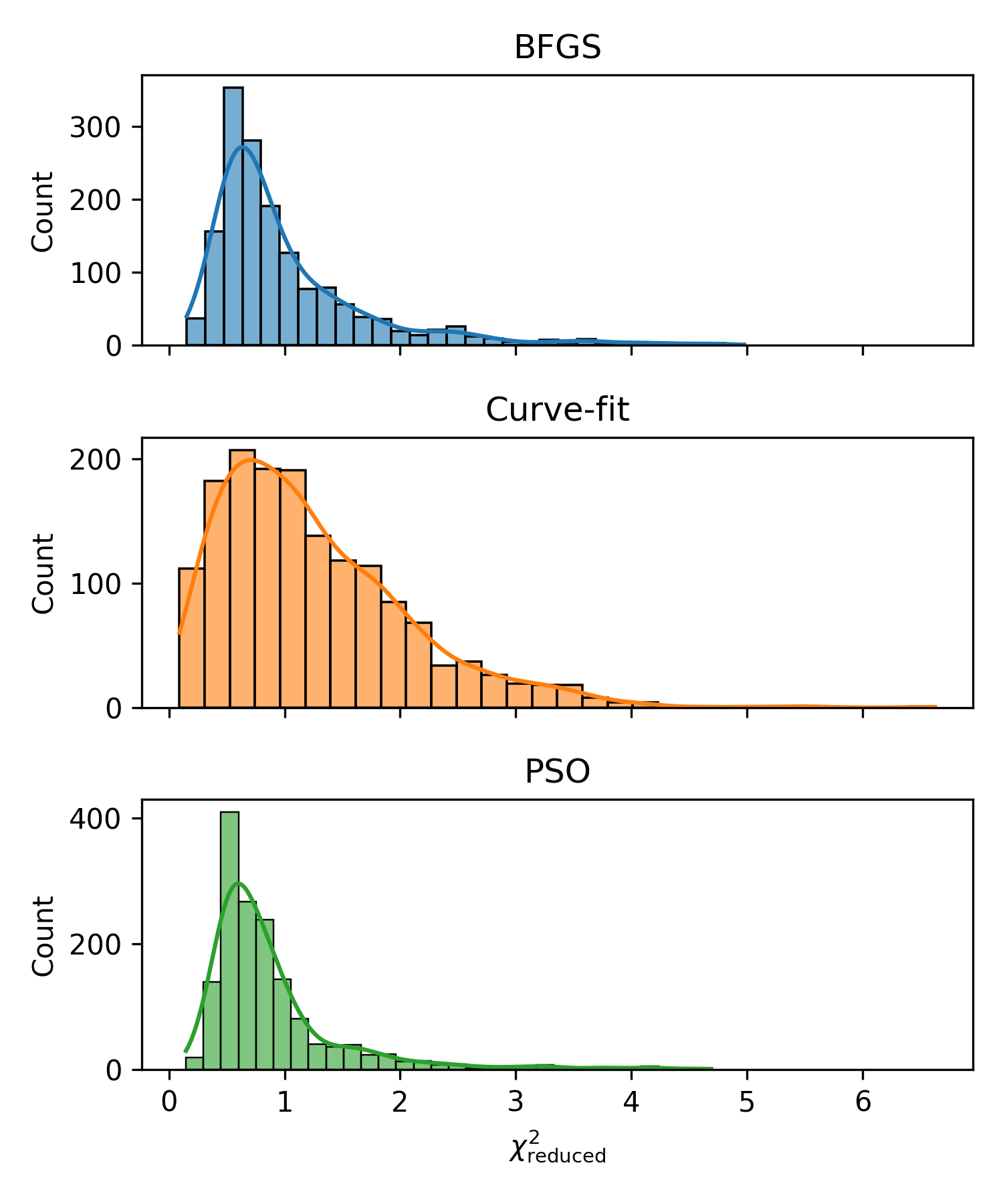}
  \caption{Algorithm performance comparisons for curve-fit (\software{scipy.optimize.curve\_fit}), L-BFGS (\software{scipy.optimize.minimize}), and PSO (\software{pyswarms}) on the same set of 10,000 random spectra taken from our sample selection. Note that  We find PSO to perform best in terms of both the median and mean $\chi^2$ distributions. Median processing times per object for each algorithm are approximately 3m (curve-fit), 15m (BFGS), and 25m (PSO) with our setup. The data were processed in parallel, with up to 1000 spectra evaluated concurrently on a pool of 50 CPU cores. With PSO this means around 100,000 spectra can be analysed in approximately 40h which we found to be acceptable given our setup.}
  \label{fig:rchi2_comparisons}
\end{figure}

Based on the lowest distribution of $\chi^2$, PSO was found to perform best (Fig. \ref{fig:rchi2_comparisons}). This method performed significantly better than \textsc{scipy.optimize.curve\_fit} and moderately better than the remaining algorithms. While some some spectra are better fit by these other algorithms, these spectra were a minority ($<5\%$). To achieve the best possible fit across all spectra, we ultimately adopted a hybrid approach where spectra that are poorly fit  ($\chi^2$ > 3) with PSO were refit with the BFGS method as the second best performing algorithm (as this had already been implemented). If this new fit leads to a smaller $\chi^2$ value, we take the BFGS parameters instead of the PSO ones as the final values of our analysis for the given spectra.

Its selection was motivated by several factors. PSO is a gradient-free approach that randomly populates the parameter space with $n$ particles initialised with random velocities. These particles move through the search space, guided by a balance between exploration (global best solution found by any particle) and exploitation (the best solution found by each individual particle). The swarm collectively works towards the optimal solution by iterating on the positions and velocities of the particles based on both the global and individual best positions. These attributes allow PSO to overcome the challenges encountered with other algorithms: avoiding local minima through extended exploration, and preventing premature convergence and degeneracies. The next step for each particle is given by

\begin{equation}
    \mathbf{v}_i(t+1) = w \mathbf{v}_i(t) + c_1 r_1 (\mathbf{p}_i^{\text{best}} - \mathbf{p}_i(t)) + c_2 r_2 (\mathbf{p}_g^{\text{best}} - \mathbf{p}_i(t)),
\end{equation}

\noindent where $\mathbf{v}_i(t+1)$ is the velocity of particle $i$ at time step $t+1$. $\mathbf{v}_i(t)$ is the velocity of particle $i$ at time step $t$, $w$ is the inertia weight, $c_1$ is the cognitive parameter, $\mathbf{p}_i^{\text{best}}$ is the individual best position of particle $i$, $\mathbf{p}_i(t)$ is the current position of particle $i$ at time step $t$, $c_2$ is the social parameter, and $\mathbf{p}_g^{\text{best}}$ is the global best position found by any particle in the swarm. $r_1$ and $r_2$ are random numbers between 0 and 1.

In our implementation we create 100 particles run over 65 iterations with $w = 0.7$, $c_1 = 0.8$, and $c_2 = 0.3$. These parameters were determined through experimentation to converge on the best median for the dataset and weight individual exploration more heavily, though still guide each particle by the global best fit. All parameters are normalised to $[0,1]$ prior to optimisation to ensure comparable step sizes across parameters spanning different physical scales (e.g., mass in $M_\odot$ versus age in Gyr). The PSO search bounds in this normalised space are then tightened to $\pm20\%$ around the GALAH DR4 initial values when the identical-twin binary model yields $\tilde{\chi}^2 < 2$; otherwise the full $[0,1]$ space is searched (see Section~\ref{sec:algo-selection} for the rationale).

\section{Uncertainty Determinations} \label{sec:appendix-unc}
For each star in our sample, we quantified parameter uncertainties by analysing the full distribution of samples generated during the PSO. Although PSO is a deterministic optimization algorithm and not a sampling method such as MCMC, it can still produce sequences of parameter values that meaningfully trace the exploration of parameter space. To isolate the portion of the PSO chain representative of stable, well-fitting solutions, and remove burn-in, two diagnostics were applied: variance stabilization and autocorrelation analysis. Each chain was divided into quarters, and the variance ($V$) of the parameter values was computed within each segment according to
\begin{equation}
    V=\frac{1}{N-1} \sum^N_{i=1}(x_i-\bar{x})^2,
\end{equation}
where $\bar{x}$ is the mean parameter value within the segment. We took a decrease greater than 75\% to indicate ongoing settling of the swarm; in such cases, the preceding segments were discarded as burn-in.

Following this, the autocorrelation length of the remaining samples was computed for each parameter to evaluate dependence across iterations. The final burn-in for each PSO run was set at the earliest iteration where the autocorrelation length reached a consistent value. The assumed stationary portion of the chain after this cutoff was used to calculate parameter uncertainties. We took the median (50th percentile) as the best-fit value for each parameter. The statistical uncertainty was defined as half the range between the 16th and 84th percentiles of the accepted samples, thereby providing a non-parametric $1\sigma$ error estimate that naturally encompasses non-Gaussian or asymmetric distributions.

We note that PSO seeks a global minimum of $\chi^{2}$ by updating each particle using information from the entire swarm via the current global-best position, whereas MCMC samples the posterior distribution through pairwise walker interactions within the ensemble rather than driving all walkers toward a single optimum. The trade-off is that PSO can efficiently find good solutions and avoid local minima with less stringent initialization, but because particles are driven towards an optimum rather than sampling around it, parameter uncertainties are more difficult to estimate in a fully probabilistic sense and several caveats apply. The inferred autocorrelation lengths do not represent probabilistic mixing rates in a strict statistical sense but serve as empirical indicators of dynamical stability within the optimization trajectory. Likewise, variance stabilization reflects convergence of swarm behaviour rather than a theoretical stationary distribution. Nevertheless, this hybrid variance–autocorrelation approach provides a consistent, data-driven framework to identify and discard transient behaviour, ensuring that uncertainty estimates reflect only the stable, well-converged portion of each PSO run. We naïvely assumed the models are perfect. In the current absence of benchmark binary measurements to test our accuracy, our uncertainties are thus only fitting uncertainties.

Finally, we ran MCMC on the final catalogue of $\sim$31,000 objects to provide a direct comparison of the uncertainties derived from our PSO methodology with a more rigorous, statistically orthodox approach. This enables us to validate our uncertainties and compare directly with \cite{traven_galah_2020}. We employed the \textsc{emcee} ensemble sampler \citep{foreman-mackey_emcee_2013} with $N_\mathrm{walkers} = 4\times N_\mathrm{dim} = 36$ walkers and 2000 production steps per walker. A burn-in phase of 500 steps was discarded prior to sampling; this comfortably exceeded the typical autocorrelation time of the chains. Walkers were initialised in a tight Gaussian ball centred on the PSO solution, with a standard deviation of $\sigma = 10^{-3}$ in the normalised parameter space. This ensures that walkers begin near the posterior peak whilst retaining sufficient spread for the differential-evolution proposal to generate diversity across the ensemble. Initialising with a large spread (e.g. $\sigma = 0.5$) was found to scatter walkers across the full prior volume, causing a significant fraction to land in undefined regions near the interpolator boundaries. 

We found that uncertainties determined through MCMC are $\sim$5–10 times smaller than those derived from our bespoke PSO-based approach. This indicates that the PSO method overestimates the true posterior width, most likely because particles retain residual momentum from the inertia term ($w = 0.7$) and do not fully converge into a tight cluster around the best-fit solution before the run terminates. Deriving reliable uncertainties from PSO therefore likely requires ~either a lower or adaptively decaying inertia weight, or a sufficiently extended run, to ensure that swarm particles have settled into the same localised region of parameter space around the solution.

\begin{figure*}
\centering
\includegraphics[width=\textwidth]{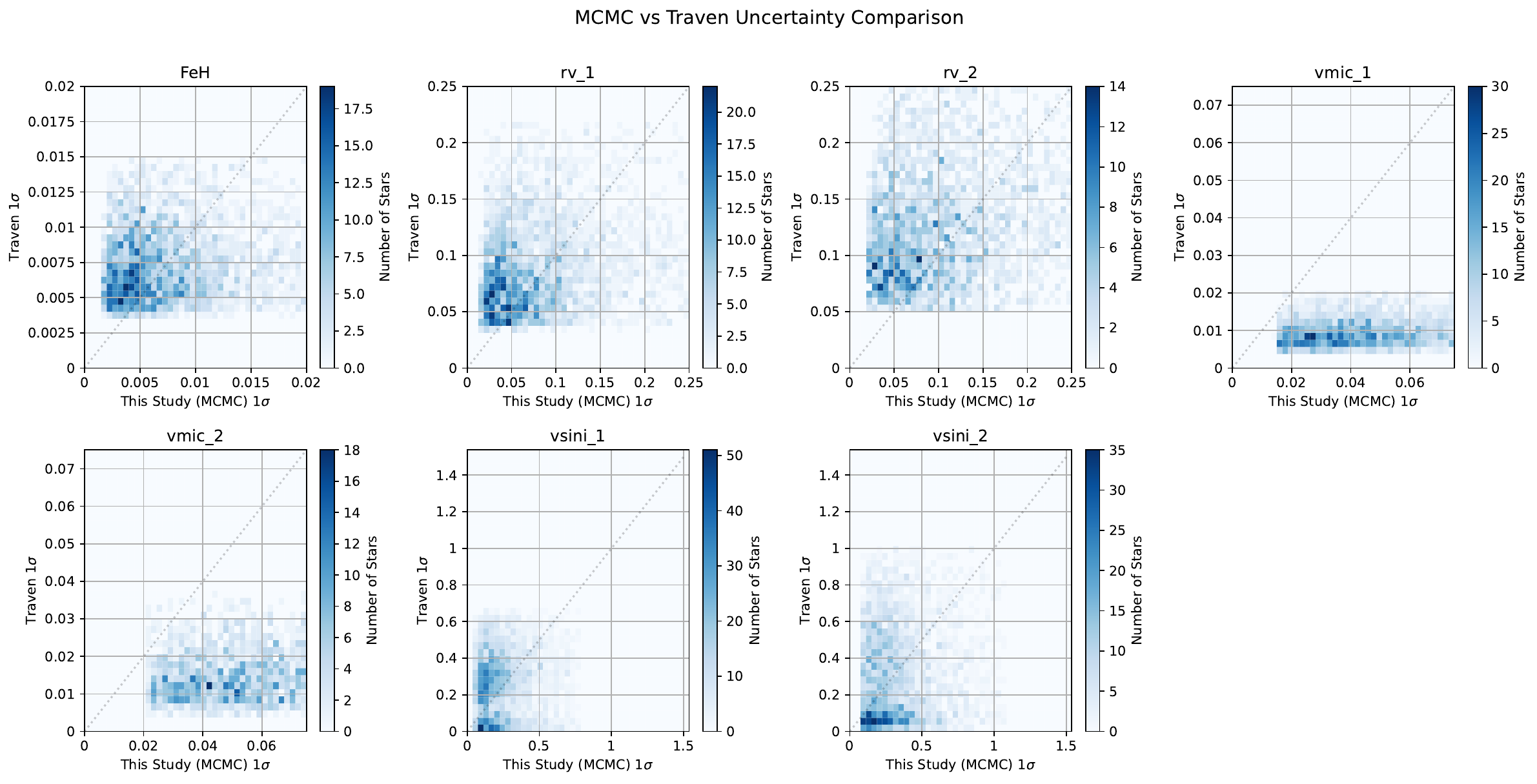}
\caption{
Two-dimensional histograms comparing the statistical uncertainty ($1\sigma$ spread) in each stellar parameter between this study and the values reported by \citet{traven_galah_2020} For each parameter, the uncertainty is defined as half the difference between the 16th and 84th percentiles of the posterior distribution, after removal of burn-in samples. Axes are matched in scale and bin size for direct comparison. A dotted one-to-one line indicates equal uncertainty estimates across methods. Colourbar shows the number of stars per bin.}
\label{fig:uncertainty_comparisons}
\end{figure*}

\section{Statistical Breakdown and Data Cuts} \label{sec:appendix-stats}

\begin{figure*}
\centering
\includegraphics[width=0.9\textwidth]{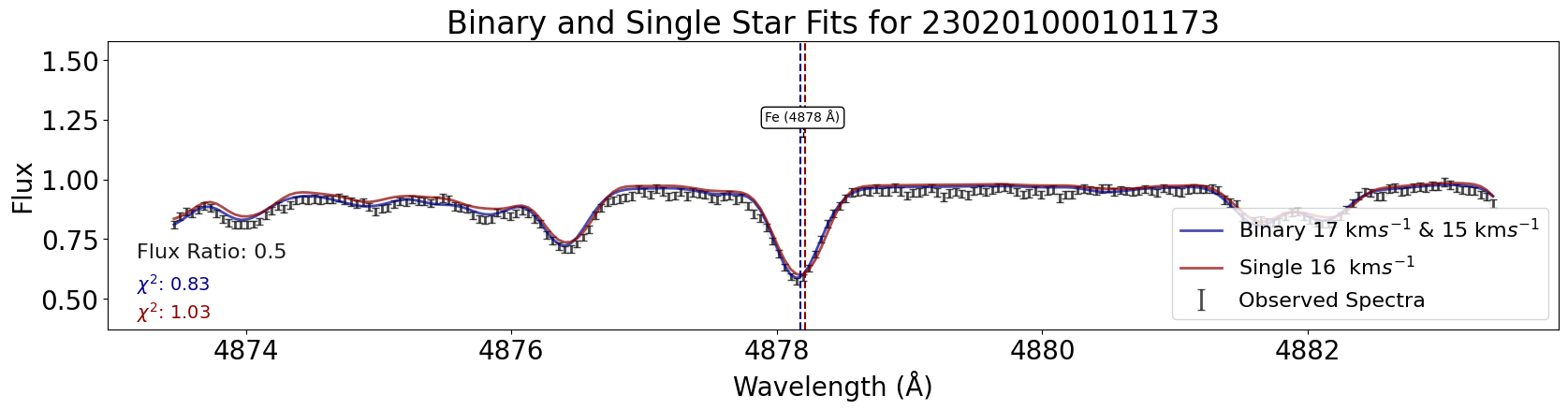}
\includegraphics[width=0.9\textwidth]{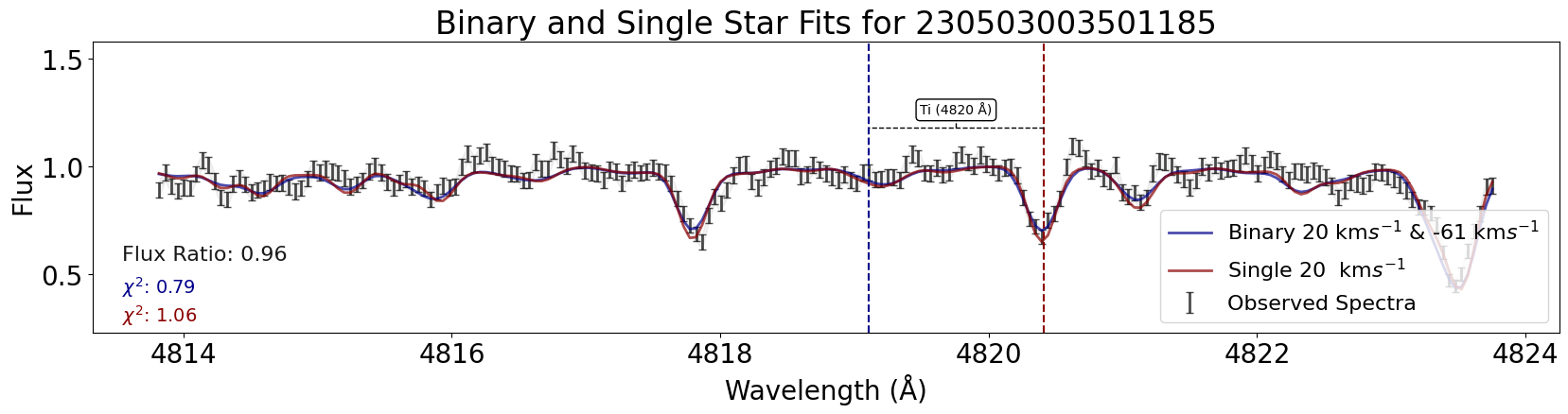} 
\caption{
Two examples of false binary detections that were removed by our data cuts. In both cases the same spectral line is shown shifted according to the radial velocities found by the analysis (red and blue lines). (\textbf{Top:}) The analysis identifies a best-fit solution with nearly equal radial velocities and flux ratios, indicating that the model is simply splitting a single star into two equally weighted components at the same velocity. This suggests that a single-star model would be more appropriate. While this could be an SB1 type binary, we have no evidence to support this and reject it as a possible false positive due to its $\Delta \textrm{RV}$ falling below the minimum threshold (Fig. \ref{fig:resolution_cut}). (\textbf{Bottom:}) A very high flux ratio ($>90\%$) is found, with one star dominating the flux; this also effectively corresponds to a single-star fit, as the secondary contributes minimally. In both cases, our binary fit yields a lower $\chi^2$ than the GALAH single-star fit, possibly because the secondary component can absorb spectral undulations caused by noise or unmodeled lines, or due to improvements in our fitting routine. 
}
\label{fig:binary_false_detections}
\end{figure*}


\begin{figure*}
\centering
\includegraphics[width=0.9\textwidth]{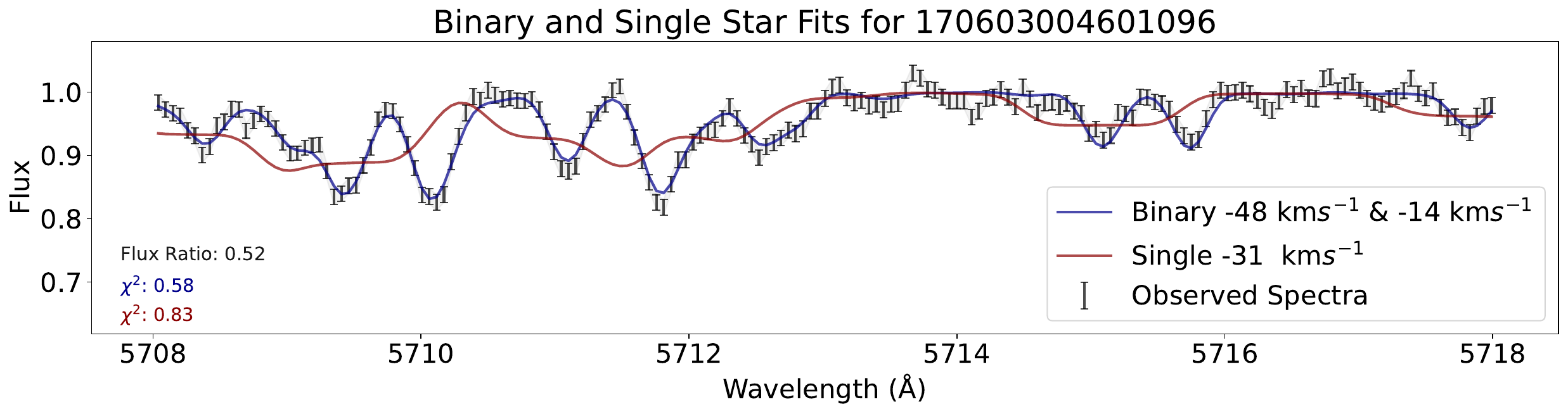}
\includegraphics[width=0.9\textwidth]{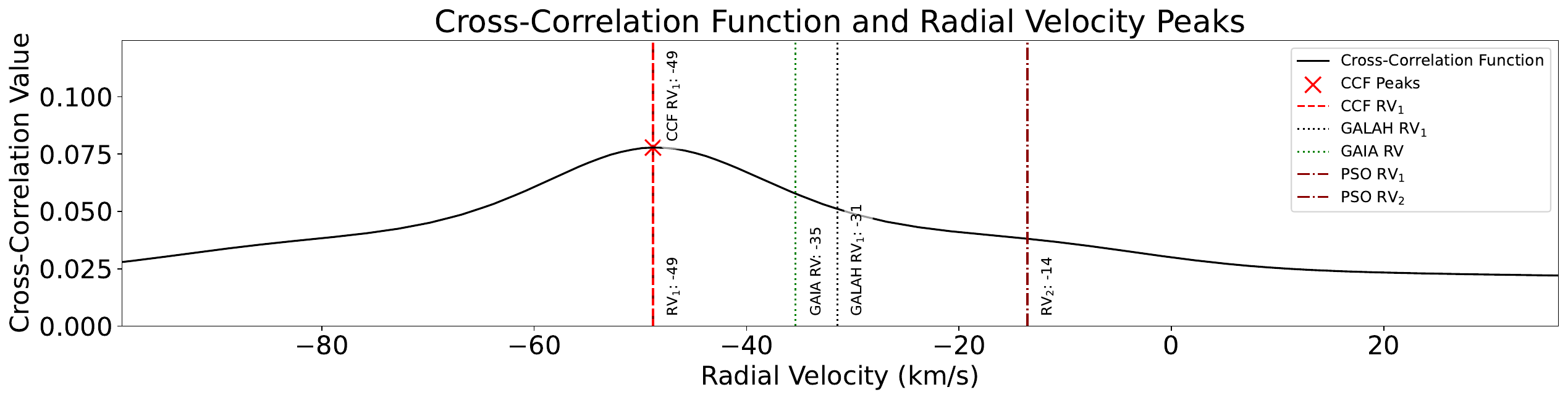}
\caption{
\textbf{Cross-correlation analysis failing to detect a binary system despite improved binary model fit.} \textbf{(Top)}~Observed spectrum (black) overlaid with single-star (red) and binary (blue) model fits. The binary model visibly provides a better fit, as indicated by a lower $\chi^2$ value. \textbf{(Bottom)}~Cross-correlation function (CCF) for the same spectrum. Only a single significant peak is identified via cross-correlation (red dashed line), while particle-swarm optimization detects two distinct radial velocity components (dark red dot-dashed lines -- CCF RV and PSO RV overlap). The radial velocity reported by both \citet{buder_galah_2025} and \citet{gaia_collaboration_gaia_2023} is shown as the grey dotted line, which lies almost exactly midway between the two component velocities. This illustrates how standard cross-correlation can miss binary signatures, even when a binary model fits the data substantially better. It also demonstrates how single-star model fits to blended spectra can yield a radial velocity close to the flux-weighted mean of the unresolved system’s components.
}
\label{fig:binary_CCF_comparison_fig}
\end{figure*}

In Table \ref{tab:cut_survival} we show the full statistic resulting from our data cuts on the full initial sample selection and the corresponding cross-matched objects from \citep{traven_galah_2020} and \textit{Gaia}. The full data cut statistics applied to each selection criteria are given in Table~S1 of the online supplementary material.

As noted in the main text, we adopt the improvement fraction $f_{\rm imp}$ as defined by \cite{el-badry_discovery_2018}:
\begin{equation}
f_{\rm imp} = \frac{\displaystyle\sum \frac{|s_{\rm mod} - s_{\rm ob}| - |b_{\rm mod} - s_{\rm ob}|}{u_{\rm ob}}}{\displaystyle\sum \frac{|s_{\rm mod} - b_{\rm mod}|}{u_{\rm ob}}}
\label{eq:fimp_original}
\end{equation}
with one modification: we replace the denominator, which in the original formulation is the summed weighted difference between the single-star and binary models, with the summed weighted single-star residuals. That is, we normalise by $\sum{|s_{\rm mod} - s_{\rm ob}|/u_{\rm ob}}$ rather than $\sum{|s_{\rm mod} - b_{\rm mod}|/u_{\rm ob}}$. This ensures that $f_{\rm imp}$ expresses the improvement of the binary fit as a fraction of the original single-star fit quality, and avoids spuriously large values in cases where the two models are nearly identical but both fit the data poorly. This relation overlaid on our data is shown in Fig.~S1 of the online supplementary material.

\begin{table*}
    \centering
    \caption{Object counts at each stage of the data cuts. \textit{Top:} shown for
             the full sample, the \citet{traven_galah_2020} cross-matched subsample,
             and the \textit{Gaia} cross-matched subsample. \textit{Bottom:} shown
             for all analysed objects (no selection criterion). Each row corresponds
             to a sequential cut applied to the surviving sample from the previous
             step. ``Uniquely removed'' denotes objects removed by a given cut that
             would not have been removed by any subsequent cut. }
    \label{tab:cut_survival}
    \renewcommand{\arraystretch}{1.2}
    %
    \begin{subtable}[t]{\textwidth}
        \centering
        \textbf{Full Sample, Traven Cross-match, and \textit{Gaia} Cross-match}\\[4pt]
        \begin{tabular}{l rrr rrr rrr}
            \hline
            \textbf{Cut}
                & \multicolumn{3}{c}{\textbf{Full Sample}}
                & \multicolumn{3}{c}{\textbf{Traven Cross-match}}
                & \multicolumn{3}{c}{\textbf{\textit{Gaia} Cross-match}} \\
            \cmidrule(lr){2-4} \cmidrule(lr){5-7} \cmidrule(lr){8-10}
                & Before & Removed & Remaining
                & Before & Removed & Remaining
                & Before & Removed & Remaining \\
            \hline
            $\tilde{\chi}^2$ quality      & 263{,}785 & 11{,}062 & 252{,}723 & 12{,}101 & 28    & 12{,}073 & 3{,}190 & 23  & 3{,}167 \\
            Flux contribution             & 252{,}723 & 75{,}397 & 177{,}326 & 12{,}073 & 725   & 11{,}348 & 3{,}167 & 103 & 3{,}064 \\
            $T_{\mathrm{eff}}$ cut-off    & 177{,}326 & 21{,}664 & 155{,}662 & 11{,}348 & 135   & 11{,}213 & 3{,}064 & 509 & 2{,}555 \\
            Signal-to-noise               & 155{,}662 & 34{,}400 & 121{,}262 & 11{,}213 & 627   & 10{,}586 & 2{,}555 & 275 & 2{,}280 \\
            Single-star $\tilde{\chi}^2$  & 121{,}262 & 93{,}530 &  27{,}732 & 10{,}586 & 4{,}981 & 5{,}605 & 2{,}280 & 1{,}299 &   981 \\
            $v\sin i$--$\Delta$RV         &  27{,}732 &  8{,}458 &  19{,}274 &  5{,}605 &   656   & 4{,}949 &   981   &    29   &   952 \\
            $f_{\rm imp}$                 &  19{,}274 &  6{,}363 &  12{,}911 &  4{,}949 &   216   & 4{,}733 &   952   &   487   &   465 \\
            \hline
        \end{tabular}
    \end{subtable}

    \vspace{1em}
    %
    \begin{subtable}[t]{0.48\textwidth}
        \centering
        \textbf{All Analysed Objects}\\[2pt]
        $N_{\mathrm{analysed}} = 263{,}785$\\[4pt]
        \begin{tabular}{l rrrr}
            \hline
            \textbf{Cut} & \textbf{Before} & \textbf{Removed} & \textbf{Remaining} & \textbf{Unique} \\
            \hline
            $\tilde{\chi}^2$ quality     & 263{,}785 & 11{,}062 & 252{,}723 &      24  \\
            Flux contribution            & 252{,}723 & 75{,}397 & 177{,}326 &     838  \\
            $T_{\mathrm{eff}}$ cut-off   & 177{,}326 & 21{,}664 & 155{,}662 &   1{,}050 \\
            Signal-to-noise              & 155{,}662 & 34{,}400 & 121{,}262 &   3{,}210 \\
            Single-star $\tilde{\chi}^2$ & 121{,}262 & 93{,}530 &  27{,}732 &   3{,}764 \\
            $v\sin i$--$\Delta$RV        &  27{,}732 &  8{,}458 &  19{,}274 &   1{,}882 \\
            $f_{\rm imp}$                &  19{,}274 &  6{,}363 &  12{,}911 &   6{,}363 \\
            \hline
        \end{tabular}
    \end{subtable}
    \renewcommand{\arraystretch}{1}
\end{table*}

\section{The \texorpdfstring{$v\sin i$}{v sin i} RV limit} \label{sec:example_spectra}
\begin{figure*}
\centering
  \includegraphics[width=0.9\textwidth]{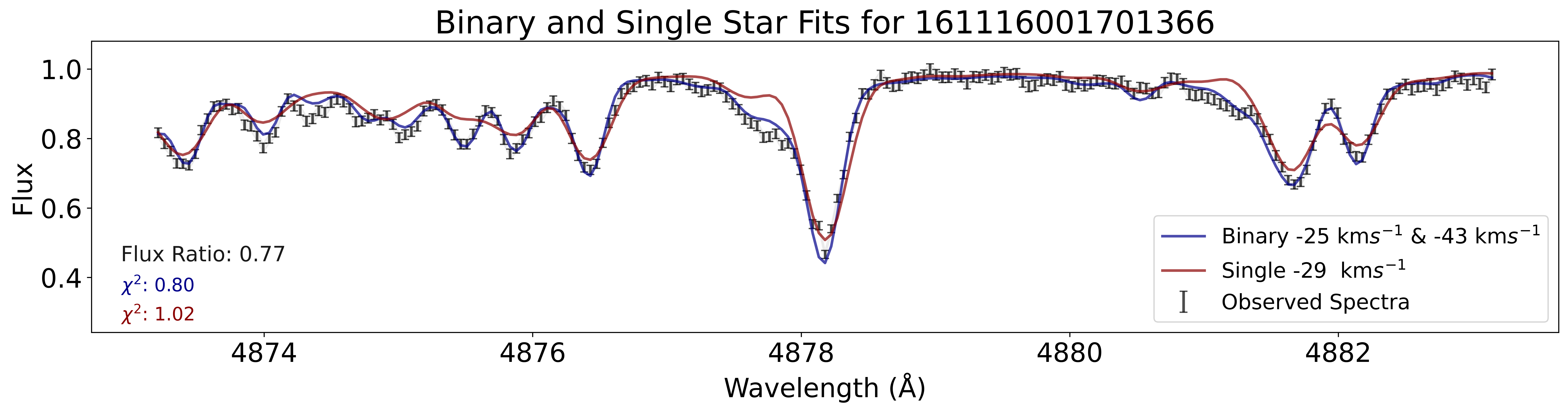}

  \includegraphics[width=0.9\textwidth]{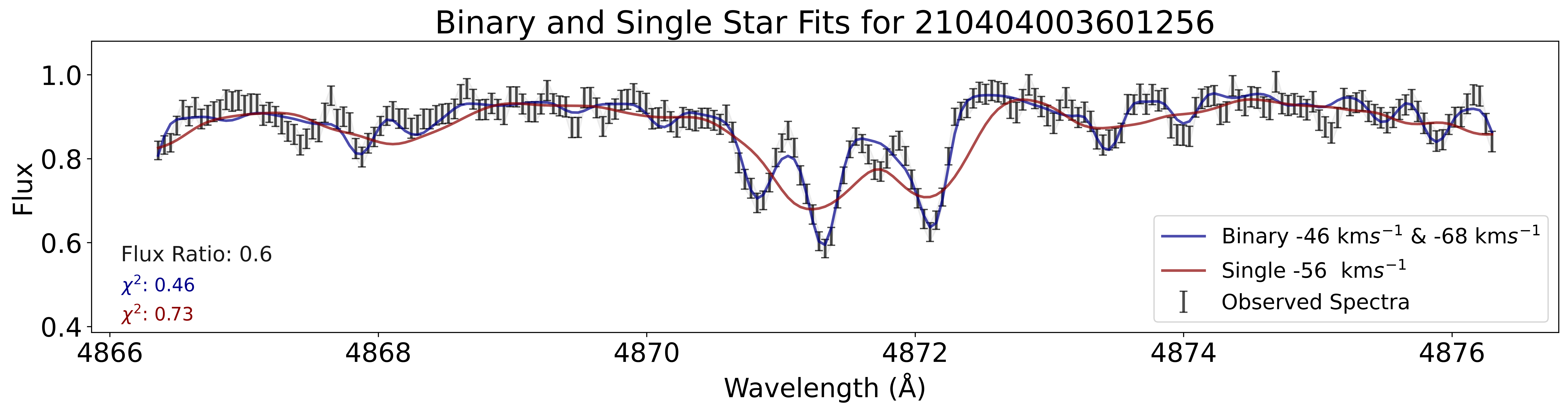}

  \includegraphics[width=0.9\textwidth]{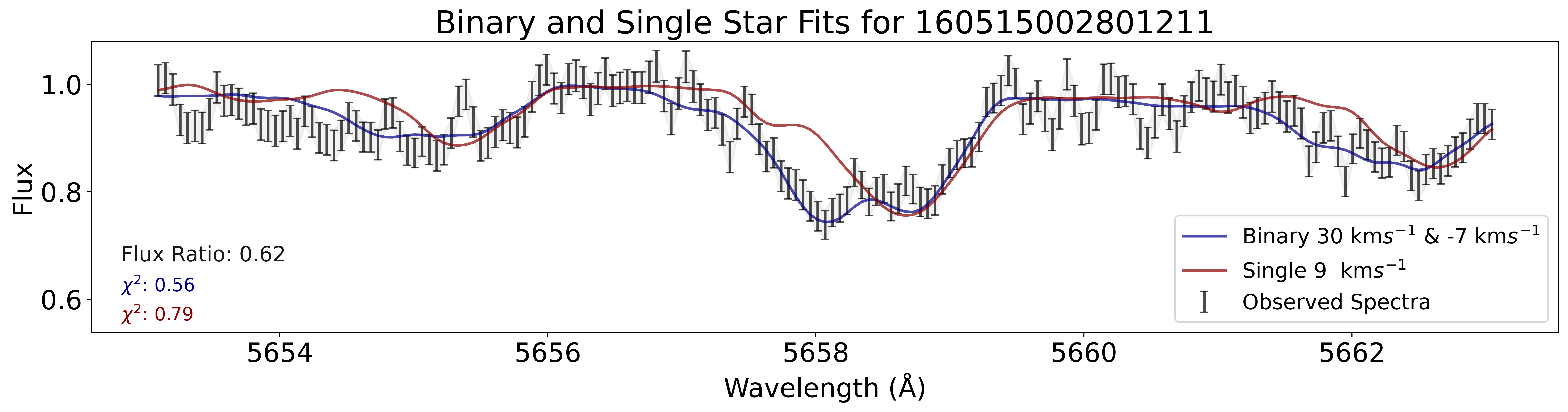}
  \caption{Example spectra along the detection threshold shown in Fig.~\ref{fig:resolution_cut}. Spectra appear in order of increasing combined $v\sin i$ from top to bottom, with the uppermost plot corresponding to point 1 in Fig.~\ref{fig:resolution_cut} and the bottommost corresponding to point 3.}
  \label{fig:examples_spec}
\end{figure*}

In Fig.~\ref{fig:examples_spec}, we show examples of spectra located along the threshold we define according to $f(\Delta \text{RV}) = 7.62\,e^{0.0775x} - 15$, above which we assume binary spectra to be adequately resolved (as illustrated in Fig.~\ref{fig:resolution_cut}).

The cyan feature in Fig.~\ref{fig:resolution_cut} reveals a cluster of objects with high $v\sin i$ and $\Delta\mathrm{RV}$ between 30--60\kms. We noted that these are present in the sample due to a variety of reasons in the main text (e.g. hotter stars rotate faster, binaries are more common amongst hotter stars, and the parent survey being magnitude limited). When examining the components of these systems we also see the analysis pipeline approaching the $v\sin i$ limit of the neural network (Fig.~S2 of the online supplementary material). Further, when examining the \vsini distribution of the final catalogue in both the resolved and unresolved spectra we note that in all cases we see the majority of objects approaching the limits of the neural network (Fig.~S3 of the online supplementary material). We also examined the cyan region for spectra with repeat observations. Of the 21,380 objects (before cuts) within the region, there are 20,235 unique objects. Of these, 305 are objects with more than one observation. We also found 884 cross-matched \textit{Gaia} binaries and 535 cross-matched from \cite{traven_galah_2020} within this region. 

We therefore assume these data to suggest a pipeline artifact rather than a genuine astrophysical over-density. Objects whose true $v \sin i$ exceeds the reliable range of the neural network are assigned values at or near the network's upper boundary, causing an artificial pile-up that would otherwise be spread across a broader range of $v \sin i$.  The cross-matches within this region support a substantial fraction as being genuine binary systems, suggesting that the cyan region preferentially selects real binaries whose rotational broadening, intrinsic or induced by unresolved multiplicity, exceeds what the pipeline can reliably characterise.

We furthermore attach the comparison of estimates from repeat observation in Fig.~S4 of the online supplementary material.

\end{appendix}

\label{lastpage}
\end{document}